\documentclass{aastex63}

\usepackage{graphicx}
\graphicspath{ {./Fractal/} }
\usepackage{amsmath}	
\usepackage{amssymb}
\newcolumntype{P}[1]{>{\centering\arraybackslash}p{#1}}
\usepackage{longtable}
\usepackage{lipsum}
\usepackage{color}
\usepackage{xcolor}

\accepted{}

\submitjournal{Icarus}

\shorttitle{Exploring Super-Earth Surfaces}
\shortauthors{Modirrousta-Galian et al.}

\begin{document}
	
	\title{Exploring Super-Earth Surfaces: Albedo of Near-Airless Magma Ocean Planets and Topography}

	\correspondingauthor{Darius Modirrousta-Galian}
	\email{darius.modirrousta@inaf.it}
	
	\author[0000-0001-6425-9415]{Darius Modirrousta-Galian}
	\affiliation{INAF – Osservatorio Astronomico di Palermo, Piazza del Parlamento 1, I-90134 Palermo, Italy}
	\affiliation{University of Palermo, Department of Physics and Chemistry, Via Archirafi 36, Palermo, Italy}
	
	\author{Yuichi Ito}
	\affiliation{Department of Physics \& Astronomy, University College London, Gower Street, WC1E 6BT London, United Kingdom}
	
	\author{Giuseppina Micela}
	\affiliation{INAF – Osservatorio Astronomico di Palermo, Piazza del Parlamento 1, I-90134 Palermo, Italy}
	
	\begin{abstract}
	In this paper we propose an analytic function for the spherical albedo values of airless and near-airless magma ocean planets (AMOPs). We generated 2-D fractal surfaces with varying compositions onto which we individually threw 10,000 light rays. Using an approximate form of the Fresnel equations we measured how much of the incident light was reflected. Having repeated this algorithm on varying surface roughnesses we find the spherical albedo as a function of the Hurst exponent, the geochemical composition of the magma, and the wavelength. As a proof of concept, we used our model on Kepler-10b to demonstrate the applicability of our approach. We present the spherical albedo values produced from different lava compositions and multiple tests that can be applied to observational data in order to determine their characteristics. Currently, there is a strong degeneracy in the surface composition of AMOPs due to the large uncertainties in their measured spherical albedos. In spite of this, when applied to Kepler-10b we show that its high albedo could be caused by a moderately wavy ocean that is rich in oxidised metallic species such as FeO, $\rm Fe_{2}O_{3}$, $\rm Fe_{3}O_{4}$. This would imply that Kepler-10b is a coreless or near-coreless body.
	\end{abstract}

\keywords{Extra-solar planets --- Terrestrial planets --- Volcanism --- Interiors --- Spectrophotometry}

\section{Introduction}

The inferred high spherical albedo values of several hot super-Earths \citep{Demory2014,Malavolta2018} are well  above what the current literature predicts for magmatic surfaces \citep{Essack2020,Rouan2011,Edgett1997}. Two main explanations have been put forward: reflective exotic magmas \citep{Rouan2011} and reflective volcanic/mineral atmospheres \citep{Hamano2015,Pluriel2019}. However, it is currently not possible to verify which mechanism applies to specific super-Earths due to observational data being either limited or degenerate. Generally, constraining the properties of super-Earth exoplanets is troublesome due to their small sizes \citep{Valencia2013,Dorn2017}. In spite of this, advances in mass-radius measurements and superior spectroscopic analyses of their atmospheres \citep[e.g.][]{Tsiaras2016,Esteves2017} has lowered their compositional degeneracies. These advancements have permitted researchers to apply geological models to specific super-Earths in order to better understand their structures \citep[e.g.][]{Zeng2013,Zeng2016,Zeng2018}. However, one aspect that remains unexplored is the surface composition and topology of these bodies. Directly analysing the geomorphology of extra-solar planets is not possible due to the observational limitations set by the \textit{Rayleigh Criterion}. For example, if one wanted to resolve a feature that occupies $10\%$ of a planetary surface (e.g. an ocean) on an Earth-sized exoplanet located $5~\rm pc$ away, a spatial resolution of a few microarcseconds would be required. However, this is beyond our present capabilities which is one of the reasons why it is troublesome to decipher which mechanisms lead to large albedos on molten super-Earths such as Kepler-10b, Kepler-21b and K2-141b \citep{Rouan2011,Demory2014,Malavolta2018}. 

The purpose of our study is to investigate the spherical albedo values of airless and near-airless, molten super-Earths with a particular focus on its dependence to the composition and ocean waviness. As this study is a first stepping-stone to theoretically estimate the albedo of these ultra-hot exoplanets, we had to set some simplifications (see Sec.~\ref{sec:limitations}) and left some tasks for future works (see Sec.~\ref{sec:future_work}). We show that the ocean's roughness will lead to a higher degeneracy in the interpretation of the spherical albedo values of these systems. The close-orbiting super-Earths relevant to our study will be tidally locked therefore containing a molten day-side and a cold night-side. Due to the high day-side temperatures, a vaporised mineral atmosphere will form that rapidly traverses into the night-side where the temperatures are low enough for the minerals to condense \citep{Schaefer2012,Ito2015,Kite2016}. Whilst the atmosphere is expected to be generally optically thin in most of the infrared spectral range, the rapidly expanding gas would carry enough kinetic energy to trigger magma ocean waves (see Sec.~\ref{sec:hurst} of the appendix and \citet{Kite2016}). In Sec.~\ref{sec:choice_of_planet} we justify our choice of using Kepler-10b as an analogue planet for our spherical albedo model. In Sec.~\ref{sec:generate_surfaces} we model the magma ocean waves using fractal mathematics. In Sec.~\ref{sec:atmospheric_effects} we explain that while mineral atmospheres may affect the spherical albedo values within the spectral range of the \textit{Kepler} telescope, for future missions with larger wavelength ranges, the effects would be negligible. In Sec.~\ref{sec:Results} we show the results of our simulations and we present an analytic approximation for the planar albedo, which we then integrate across all angles to determine the spherical albedo. Our analytic functions model the spherical albedo values of molten, airless magma ocean super-Earths for different roughnesses and surface compositions. Finally, in Sec.~\ref{sec:Discussion} and \ref{sec:conclusion} we discuss and present our results respectively.

Note that many of the terms used within this manuscript to describe the reflective properties of super-Earths are similar but not identical so it is necessary to define them to avoid confusion. The reflectance, $R$, is a directional quantity that depends on the incident and emerging angles, as well as the wavelength of the light and the composition (via the refractive indices) and roughness of the surface. The spherical albedo, $A_{S}$, is the total fraction of incident radiance scattered by a body into all direction. The planar (or plane) albedo, $A_{pla}$, is the spherical albedo of a planar surface of the planet. Finally, the bolometric Bond albedo, $A_{B}$, is the spherical albedo integrated across all wavelengths.

\section{Our Choice of Planet}
\label{sec:choice_of_planet}

\begin{figure}[h]
	\centering
	\includegraphics[scale=0.8]{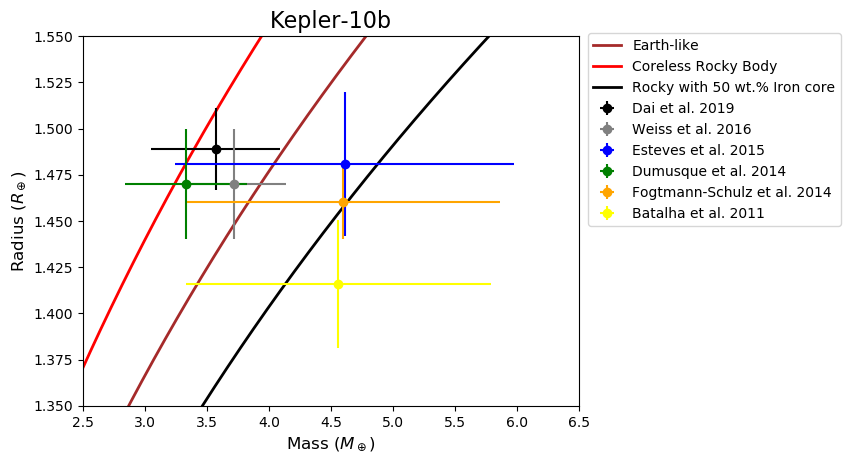}
	\caption{The mass and radius of Kepler-10b according to different studies. The planetary composition model curves are from \citet{Zeng2013} and \citet{Zeng2016}. The variations in the measurements are due to a combination of uncertainties in the stellar parameters and using different observational raw data.}
	\centering
	\label{fig:kepler10b}
\end{figure}

Throughout the rest of this manuscript we will use Kepler-10b as an analogue for molten, airless super-Earths within our spherical albedo models. We do this for two reasons: first we want to input realistic parameters into our simulations instead of speculative idealised ones. Secondly, we want to show how our analytic function for the spherical albedo values can be applied with ease to any molten, airless super-Earth of choice. Therefore, before we progress into an explanation of our models we would first like to justify our choice of using Kepler-10b as our test planet. We believe this body is an adequate candidate for our method because most mass and radius measurements are consistent with a rocky composition \citep[e.g.][see Fig.~\ref{fig:kepler10b}]{Batalha2011,Fogtmann2014,Dumusque2014,Esteves2015,Weiss2016,Dai2019}.
Furthermore, the strong temperature contrast inferred from the observed phase curve between the day-side temperature of $\simeq 2750$~K and night-side temperature $\simeq 50$~K implies that a thick atmosphere is improbable \citep{Rouan2011}. In addition, the large thermal gradient is suggestive that Kepler-10b is tidally locked. Hence, the vaporised mineral gases formed on the day-side \citep{Schaefer2012,Ito2015} will rapidly expand into the night-side and condense, forming waves in the magma ocean beneath as they travel \citep{Kite2016}. Whilst there is still an ambiguity in the measured data, in this paper we focus on the scenario where Kepler-10b is a rocky exoplanet hosting a fully magmatic day-side that is lacking a significant atmosphere. Currently, the best fit value for the bolometric Bond albedo is $0.48 \pm 0.35$ \citep{Rouan2011}, which is the one we will reference throughout this paper.
This study investigates the spherical albedo values for various magma ocean compositions and surface roughnesses. The spherical albedo is useful to consider because it is defined to be a monochromatic property, so it is independent of the irradiation spectrum of the host star. From our spherical albedo values and an estimation of the stellar spectra, the bolometric Bond albedo values could be inferred \citep[see][for an example of a similar concept but for planets with atmospheres]{Marley1999}. It is also important to note that the primary objective of this study is to present our model for constraining the composition of exoplanetary magma oceans; we are therefore not focusing on formational or observational limitations.

\section{Models}
\subsection{Generating our Magma Ocean Fractal Surfaces}
\label{sec:generate_surfaces}

One commonly used, although incomplete, definition of a fractal was given by Benoit Mandelbrot as a shape whose fractal dimension (also called Hausdorff dimension) is greater than its topological dimension. The fractal dimension is a measure of the geometric complexity of a system so that from a geological perspective, this parameter would provide information on the roughness of a surface. For instance, Earth’s oceans have calculated fractal dimensions of $\approx2.3$ \citep{Stiassnie1991,Stiassnie1991b} which means that due to surface inhomogeneities Earth’s true surface area could be several orders of magnitude greater than what is predicted from a perfect spherical model. Similar analyses have also been performed on the fractal-nature of Mars \citep{Deliege2017,Demin2017} and Venus \citep{Demin2018}. Roughness has also been shown to affect the reflective properties of non-fluid surfaces with coarser systems having lower ones than their smoother counterparts \citep[e.g.][]{Bennett1961}. This is because with a rough surface photons can get trapped in crevices and consequently absorbed. In contrast, with smooth surfaces photons are absorbed less as the albedo values are only dependent on the material properties. For example, soil \citep{Matthias2000} and ice \citep{Lhermitte2014} have been found to be almost $\sim 50\%$ less reflective than their standard values when they are rough. This is explained by a difference in the phase function of a rough surface and that of a smooth one with the same composition. In addition, several studies have found strong linear relationships between soil's albedo values and its roughness \citep{Oguntunde2006,Cierniewski2013}. In contrast, when dealing with fluids their bolometric Bond albedo values do not scale proportionally to the roughness of the system \citep{Haltrin2001}. This is because liquid waves are topologically distinct from the types of concavities one would find in solids (e.g. cracks, holes etc.). Waves are not as efficient in trapping photons so other effects such as the angle of incidence of the light ray relative to the immediate surface, and the number of internal reflections are important as they determine how much energy is lost. Since this cannot be modelled analytically, in this paper we perform ray tracing simulations in order to determine the reflective properties of synthetic magma ocean surfaces. 

\begin{figure}[h]
	\centering
	\includegraphics[scale=0.70]{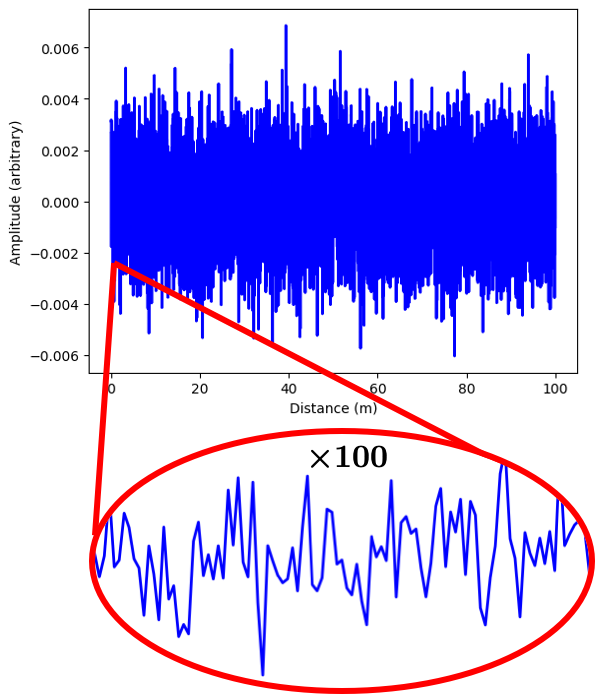}
	\caption{The unsmoothed fractal surface of an ocean whose wave heights are yet to be adjusted for. This surface was generated for a Hurst exponent of 0.7. The circled area is a magnification of a $\times 100$ for the region $0-1~\rm m$}
	\centering
	\label{fig:fractalsurfacesharp}
\end{figure}

\begin{figure}[h]
	\centering
	\includegraphics[scale=0.80]{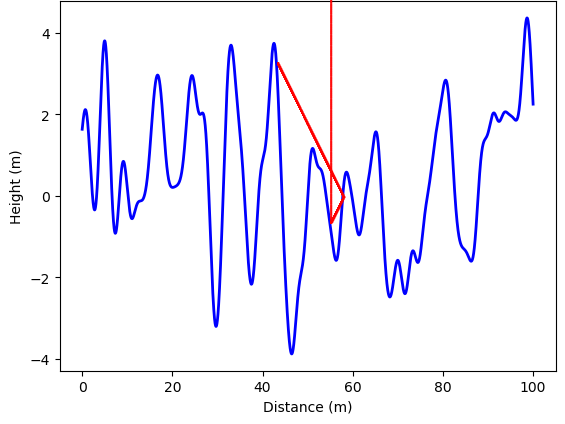}
	\caption{An example of a single light ray (red line) interacting with a smoothed height-corrected synthetic magma ocean surface. Take note that the x-axis and y-axis do not have the same scale so the bouncing angle may seem distorted.}
	\centering
	\label{fig:fractalbouncing}
\end{figure}

To keep the simulations inexpensive we will use Schlick's approximation, Eq.~\ref{eq:Schlicks}, for the reflectance at a liquid boundary \citep{Schlick1994}. Schlick's approximation is not an exact solution, but rather a method of keeping the ray tracing simulations inexpensive, whilst still maintaining physical accuracy. This method only works when the ray of light is traversing from a medium with a low refractive index to that of a higher one, which is relevant to our model. We are aware that this approximation will inevitably result in a decrease of accuracy but we unfortunately had to make this compromise in order to cover the large parameter space within a reasonable time-frame. In Sec.~\ref{sec:comparison_of_models} of the appendix we compare Schlick's approximation with other models and the exact Fresnel equations in order to quantify their differences. We define the parameter $n_{sy}$ as the synthetic refractive index of the magma, which accounts for the effects of the real ($n$) and imaginary ($k$) indices. The derivation of the synthetic refractive index, $n_{sy}$, is shown in Sec.~\ref{sec:derivation_n_sy} of the appendix. Therefore, $n_{sy}$ is dependent on the geochemical composition of the medium and the wavelength of the light as shown by the best-fit equations listed in Table~\ref{tab:bestfitequations}:
\begin{subequations}
	\begin{equation}
	R\left(\theta \right) \approx R_{0} + \left( 1 - R_{0} \right) \left(1 - \cos{\theta} \right)^{5}
	\label{eq:Schlicks}
	\end{equation}
	\begin{equation}
	R_{0} = \left(\dfrac{n_{sy} - 1}{n_{sy} + 1} \right)^{2} 
	\label{eq:R0}
	\end{equation}
	\begin{equation}
	\ n_{sy} = \frac{ \left(\frac{ \left|n - 1 \right|^{2} + \left|k \right|^{2} }{ \left|n + 1 \right|^{2} + \left|k \right|^{2} } \right)^{1/2} +1 }{1 - \left(\frac{ \left|n - 1 \right|^{2} + \left|k \right|^{2} }{ \left|n + 1 \right|^{2} + \left|k \right|^{2} } \right)^{1/2} }
	\label{eq:nsynth}
	\end{equation}
\end{subequations}
Where $ \theta $ is the angle of incidence and $R$ is the reflectance. Having chosen our equations, the next step was to model the fractal ocean surfaces. We used the \textit{fbm 0.3.0} python package with the \citet{Davies1987} fractional Gaussian noise (fGn) method. This technique is theoretically exact in generating discretely sampled fGn numerical values for a given fractal dimension. A thorough explanation of the Davies and Harte method can be found in page 412 of \citet{Wood1994}. We chose to model the magma ocean surfaces as a collection of different points with elevations given by fGn in order to approximate the random motions induced from the vapourised atmospheric winds (see Sec.~\ref{sec:hurst} of the appendix for a mathematical relationship of the wind energies and the surface roughness). However, this approach assumes that each point is independent of its neighbours which would only be true for a zero viscosity fluid (i.e. a superfluid) that also lacks surface tension. Clearly magmas have viscosities so the non-continuous, sharp terrains are unrealistic. Fig.~\ref{fig:fractalsurfacesharp} is an example of a pre-smoothed, unrealistic surface.
In order to smooth these surfaces we applied a Gaussian filter with a standard deviation of 100 pixels (equivalent to 1~m) given by the \textit{Astropy} package \citep{Astropy2013,Astropy2018}. The process of smoothing the surface is analogous to adding a viscosity which, by definition, will act against the fluid flow. We chose to smooth out perturbations smaller than 1~m as this closely matches the waves we observe on Earth's oceans (see Sec.~\ref{sec:limitations} for a justification of why the waves of Earth's oceans are an appropriate analogue for the roughness of the magma ocean on Kepler-10b and other ultra-hot super-Earths). From a mathematical perspective, the act of smoothing perturbations smaller than $1~\rm m$ means that fractal behaviour seizes at very small scales which is expected for natural entities instead of idealised mathematical constructs. It must be noted that maximising the Gaussian filter will produce a smooth surface so that the planar albedo will be given by Eq.~\ref{eq:R0}. If instead no filter is applied then the planar albedo will converge to 1 as the individual waves will be very thin and tall (i.e. the waves will be steep so the angle $\theta$ will approach $90^{\circ}$; see Fig.~\ref{fig:fractalsurfacesharp}). These two scenarios are unrealistic as they require extreme viscosities which is unlikely given the thermodynamic conditions and magma compositions considered in this study. Concerning with the average wave peak amplitude, this required finding a relationship between the Hurst exponent, which measures the coherence of the surface, and the root mean squared (RMS) roughness. Experimental results from \citet{Durst2011} found that the fractal dimension and the RMS were related by the following equation:
\begin{equation}
RMS \approx h_{0}\left(F - 2 \right) 
\label{eq:Durst}
\end{equation}
Where $h_{0}$ is a height constant and $F$ is the fractal dimension. For self-similar processes where local and global properties are interlinked, the fractal dimension is related to the Hurst exponent by $F = 3 - H$. Therefore \textit{H} can be substituted in:
\begin{equation}
RMS \approx h_{0}\left(1 - H \right) 
\label{eq:DurstH}
\end{equation}
We decided to work with the Hurst exponent ($H$) instead of the fractal dimension ($F$) in order to make our code easier to adopt. This is because the Hurst exponent is independent of the topological dimension, whilst the fractal dimension is not. Therefore, if one wanted to implement our code into a 1-D, 2-D or 3-D radiative transfer model, the same Hurst exponent could be used resulting in the same spherical albedo value for each scenario. The constant $h_{0}$ is calculated experimentally, which can be done using Earth's oceans as an analogue. We know that these have a Hurst exponent of $\approx 0.7$ \citep{Stiassnie1991,Stiassnie1991b} with a root mean squared average waveheight of $\approx 1.5$~m \citep{MetOffice2010}. Consequently, solving Eq.~\ref{eq:DurstH} for $H = 0.7$ and $\rm RMS = 1.5$~m gives $h_{0} \approx 5$~m, yielding the final form of our equation:
\begin{equation}
	RMS \approx 5\left(1 - H \right) 
\label{eq:DurstH2}
\end{equation}
In order to correct for the height of our magma ocean waves we implemented Eq.~\ref{eq:DurstH2}. This was done by taking the smoothed (i.e. the Gaussian filter had been applied), unadjusted surface and finding the average (absolute) height. The RMS value (calculated with Eq.~\ref{eq:DurstH2}) was then divided by the average height of the smoothed surface to find the average correction factor. Finally, we multiplied the average correction factor to each point in the smoothed, unadjusted surface in order to get our more realistic magma ocean surface as shown by Fig.~\ref{fig:fractalbouncing}. Note that we are assuming that the parameter $h_{0}$ is constant and equal for the Earth and Kepler-10b. This assumption has been made because the viscosity of magma at very high temperatures is fortuitously similar to that of water. We explain this point more thoroughly in Sec.~\ref{sec:limitations}.

Once we had the fractal surface generator we created a basic simulator that would throw 10,000 light rays whose initial locations were uniformly sampled. Each of these light rays interacted specularly with the surface according to the approximate Fresnel equations (\ref{eq:Schlicks},\ref{eq:R0}) and the refractive index equations listed in Table~\ref{tab:bestfitequations}. By recording how much of the incident light was reflected we constrained the planar albedo of the surfaces for different compositions and roughnesses. We set the boundary condition that if a light ray lost more than $99.99\%$ of its strength we stopped any further interactions and assumed it was fully absorbed. This was done in order to increase the efficiency of our code.

Fig.~\ref{fig:fractalbouncing} shows an example of a light ray reflecting specularly on our synthetic smoothed, height-corrected magma oceans surface. The bouncing angles may seem distorted because of the scale difference between the x- and y-axes. Fig.~\ref{fig:fractalbouncing} is also a good example of how a light ray may get trapped in a crevice resulting in it bouncing multiple times so a large portion of its original intensity is lost through absorption.
With the simulator ready we began taking measurements. For each synthetic refractive index ($n_{sy}$) ranging from $1.5-6.0$ we generated multiple fractal terraines for Hurst exponents ranging from $0.95 \: \rm (very \: smooth) \rightarrow 0.01 \: (very \: rough)$. For each Hurst exponent value we simulated 10 morphologically different fractal surfaces where in each case we threw 10,000 light rays and recorded the average reflected component to estimate the planar albedo. However, planets are spherical so the planar albedo could not accurately describe their reflective properties (especially at high latitudes). Because of this, it is necessary to convert the planar albedo into the spherical albedo. By definition, the spherical albedo is the ratio of the monochromatic power reflected by a planet to the monochromatic power incident upon that planet, which is given as follows:
\begin{subequations}
\begin{equation}
    A_{S} = \frac{P_{refl}(\lambda)}{P_{in}(\lambda)}
\end{equation}
Where $P_{refl}$ and $P_{in}$ are the reflected and incident monochromatic powers respectively. The power incident upon the planet is given by:
\begin{equation}
\label{eq:P_in}
    P_{in} = F_{in}(\lambda) \cdot \pi R^{2}_{p}
\end{equation}
Where $F_{in}(\lambda)$ is the monochromatic energy flux and $R_{p}$ is the planetary radius. The power reflected by the planet is more complicated to calculate as one must determine how much light is reflected by each area segment of the planet, as a function of the sub-stellar latitude (or longitude, as for tidally locked planets they are the same). The equation goes as follows:
\begin{equation}
\label{eq:Power_reflected}
    P_{refl} = \int^{\pi/2}_{0} F_{in}(\lambda) \cos{\phi} \times q(\phi) \times 2 \pi R^{2}_{p} \sin{\phi} ~ d\phi
\end{equation}
Where the first section is the monochromatic energy flux received from the star at a given sub-stellar latitude, the second part tells us how much light is reflected at a given sub-stellar latitude, and the third is the area at a given sub-stellar latitude. Although it may seem necessary to multiply Eq.~\ref{eq:Power_reflected} by two in order to account for the whole hemisphere ($\phi = -\pi/2 \rightarrow \pi/2 $), one would also be required to multiply by a half to normalise the integral so that the total power arriving at the hemisphere is $F_{in}(\lambda) \cdot \pi R^{2}_{p}$, hence cancelling each other out. Nevertheless, we can now rearrange as follows:
\begin{equation}
    P_{refl} =  F_{in}(\lambda) \cdot \pi R^{2}_{p} \int^{\pi/2}_{0} 2 \sin{\phi} \cos{\phi} \cdot q(\phi) ~ d\phi
\end{equation}
Using the trigonometric identities,
\begin{equation}
\label{eq:P_refl}
    P_{refl} =  F_{in}(\lambda) \cdot \pi R^{2}_{p} \int^{\pi/2}_{0} \sin{2\phi} \cdot q(\phi) ~ d\phi
\end{equation}
\end{subequations}
We can now divide Eq.~\ref{eq:P_refl} by \ref{eq:P_in} to get the spherical albedo:
\begin{equation}
\label{eq:A_S}
    A_{S} = \int^{\pi/2}_{0} \sin{2\phi} \cdot q(\phi) ~ d\phi
\end{equation}
With the exception of idealised constructs like a Lambertian surface, $q(\phi)$ is never a constant. In order to determine $q(\phi)$, we ran another set of ray-tracing simulations for different refractive indices ($n_{sy}$), Hurst exponents ($H$), and sub-stellar latitudes ($\phi$). The analytic approximation of $q(\phi)$, which incorporates the planar albedo, can be found in Sec.\ref{sec:Results}. By solving Eq.~\ref{eq:A_S} one can determine the spherical albedo, but before this is explored it is necessary to briefly discuss the possibility of atmospheric effects.

\subsection{Atmospheric Effects on the spherical albedo}
\label{sec:atmospheric_effects}

In this paper we focus on how the surface properties of a magma ocean effect the spherical albedo values of airless or near-airless, molten super-Earths. However, one could argue that the vaporised mineral atmospheres resulting from highly irradiated magmas could also affect the spherical albedo values. Fundamentally, this will depend on the composition of the magma and the resultant vaporised atmosphere. For volatile-free magmas, the secondary atmospheres would be thin due to the low refractory nature of rocky species \citep{Schaefer2009,Ito2015}. Furthermore, 1-D atmospheric models show that the high temperatures found on some of these hot super-Earths could prevent the formation of clouds within their daysides \citep{Mahapatra2017}. Therefore, one can consider the case of a thin, cloudless, mineral atmosphere on a hot molten super-Earth. Within this framework, two effects need to be considered. First, the Rayleigh scattering of gases should increase the planetary spherical albedo. Second, the spectroscopic absorption by the gases would decrease it.

The effects of Rayleigh scattering can be understood by considering Eq.~\ref{eq:rs}:
\begin{equation}
\sigma_{\lambda} = \frac{128\pi^5}{3\lambda^4}\alpha^2,
\label{eq:rs}
\end{equation}
where $\sigma_{\lambda}$ is the Rayleigh scattering cross section for a given species, $\lambda$ is the wavelength of light, and $\alpha$ is the static average electric dipole polarisability of the gaseous species being considered. According to the \citet{CRC92}, the $\alpha$ values for the following species (in units of 10$^{-24}$ cm$^3$) are: 24.11 for Na, 43.4 for K, 8.4 for Fe, 5.53 for Si, 10.6 or 11.1 for Mg, 0.802 for O, and 1.5689 for O$_{2}$ \citep[for more information on these species, see][ references therein]{CRC92}. Due to having a large $\alpha$ and being one of the most abundant gases in mineral atmospheres, Na is expected to be the most scattering gas within these extreme environments \citep{Schaefer2009,Ito2015}. The effects of a pure Na mineral atmosphere is illustrated in Fig.~\ref{fig:AS} of Sec.~\ref{sec:Atmosphere} in the appendix where the optical depth from Rayleigh scattering is given for the case of Kepler-10b. It becomes clear how even for a strongly scattering mineral atmosphere, the influence on the spherical albedo becomes negligible for $\lambda\gtrsim$1~$\mu$m.

Conversely, if the magma were to include several volatile elements such as H, C, N, S and Cl, then it would most probably be thick \citep{Schaefer2012}. In that case the scattering of the atmosphere would affect the spherical albedo value. However, the stability of highly volatile elements on close-in exoplanets against photoevaporation is still not well understood \citep{Lopez2017}. For simplicity, in our paper we will only focus on the case where there is a volatile-free mineral atmosphere. However, we are aware that thick, cloud-free, atmospheres would have high albedo values in the optical range \citep{Hamano2015}.

Regarding Kepler-10b, whether the mineral atmosphere affects the albedo or not will depend on its composition, which we do not know with certainty. We give the case of sodium as it is the most optically thick of all non-volatile mineral atmospheric components (and hence, it is an interesting gas to consider). Sodium would affect the albedo at $\lambda= 0.5~\rm \mu m$ as its optical depth is $\simeq 1$. Nevertheless, our code is best adapted for larger wavelengths, primarily $\lambda > 1~\rm \mu m$, so the effects of a mineral atmosphere should not be large. We are aware that the \textit{Kepler} telescope operates at shorter wavelengths so one would expect the mineral atmosphere to contribute towards the spherical albedo. However, the high bolometric Bond albedo values inferred for Kepler-10b cannot fully be accounted for by a mineral atmosphere \citep{Rouan2011}, which is why we argue that an exotic magma must also be present.

We believe that in future works our code could be implemented into radiative transfer models where one could include clouds or a scattering atmosphere (see Sec.~\ref{sec:future_work}). Nonetheless, it is beyond the scope of our study to include atmospheric effects on the planetary spherical albedo as our paper focuses primarily on surfaces.

\section{Results}
\label{sec:Results}

\begin{figure}[h]
	\centering
	\includegraphics[scale=1]{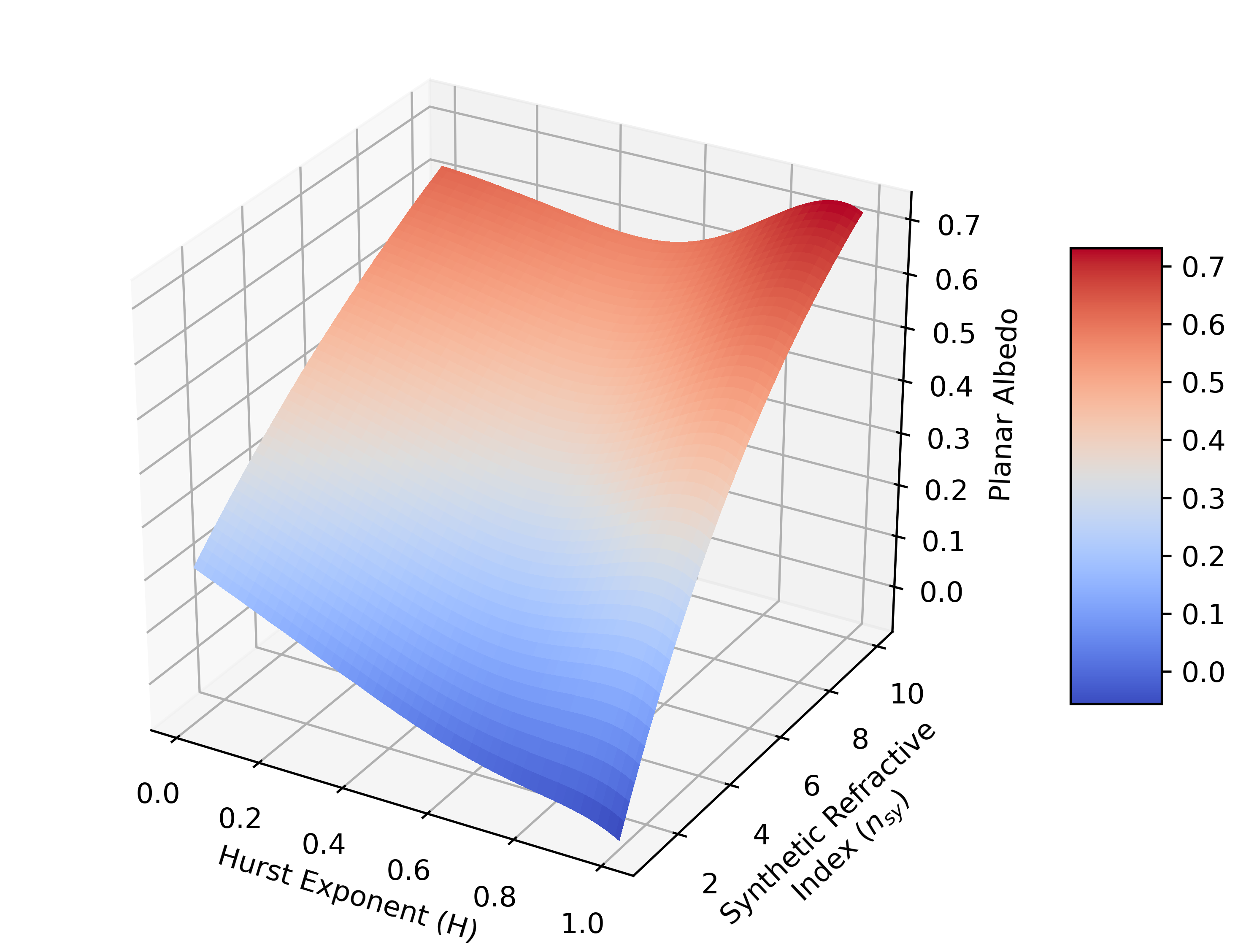}
	\caption{The planar albedo as a function of the synthetic refractive index ($n_{sy}$) and the Hurst exponent ($H$). This figure was generated with Eq.~(\ref{eq:Albedo}) and (\ref{eq:psiconstant}).}
	\centering
	\label{fig:Albedo_H_n}
\end{figure}

We were able to find an analytic relationship between the planar albedo ($A_{pla}$) of a molten surface and the Hurst exponent (\textit{H}) as shown in Eq.~(\ref{eq:Albedo}) and (\ref{eq:psiconstant}). These equations were found by fitting the numerical data from our simulations. The raw data and its comparison to the planar albedo values predicted by Eq.~(\ref{eq:Albedo}) and (\ref{eq:psiconstant}) can be found in the Supplementary File 1. Within Supplementary File 1, we also show that the $\chi^{2}$ values of our analytic approximation are small and hence we recommend its use over the raw data. For Hurst exponents greater than 0.95 the effects of roughness become negligible so Eq.~\ref{eq:R0} is an appropriate approximation for the planar albedo.
\begin{table}
	\centering
	\caption{Parameters for Eq.~(\ref{eq:Albedo}) and (\ref{eq:psiconstant})}
	\label{tab:parametervalues}
	\begin{tabular}{P{1.5cm}P{1.5cm}} 
		\hline
		\hline                     
		Parameter & $H \leq 0.95$ \\
		\hline
		$\alpha_1$ & -0.397910\\
		$\alpha_2$ & 0.149320\\
		$\alpha_3$ & 0.388928\\
		$\alpha_4$ & 0.852872\\
		$\beta_1$ & 1.594982\\
		$\beta_2$ & -3.375571\\
		$\beta_3$ & 2.437896\\
		$\beta_4$ & -0.881001\\
		$\beta_5$ & 0.194354\\
		$\beta_6$ & -0.041750\\
		$\beta_7$ & -0.081323\\
		\hline  
	\end{tabular}
\end{table}
\begin{subequations}
	\begin{equation}
	A_{pla} \approx (\alpha_1 H + \alpha_2 ) + (\alpha_3 H + \alpha_4 )(1 - e^{\Psi \cdot n_{sy}})
	\label{eq:Albedo}
	\end{equation}
	with
	\begin{equation}
	\Psi = \beta_1 H^{6} + \beta_2 H^{5} + \beta_3 H^{4} + \beta_4 H^{3} + \beta_5 H^{2} + \beta_6 H + \beta_7	
	\label{eq:psiconstant}
	\end{equation}
\end{subequations}
Where $\alpha_{1\rightarrow4}$ and $\beta_{1\rightarrow7}$ are the coefficients listed in Table~\ref{tab:parametervalues}. Fig.~\ref{fig:Albedo_H_n} is a plot of Eq.~(\ref{eq:Albedo}, \ref{eq:psiconstant}).
Combining this result with the refractive indices from \citet{Polyanskiy2008} and \citet{Grainger2008} gave us the planar albedo values for different materials at varying roughnesses and wavelengths. To convert the planar albedo into the spherical albedo, the dependence of the albedo on the sub-stellar latitude, $q(\phi)$, was required. From our simulations (see Supplementary File 2 for the raw data and $\chi^{2}$ values) we found that $q(\phi)$ is well approximated as:
\begin{subequations}
\begin{equation}
\label{eq:phase_function}
    q(\phi) \approx 1 - \frac{1 - A_{pla}}{1 + \left(\frac{\phi}{ \mathcal{C}_{1}(H) } \right)^{ \mathcal{C}_{2}(H) }}
\end{equation}
Where,
\begin{equation}
\label{eq:phase_function_constants}
\begin{split}
    &\mathcal{C}_{1}(H) = 0.320717H + 1.14083 \\
    &\mathcal{C}_{2}(H) = 8.54326 + 82.6376H^{3.88648}
\end{split}
\end{equation}
\end{subequations}
From Eq.~\ref{eq:phase_function} it follows that as $\phi \rightarrow \pi/2$, $q(\phi) \rightarrow 1$ because at the terminator the planet is fully reflecting (at right angles there is no absorption). Conversely, when $\phi \rightarrow 0$, the local albedo becomes identical to the planar albedo as the light is interacting with the sub-stellar point. Combining Eq.~\ref{eq:A_S} with Eq.~\ref{eq:Albedo}, (\ref{eq:psiconstant}), (\ref{eq:phase_function}), and (\ref{eq:phase_function_constants}) gives the spherical albedo of a planet. Hence, it is now possible to explore the reflective properties of planets with different magma ocean properties.

We begin by considering the spherical albedo of simple end-member magma compositions (see Fig.~\ref{fig:materials} in the appendix). We adopted a Hurst exponent of 0.8 which is consistent with the type of waves expected on Kepler-10b (see Sec.~\ref{sec:hurst} in the appendix for more information). The best-fit equations for the numerical data of \citet{Polyanskiy2008} and \citet{Grainger2008} are shown in Table~\ref{tab:bestfitequations} of Sec.~\ref{sec:listofequations} in the appendix. A brief summary of our results for the different materials is explained below (for a Hurst exponent of 0.8). The graphs showing the spherical albedo are shown in Sec.~\ref{sec:spectral_albedo_materials} of the appendix.

Planets composed of common Earth minerals and rocks typically have spherical albedo values of $\sim 10\%$. They also tend to drop as the wavelength increases from $\rm 0.5~\mu m$ to $\rm 7.8~\mu m$. The only exceptions are $\rm SiO_{2}$ (i.e. rhyolites) and komatiites which have sudden increases in their spherical albedos at $\rm \sim 7.2~\mu m$.

Pure metal planets would have very high spherical albedos, especially if they were made of Ag, Al, Au, Cu, Fe, Mg, Mo, or Ni as they would reach values close to $100\%$ at larger wavelengths. Metalloids (Ge and Si) would produce lower spherical albedos and within the wavelengths of interest, they appear to stabilise at $30-50\%$.

The reflective properties of metal oxides vary strongly depending on the species. FeO is the shiniest oxidised material analysed in this study with values reaching as high as $45\%$. Conversely, $\rm Al_{2}O_{3}$ and MgO are the least reflective oxidised metallic species with spherical albedo values close to common Earth minerals and rocks such as the ones explained above.

We present two carbonaceous minerals; $\rm CaCO_{3}$ and SiC. SiC's spherical albedo drops from $\sim 20 \%$ to $\sim 15 \%$ from $\rm 0.5-7.8~\mu m$ respectively. Conversely, $\rm CaCO_{3}$ has a spherical albedo $\lesssim 10\%$ for wavelengths below $\rm \sim 6.2~\mu m$ but beyond this it quickly increases to a maximum of almost $\sim 90\%$ at $\rm \sim 6.8~\mu m$. This peak then drops to $\sim 15\%$ for wavelengths greater than $\rm \sim 7.5~\mu m$.

Whilst Fig.~(\ref{fig:materials}) are useful for understanding the spherical albedo spectrum of pure materials; AMOPs are composed of multiple chemical species. It is therefore important to find geochemically self-consistent mixtures of the aforementioned materials so that from an observed albedo spectrum one can categorise the composition of an exoplanet. Hence, we propose the following categorisations for constraining exoplanet compositions from their spherical albedos: evolved molten bulk-silicate Earth (BSE) planets, metallic planets, coreless terrestrial planets (CTPs), reduced planets, and carbon-rich planets.

\subsection{Evolved Magmatic BSE Planet}
\label{sec:evolvedbse}

\begin{figure}[h]
	\centering
	\includegraphics[scale=0.9]{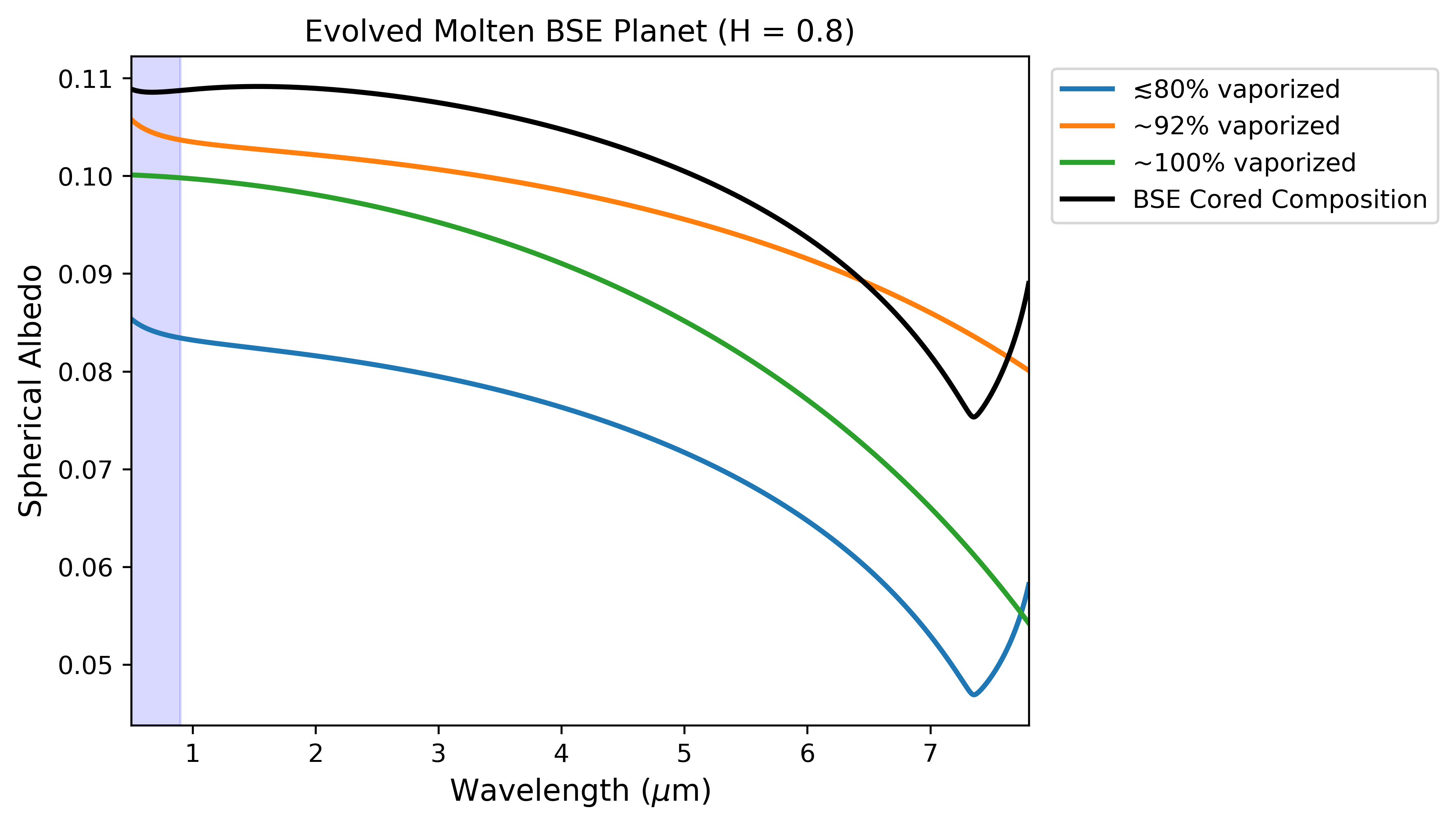}
	\caption{The spherical albedo for a terrestrial planet with an initial composition corresponding to the BSE that has been evolved due to chemical interactions between the magma and the mineral atmosphere. Four different spherical albedo spectrums are shown corresponding to varying levels of vaporisation. The BSE composition is from \citet{Oneill1998} whilst the vaporised compositions were taken from \citet{Kite2016}. The light-blue section is \textit{Kepler's} band-pass. This figure was generated with Eq.~(\ref{eq:A_S}), (\ref{eq:Albedo}), (\ref{eq:psiconstant}), (\ref{eq:phase_function}), and (\ref{eq:phase_function_constants}).}
	\centering
	\label{fig:evolvedterrestrial}
\end{figure}

When a terrestrial planet with a BSE composition is subjected to temperatures above $\rm\sim1500~K$, the minerals found within the magma begin to vaporise \citep[e.g.][]{Schaefer2009,Schaefer2012,Ito2015}. This vaporisation could lead to a strong evolution in the chemical composition of the magma \citep{Kite2016}. It therefore follows that the physical and optical properties of the lava should also change with time. In Fig.~\ref{fig:evolvedterrestrial} we show our predicted spherical albedo spectrums for a BSE airless planet under 4 different levels of vaporisation: $0\%$, $\lesssim80\%$, $\sim 92\%$, and $\sim 100\%$. The first has a typical BSE composition \citep{Oneill1998} whilst the latter three approximately correspond to $60:40$ of $\rm MgO-SiO_{2}$, $50:50$ of $\rm CaO-Al_{2}O_{3}$, and pure $\rm Al_{2}O_{3}$ respectively.

Our synthetic spectrums predict spherical albedos lower than $\sim 11\%$. Due to current limitations in our astronomical instrumentation there appears to be no clear way of distinguishing between different types of Earth-like compositions regardless of how evolved the magma is. Notwithstanding, our findings suggest than a spherical albedo value of $\sim 11\%$ or lower, with a decrease at longer wavelengths, is suggestive of a composition similar to the Earth's.

\subsection{Metallic Planet}
\label{sec:metallic}

Prior to considering the spherical albedo values of metallic planets it is important to understand their formations. One can begin by analysing Mercury that has an estimated bulk iron abundance of $\sim70\%$ \citep[e.g.][]{Riner2008,Malavergne2010,Hauck2013,Chabot2014}. There are three main explanations for this super-ferruginous composition: thermally-induced surface vaporisation \citep[e.g.][]{Cameron1985}, collisional stripping \citep[e.g.][]{Benz1988}, and chemical/thermal gradients within the primordial solar nebula triggering an unusual formation \citep[e.g.][]{Lewis1972,Lewis1974}. For larger bodies such as super-Earths, surface vaporisation would be unable to trigger sufficient mass-losses as the required escape velocities of silicate species are too large (Ito et al. in prep). Regarding collisional stripping, \citet{Marcus2010} showed that the maximum amount of iron that a super-Earth could have is $\sim70\%$ which is compatible with the abundance in Mercury. Finally, forming iron-rich exoplanets due to chemical/thermal gradients has been shown to be possible in our solar system \citep[e.g.][]{Lewis1972,Lewis1974}, and considering how iron is the sixth most abundant element in the universe (by mass) there does not appear to be a strong argument against this mechanism either. We are therefore left with the latter two mechanisms for the formation of metal-rich super-Earths.

Before one can explore the outcomes that could arise from these two dissimilar formation mechanisms, it is necessary to issue the following caveat: a bulk iron composition is not implicative of a surface also containing iron. For example, Mercury's surface composition is very iron-poor $\rm 1-2~wt.\%$ \citep{Evans2012,Weider2014} which may seem paradoxical at first. There is a straightforward explanation for this; Mercury formed from very reduced materials. We will discuss reduced planets more in-depth in section \ref{sec:reduced}. Notwithstanding, if a terrestrial exoplanet were to form under very oxidised conditions, a large portion of its total iron would bind with oxygen to form the usual oxides (FeO, $\rm Fe_{2}O_{3}$, and $\rm Fe_{3}O_{4}$). These compounds are less dense than pure metals or alloys so they would become incorporated into rock-forming minerals such as olivine and pyroxenes that would then form the mantle and crust \citep{Elkins2008}. Conversely, the pure iron and iron-nickel alloys would sink to the centre of the planet thus forming a core. Following this line of thought, after a certain critical level of oxygen, all of the iron would form oxides therefore leaving the planet coreless (see Sec.~\ref{sec:coreless}). It is therefore very important not to assume that an iron-rich world will host an iron-rich surface as the stability of metals in the crust and mantle are strongly dependent on the presence of oxygen. Keeping in mind the above caveat, one can now refer back to the two distinct formation mechanisms and show how they can lead to two geochemically distinct magmas.

According to the collisional-stripping model, a planet becomes Fe-rich \textit{after} its formation due to the partial removal of the mantle, while its core remains relatively intact. Therefore, the abundance of iron in the mantle and crust (from which the magmas are formed) is approximately independent of any collisions (excluding exogenous deposits). It then follows that the Fe-richness of metallic exoplanets' magmas is more strongly dependent on the presence of oxygen than the bulk abundance of iron. Conversely, if iron planets form due to thermal/chemical gradients in the primordial stellar nebula then there could be a relationship between being oxygen-poor and iron-rich. This is best explained by briefly summarising the requisites for both of these outcomes:
\begin{itemize}
	\item At very high temperatures $\rm MgSiO_{3}$ is weakly captured whilst iron condenses with relative ease \citep[e.g.][]{Lewis1972,Lewis1974}. Consequently, under these conditions the minerals that form are richer in iron-bearing minerals and poorer in silicates than elsewhere in the stellar nebula.
	\item At very high temperatures the chondrules that form enstatite chondrites are created from supercooled, fluid, silicate, rain-like droplets \citep{Blander2009}. These reduced chondrites are believed to be the materials that form reduced planets such as Mercury \citep[e.g.][]{Nittler2011,Malavergne2014,Nittler2019}.
\end{itemize}
In contrast to the collisional stripping model, if iron-planets form from chemical/thermal gradients, having a high abundance of iron might coincide with being reduced. This would result in most of the iron being captured within the core and little being left within the mantle. Therefore, according to this mechanism, metallic exoplanets could have felsic magmas that are sulfur-rich \citep[e.g.][]{Haughton1974,Mccoy1999,Namur2016}.

To summarise, the magma composition of metallic planets is strongly dependent on the formation mechanisms and history of the planet. Hence, we believe that the bulk iron mass-fraction cannot be constrained from the measured spherical albedo spectrum of these types of bodies.

\subsection{Coreless terrestrial Planet}
\label{sec:coreless}

\begin{figure}[h]
	\centering
	\includegraphics[scale=0.9]{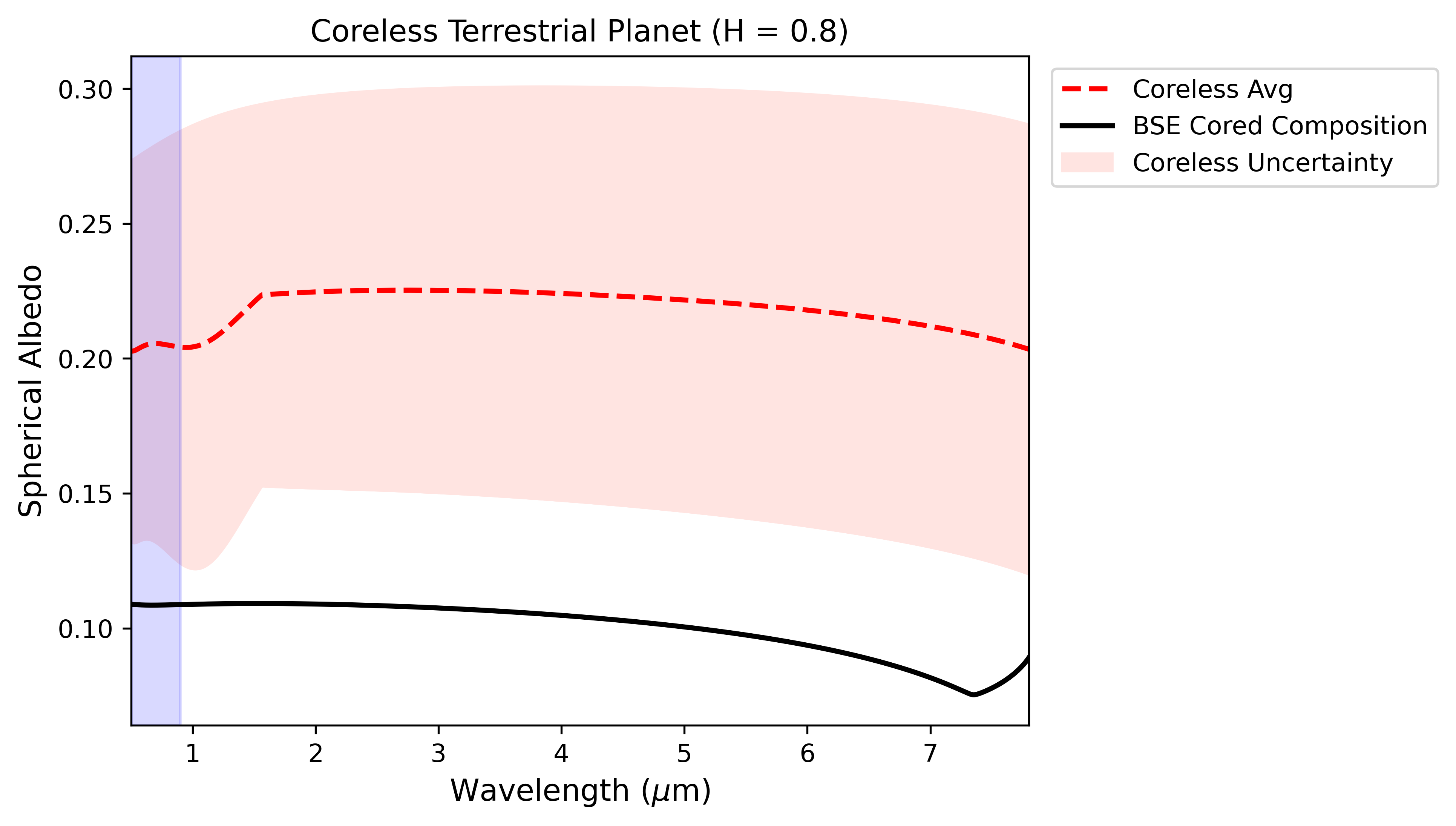}
	\caption{The spherical albedo of a coreless terrestrial planet with the composition given by \citet{Elkins2008}. The red dotted line is the average coreless spherical albedo value for $H = 0.8$, the red shaded region represents the uncertainty, and the black line is for the spherical albedo value of a cored BSE planet \citep{Oneill1998}. The light-blue section is \textit{Kepler's} band-pass. This figure was generated with Eq.~(\ref{eq:A_S}), (\ref{eq:Albedo}), (\ref{eq:psiconstant}), (\ref{eq:phase_function}), and (\ref{eq:phase_function_constants}).}
	\centering
	\label{fig:corelessterrestrial}
\end{figure}

In order to model coreless terrestrial planets (CTPs) we first had to set a planetary bulk composition and then subject it to the relevant thermodynamic conditions. In order to do so we used the elemental composition listed in Table~\ref{tab:chemicalcomposition} with the temperature and pressure shown in Table~\ref{tab:Hurstcalculation} (see appendix) and then applied the GG\textsc{chem} geochemical code \citep{Woitke2018}. Our resultant magma composition is well approximated as $60:40$ of iron oxides to enstatite which is much more iron-rich than the BSE ($\rm \lesssim 10\%$ iron oxides). However, because GG\textsc{chem} simulates only FeO as the Fe-bearing oxide in liquid phases, it does not specify the ratio of $\rm FeO:Fe_{2}O_{3}:Fe_{3}O_{4}$. As shown in Fig.~\ref{fig:materials}, the spherical albedo values of these species differ moderately so one cannot treat all of them as FeO. We will therefore present the upper- and lower-bounds with the average spherical albedo for a given Hurst exponent (Fig.~\ref{fig:corelessterrestrial}).

\begin{table}
	\centering
	\caption{Predicted Bulk Chemical Composition of Coreless Terrestrial Planets (CTPs)}
	\label{tab:chemicalcomposition}
	\begin{tabular}{P{2cm}P{3cm}} 
		\hline
		\hline                     
		Chemical & Abundance ($\%$) \\
		\hline
		O &	35.2\\
		Si & 13.5 \\
		Fe & 37.8\\
		Mg & 11.3\\
		Ca & 1.1\\
		Al & 1.1 \\
		\hline  
	\end{tabular}
	\tablecomments{Based on the findings from \citet{Elkins2008}}
\end{table}

Our results show that coreless bodies have much higher spherical albedos than their differentiated counterparts with similar compositions. This is due to iron-oxides being shinier (see Fig.~\ref{fig:materials}) and more abundant in CTP magmas. The uncertainty in the spherical albedo is due to the different possible ratios of $\rm FeO:Fe_{2}O_{3}:Fe_{3}O_{4}$, which is dependent not only on the Gibbs energies but also on how much oxygen was present during the formation of the body. Whilst one might not be able to differentiate between which iron oxides are present within the magma, one could make a strong argument for the presence of iron within these bodies. Adjusting the surface roughness parameter (\textit{H}) and maximising the FeO abundance allows for a theoretically maximum spherical albedo of $\sim 55\%$.

\subsection{Reduced Planet}
\label{sec:reduced}

In order to understand the formation mechanism of reduced planets, one can analyse Mercury which is the most reduced planet in our solar system \citep[e.g.][]{Nittler2011,Malavergne2014,Nittler2019}. This is probably because it formed from enstatite chondrites which are extremely reduced \citep{Blander2009}. These rocks can be classified into two groups; EH (high enstatite) and EL (low enstatite) chondrites. Each group has slightly different chemical makeups \citep{Wasson1988} but their close similarities allow for an approximate average composition to be deduced \citep[see Table~\ref{tab:chemicalcomposition2},][]{Javoy2010}.
\begin{table}
	\centering
	\caption{Predicted Bulk Chemical Composition of Reduced Planets}
	\label{tab:chemicalcomposition2}
	\begin{tabular}{P{2cm}P{3cm}} 
		\hline
		\hline                     
		Chemical & Abundance ($\%$) \\
		\hline
		O &	33.3 \\
		Si & 20.2 \\
		Fe & 28.3 \\
		Mg & 14.0 \\
		Ca & 1.0 \\
		Al & 0.97 \\
		Ni & 1.73  \\
		Ti & 0.06 \\
		Cr & 0.36 \\
		Co & 0.08 \\
		\hline  
	\end{tabular}
	\tablecomments{Based on the findings from \citet{Javoy2010}}
\end{table}

By applying the GG\textsc{chem} code \citep{Woitke2018} with the elements listed in Table~\ref{tab:chemicalcomposition2}, the temperature and pressure shown in Table~\ref{tab:Hurstcalculation}, we predict a magma composed of over $\gtrsim 90\%$ pure Fe with the rest mostly comprising typical terrestrial silicate minerals such as forsterite and spinel. The reason why the Fe concentration in the magma is much higher than the one shown in Table~\ref{tab:chemicalcomposition2} is because one also has to consider the composition of the emitted gas. This vaporised atmosphere contains many of the volatile elements that used to be present in the melt. Nevetheless, because of gravity, the pure iron would sink to the planet's core thus resulting in an iron-poor mantle. These predictions are strongly corroborated by observational \citep[e.g.][]{Evans2012,Weider2014} and geophysical \citep[e.g.][]{Riner2008,Hauck2013} data of Mercury. Furthermore, spinel, fosterite and many of the other silicate minerals that are predicted by the GG\textsc{chem} geochemical code are the same materials that makeup large portions of Earth's mantle. These silicates would be molten within the high temperature ranges we are focusing on in this paper. In other words, from a reduced planet made of enstatite chondrites (or from similarly reduced rocks) we would expect silicate-rich magmas that are iron-poor. A rhyolitic magma could be a good contender due to being very $\rm SiO_{2}$-rich \citep[typically $\gtrsim 69\%$;][]{Lemaitre2002}. If rhyolites were to be the most common magmas on reduced super-Earths then from Fig.~\ref{fig:materials} it can be seen that spherical albedo values less than $10\%$ would be expected. However, other exotic magmas that are oxygen-poor could also be possible. For instance, if a planet were carbon-rich then carbonaceous magmas might exist. We discuss this concept in Sec.~\ref{sec:carbonplanet}.

Finally, we would like to briefly discuss one potential way of forming a reduced, airless, molten planet with pure iron deposits on its surface. Suppose a planet were to form under very cold conditions, this would result in most of the oxygen bonding with silicon, carbon, hydrogen and magnesium. Considering how iron is not a very aerophilic element, some of it may be left in its elemental form. If this planet were then to migrate very close to its host star, XUV-irradiation may fully erode away its atmosphere leaving behind a denuded core \citep[e.g.][]{Lecavelier2007, Ehrenreich2011, Lammer2013, Owen2013, Jin2014, Owen2017, Jin2018, Kubyshkina2018(2),Locci2019,Modirrousta2020b}. Due to iron being relatively heavy, it may be resilient to hydrodynamic escape \citep[e.g.][]{Zahnle1986,Hunten1987,Luger2015} which could result in pure iron deposits being left behind on the bare core. Pure iron is very reflective (see Fig.~\ref{fig:materials}) so it is possible for these planets to have very high spherical albedo values if there were to be a copious abundance of iron in its elemental form. Undeniably, the stability of pure iron within a lava matrix would be subjected to the usual caveats of the circulation and sinking timescales. These systems are highly speculative and warrant more research.

In summary, the spherical albedos of reduced planets that form from enstatite chondrites should be very low ($\lesssim 10\%$) as their surfaces are expected to be rich in silicate species. We find that the amount of light they reflect is similar to that of BSE composition planets, albeit with slighly different spectrum shapes. However, one cannot rule out the possibility of more unusual formation mechanisms which could lead to an ambiguity in the predicted spherical albedos.

\subsection{Carbon-rich Planet}
\label{sec:carbonplanet}

There are several arguments for carbon-rich super-Earths such as the existence of carbonaceous bodies in our solar system (e.g. c-type asteroids) and the presence of main sequence stars with high carbon-to-oxygen ratios (C/O). These stars could host these exotic worlds, especially since theoretical modelling predicts that even with moderate C/O ratios several carbon-rich worlds could form \citep{Moriarty2014}. However, as of February 2020 there are no known carbon-rich super-Earths. Because of this limitation we will base our magma composition on theoretical principles. We will avoid using pure carbon \citep{Madhusudhan2012} as the minimum melting points of carbon allotropes tend to be $\rm > 4000~K$ \citep[e.g.][]{Bundy1989,Bundy1996,Correa2006}, which is generally beyond the equilibrium temperature of most super-Earths. Hence, we will adopt a more chemically diverse composition such as the one delineated in \citet{Miozzi2018}.

Whilst SiC has been found in presolar grains \citep[e.g.][]{Lodders1995,Hoppe2010}, it has a tendency to decompose due to the high affinity between oxygen and silicon. For instance, if a planet were to be composed of equal parts Si-O-C, most of the silicon would bind with the oxygen to form silicates such as $\rm SiO_{2} $ and SiO. The carbon would then bond with itself to form graphite in the outer layers of the planet and diamond at greater depths. Due to graphite being less dense than $\rm SiO_{2} $ and having a higher melting temperature, it would rise to the top layers of the mantle in the form of solid graphitic grains. Our simple analogy is strongly supported by high-pressure, high-temperature (high P-T) laboratory experiments \citep[e.g.][]{Hakim2018,Hakim2019} that arrived at the same conclusion, albeit in more detail. These high P-T experiments predict that extremely reduced conditions would be required in order to maintain SiC in a geochemically stable condition. Therefore, in spite of SiC planets being theoretically plausible, they may be unlikely. Notwithstanding, we included the spherical albedo of SiC in Fig.~\ref{fig:materials}.

On that account, it seems more probable that carbon-rich worlds will be composed of silicate mantles covered in solid graphite. The spherical albedo values of these planets would therefore be dominated by the graphitic layer which we cannot model with our code due to it being adapted for fluid systems. If one ignores surface roughness, the spherical albedo of a graphite layer within the range of $\rm 0.5-7.8~\mu m$ would vary from $30-80\%$ \citep[e.g.][]{Taft1965,Philipp1977,Querry1985,Djurisic1999,Kuzmenko2008,Papoular2014}, which would make carbon-rich super-Earths the brightest airless bodies listed in our paper.

\section{Discussion}
\label{sec:Discussion}

Measuring, constraining, and interpreting the spherical albedo of exoplanets is an active area of research which is mostly limited by observational constraints. With the JWST expected to launch in 2021, the \textit{Atmospheric Remote-sensing Infrared Exoplanet Large-survey} (ARIEL) in 2028, a more in-depth analysis could be possible. Because of this we now provide two tests that could be used to constrain the magma composition of molten super-Earths from their spherical albedo values:

\subsection{Optimistic Test: High Precision High Accuracy Measurements (HPHA)}

In the ideal case scenario, the extracted albedo spectra from the aforementioned missions would be robust enough to perform statistical tests on them (e.g. MCMC algorithms). The shape of the spectrum would be the major source of information for the surface composition whilst the amplitude would mostly determine the surface roughness (i.e. the Hurst exponent). The experimental data that relates a material's refractive index to the wavelength of light began almost a century ago and it is ongoing. From these relationships, Eq.~(\ref{eq:nsynth}), (\ref{eq:A_S}), (\ref{eq:Albedo}), (\ref{eq:psiconstant}), (\ref{eq:phase_function}), and (\ref{eq:phase_function_constants}) could be used to constrain the surface compositions of molten, airless super-Earths. Because of this, regardless of whether ARIEL or the JWST missions could extract HPHA measurements, it is certainly possible that in the future the spherical albedo spectrum of a planet could be matched to a certain composition in a similar fashion to how exoplanet atmospheres are being constrained from spectroscopic data.

\subsection{Conservative Test: Low Precision Low Accuracy Measurements (LPLA)}

A more realistic scenario is that a few albedo measurements at varying wavelengths with considerable uncertainties are extracted. For this scenario we propose a test that will most probably be possible for ARIEL and the JWST missions once an optically thick atmosphere is excluded (i.e. the planet is airless or near-airless). We propose that based on the mass and radius of an observed AMOP one limits the possible internal compositions. From this information, the observed albedo spectrum is compared to our five compositional categories described in sections \ref{sec:evolvedbse}, \ref{sec:metallic}, \ref{sec:coreless}, \ref{sec:reduced}, and \ref{sec:carbonplanet}. For instance, in the case of Kepler-10b the mass and radius measurements are consistent with a coreless or near-coreless interior which is illustrated in Fig.~\ref{fig:kepler10b}. Therefore, whilst the spherical albedo is very uncertain, an extremely ultramafic magma is compatible with the values shown in Fig.~\ref{fig:corelessterrestrial}.

\subsection{The Effects of Changing the Hurst Exponent (H)}

\begin{figure}[h]
	\centering
	\includegraphics[scale=0.8]{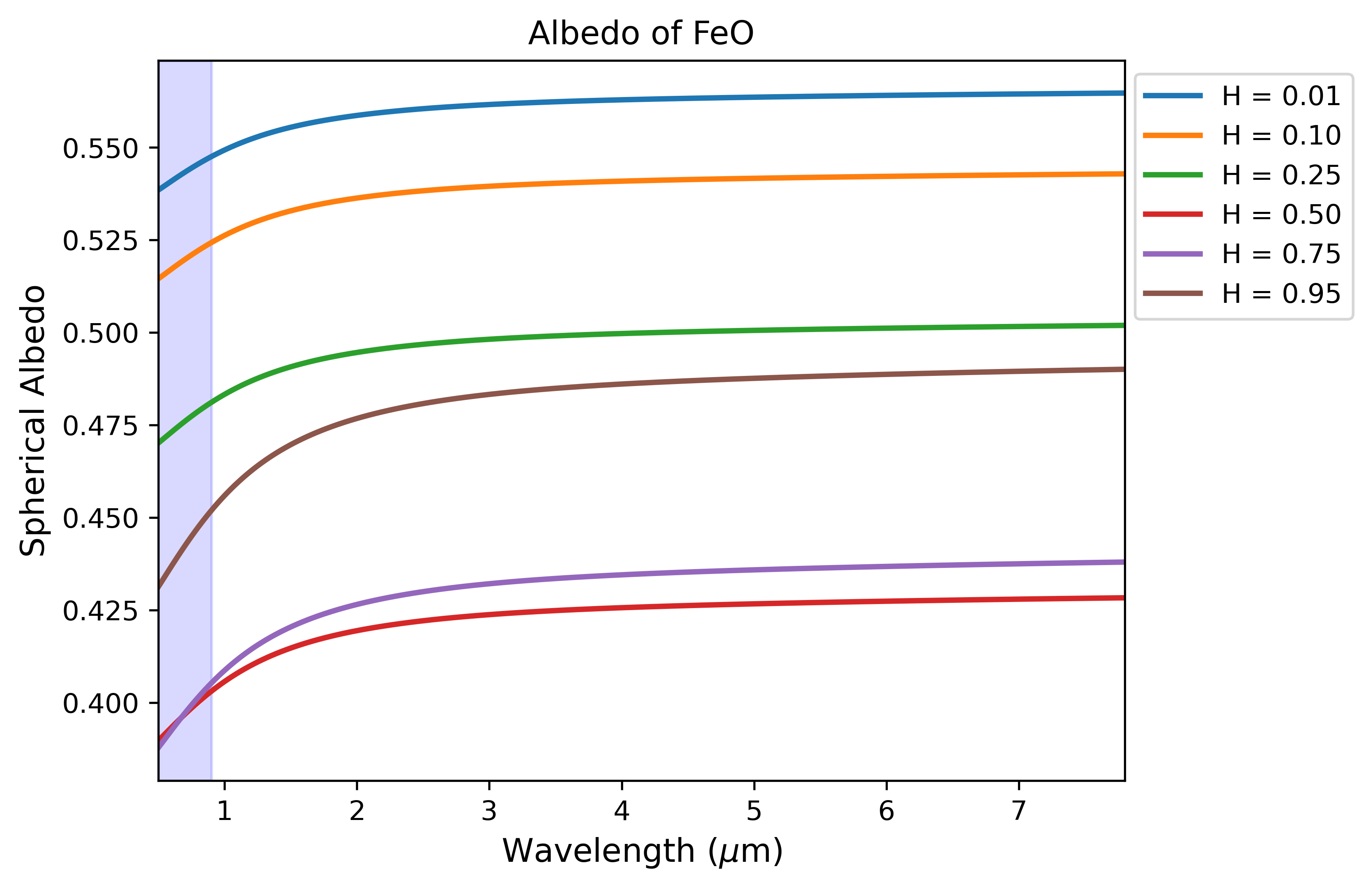}
	\caption{The spherical albedo of FeO for different Hurst exponents and wavelengths. The light-blue section is \textit{Kepler's} band-pass. This figure was generated with Eq.~(\ref{eq:A_S}), (\ref{eq:Albedo}), (\ref{eq:psiconstant}), (\ref{eq:phase_function}), and (\ref{eq:phase_function_constants}).}
	\centering
	\label{fig:FeOH}
\end{figure}

In Fig.~\ref{fig:Albedo_H_n} we show how the Hurst exponent and the synthetic refractive index affect the planar albedo of a system. However, understanding the isolated influence that the \textit{H} parameter has may be difficult to picture so we have included Fig.~\ref{fig:FeOH}. Although the shape of the spectrum is slightly affected by varying \textit{H}, the most prominent change is in the amplitude. This is an important point to consider as it implies that strong deviations in the shape of the retrieved albedo spectrum are most probably due to compositional differences between the adopted (predicted) magmas and the ones that are observed. From an observational perspective, this lowers the degeneracy of the inferred surface properties from the observed spherical albedo as the effects from the surface roughness (\textit{H}) and the composition can be partially isolated. 

\subsection{Comparing our Results with \citet{Hu2012}}

Our study focuses on molten airless bodies whilst \citet{Hu2012} focuses on solid ($\rm T_{eff} = 300 - 880$ K) planetary surfaces so there will be inevitable differences in our results. For instance, for solid systems the surface roughness is affected by grain configurations and sizes \citep[e.g.][]{Bennett1961}, which is very hard to model mathematically. In their study, fine grained surfaces were assumed when constructing their synthetic spectra. This will lead to an overestimation of the albedo value as finer grains typically have higher albedos than their coarser-grained counterparts. Because of this they point out that their results are “upper limits”. This is further emphasised in our study, as in Fig.~\ref{fig:FeOH} we show that differences in the surface roughness could change the spherical albedo values by $\sim 15\%$ for certain materials. However, this alone might not fully explain the different albedo values calculated by our study and theirs, so below we will explore other possibly reasons:
\begin{itemize}
	\item\textbf{Basaltic:} Their basaltic values are much higher than our results as well as \citet{Pollack1973}, \citet{Egan1975}, \citet{Shestopalov2013}, and \citet{Fornasier2016}. One possibility is that they used very young basaltic powder which can be several times more reflective than older specimens. The reason why there is a darkening with age is due to interactions between the basalt grains and the H-ions from solar winds \citep{Shestopalov2013,Fornasier2016}.
	\item\textbf{Fe-oxydised}: In their paper, they adopt a composition of $50\%$ $\rm Fe_{2}O_{3}$ (haematite) and $50\%$ basalt, with which they predict an albedo of $10-20\%$. This is consistent with our results as we predict a basaltic albedo of $2-7\%$ and a value of $20-30\%$ for $\rm Fe_{2}O_{3}$, so a $50:50$ mixture could reproduce their results.
	\item\textbf{Metal-rich:} We consider planets that are rich in metals (bulk), which as explained in Sec.~\ref{sec:metallic} would result in metal-poor surfaces. Conversely, in \citet{Hu2012} they adopt an $\rm FeS_{2}$ composition.
	\item\textbf{Ultramafic:} Their ultramafic albedo is large, but it could be explained by a very high concentration of metal oxides (see Sec.~\ref{sec:coreless}). However, if we assume that \citet{Hu2012} used a moderate metal oxide abundance then their results would be substantially larger than ours (see Fig.~\ref{fig:corelessterrestrial}). This may be explained by their fine-grained assumption and the possibility that they used very young basaltic specimens.
\end{itemize}
Whilst the results from \citet{Hu2012} differ moderately from ours, we are working in two different regimes: molten versus solid. Thus, we do not feel like there is a contradiction between our spherical albedo values and theirs. However, we are aware that our model has its own limitations that could be improved upon in the future. We list these limitations in Sec.~\ref{sec:limitations}.

\subsection{Limitations of our Model}
\label{sec:limitations}

Our model has limitations that need to be noted by anyone who wishes to use it. Here we present a list of shortcomings and how they could be improved upon by future research:
\begin{itemize}

    \item\textit{We used Earth's oceans as an analogue for the roughness of the magma oceans of airless or near-airless molten planets.} We used this approximation twice: when applying the Gaussian filter to the pre-smoothed synthetic magma ocean surfaces (i.e. Fig.~\ref{fig:fractalsurfacesharp}), and when solving for the constant $h_{0}$ in Eq.~\ref{eq:DurstH}. The question one might ask is whether the viscosity of lava is comparable to that of water in order for surface perturbations (such as waves) to be similar. Using the Vogel–Fulcher–Tammann equation \citep{Vogel1921,Fulcher1925,Fulcher1925b,Tammann1926} the similar dynamical viscosities of these two fluids can be shown:
    \begin{equation}
        \log_{10}\left(\mu \right) = A + \frac{B}{T - C}
    \end{equation}
    Where $\mu$ is the dynamic viscosity at $T$ which is the temperature of interest; $A$, $B$, and $C$ are experimentally-determined constants, which depend on the composition of the magma. If one were to set the temperature high enough one can see how the viscosity drops precipitously. Let $T$ be the day-side temperature of Kepler-10b $\approx 2750~\rm K$ and for the experimental constants we will adopt the values for peridotite \citep{Dingwell2004}: $A \simeq -4.31$, $B \simeq 3703$, $C \simeq 761.7$. These values give $\mu \approx 10^{-2.5}~\rm Pa~s$ which is only twice as viscous as water \citep{Viswanath1989}. This result is compatible with other studies which are consistent with water-like lava viscosities at very high temperatures \citep[e.g.][]{Russell2004,Giordano2008}. A similar calculation can be done for metals to show that at high temperatures they have very low viscosities \citep[e.g.][]{Alfe1998}.
    
    It is important to note that our model will only work for fluid surfaces. For example, for silicate planets effective temperatures $\gtrsim 1500~\rm K$ will result in magma oceans. Moreover, higher temperatures will increase the accuracy of our model as the melt viscosities will approach that of water. For other materials the conditions will differ, such as a carbonatite magma that at $\simeq 800~\rm K$ will be molten and have a viscosity similar to that of water. Nevertheless, we are aware that the viscosity of the magma oceans depends on many parameters such as the composition, the spherical albedo, and the effective temperature of the planet. These are important parameters that should be investigated further but they are beyond the aims of our study.
    
	\item\textit{We assumed that the refractive index is independent of the temperature of the system}. We are aware that this is a strong simplification as, for example, Ge and Si's refractive indices grow by approximately $\sim 5\%$ as their temperatures increase from 100 K to 600 K \citep{Li1980}. We also lack experimental data on how the refractive indices evolve when there is a phase change which could be problematic as we use the refractive indices for the solid phases whilst magmas are fluids. However, as shown by \citet{Essack2020}, the albedo of $100\%$ molten and $100\%$ solidified basaltic glass remains relatively constant which implies that the refractive indices might not vary too strongly. In any case, the dependence of the refractive indices on the temperature and phase of the materials are poorly constrained and warrant more research. This limitation can be fixed with relative ease; whenever a better data set is discovered for a given material one can replace the $n$ and $k$ or $n_{sy}$ values given in table~\ref{tab:bestfitequations} with the improved ones. This new data could then be used within Eq.~(\ref{eq:A_S}), (\ref{eq:Albedo}), (\ref{eq:psiconstant}), (\ref{eq:phase_function}), and (\ref{eq:phase_function_constants}).
	
	\item\textit{We did not account for the albedo contribution from small, solid, undissolved particles i.e. single-scattering albedo.} The presence of solid undissolved particles could strongly influence the spherical albedo of a planet. As explained by \citet{Rouan2011}, molecules such as $\rm UO_{2} $ and $\rm ThO_{2} $ would not dissolve in certain melts such as an $\rm Al_{2}O_{3}-CaO $ matrix and they would be solid. Other molecules such as MgO and CaO also have very high fusion temperatures but they are more commonly dissolved. Notwithstanding, spherical albedos as high as $\sim 50\%$ may be caused by such minerals, so planets like Kepler-10b may be very reflective because of their presence. Our model does not account for these particles but one could separately include their contribution therefore overcoming this limitation.
	
	\item\textit{We assume that the spherical albedo becomes independent of the Hurst exponent for $\rm H>0.95$.} In reality the spherical albedo would change for surface roughnesses ranging from 0.95 to 1.0. However, it becomes exceedingly difficult to model such surfaces as the ocean stops behaving fully like a fractal and starts having properties more akin to simple geometrical shapes. Because of this, the Davies-Harte \citep{Davies1987} method cannot be used. Hence, considering how the spherical albedo values of surfaces with Hurst exponents of 0.95 are very close to that of $H = 1.0$; we approximated the surfaces as being flat for $0.95 < H < 1.0$ (i.e. Eq.~\ref{eq:R0}).
	
\end{itemize}

\subsection{Future Work}
\label{sec:future_work}

There is a bimodal distribution of exoplanet radii with one peak at $\rm \sim 1.3R_{\oplus}$, the other at $\rm \sim 1.75R_{\oplus}$, and the minimum at $\rm \sim 2.4R_{\oplus}$ \citep{Fulton2017}. According to theoretical modelling \citep[e.g.][]{Owen2013,Owen2017,Jin2018,Modirrousta2020b} and observational data \citep[e.g.][]{Swain2019} most planets in the first peak are lacking a hydrogen atmosphere whilst most planets in the second peak have large hydrogen envelopes. On the one hand, for planets with hydrogen atmospheres it is much easier to constrain their compositions as atmospheric species can be detected \citep{Tsiaras2016,Ridden2016,Esteves2017} and then used to infer some interior properties. On the other hand, if a planet lacks a large hydrogen component then spectroscopy may not be feasible which leaves one with less data to deduce interior properties. In the case of molten, airless super-Earths (located mostly within the first peak) the spherical albedo could be used in an analogous manner to spectroscopic data in order to deduce the surface composition and hence constrain the interior structure. Hence, one could test exotic bodies like 55 Cancri e which could either be a coreless body \citep{Bourrier2018(2)}, a carbonaceous planet \citep{Madhusudhan2012,Miozzi2018}, an icy planet \citep{Zeng2013,Zeng2016}, or a planet with a hydrogen atmosphere partially confined to the nightside and with a dayside that is near-airless \citep{Modirrousta2020}.

In addition, due to our code being analytic, it could be implemented into atmospheric general circulation models (GCMs) or spectral retrieval codes. This might provide a more holistic analysis of a planet since the effects of an atmosphere and those of the surface could be better coupled. Furthermore, it would be interesting to expand our work to non-molten surfaces such as desert planets. In theory, solid surfaces could also be explained using fractal mathematics \citep[e.g.][]{Mandelbrot1975,Shelberg1983}. We would also welcome more experimental data on the refractive indices of materials at different temperatures, pressures, and phases in order to better model the spherical albedo values of exoplanetary surfaces.

\section{Conclusions}
\label{sec:conclusion}

In this paper we present an analytic model for the spherical albedo of molten, airless or near-airless super-Earths. We developed this model by fitting the results from the simulations of photons interacting with magma ocean surfaces that had varying reliefs and geochemical compositions. Our list shown in Table~\ref{tab:bestfitequations} is not comprehensive as there are many potential chemicals that could be added to the collection. Due to being analytic in nature, our model could be adopted in a computationally inexpensive manner in order to constrain the surface composition of some exoplanets.

\section*{Acknowledgements}

We acknowledge the support of the ARIEL ASI-INAF agreement n. 2018-22-HH.0. We also thank the researchers and staff at the Telescopio Nazionale Galileo, La Palma (Spain) for introducing us to Kepler-10b that inspired us to carry out this study. We are grateful to Podolak M. and the anonymous referees for their useful comments.

\bibliography{bibliography.bib}

\begin{thebibliography}{151}
\providecommand{\natexlab}[1]{#1}
\providecommand{\url}[1]{\texttt{#1}}
\expandafter\ifx\csname urlstyle\endcsname\relax
  \providecommand{\doi}[1]{doi: #1}\else
  \providecommand{\doi}{doi: \begingroup \urlstyle{rm}\Url}\fi

\bibitem[Abraham and Becker(1950)]{Abraham1950}
M.~Abraham and R.~Becker.
\newblock \emph{The Classical Theory of Electricity and Magnetism}.
\newblock Blackie and Son, 2 edition, 1950.

\bibitem[{Amotchkina} et~al.(2020){Amotchkina}, {Trubetskov}, {Hahner}, and
  {Pervak}]{Amotchkina2020}
T.~{Amotchkina}, M.~{Trubetskov}, D.~{Hahner}, and V.~{Pervak}.
\newblock {Characterization of e-beam evaporated Ge, YbF3, ZnS, and LaF3 thin
  films for laser-oriented coatings}.
\newblock \emph{\ao}, 59\penalty0 (5):\penalty0 A40, Feb 2020.
\newblock \doi{10.1364/AO.59.000A40}.

\bibitem[{Astropy Collaboration} et~al.(2013){Astropy Collaboration},
  {Robitaille}, {Tollerud}, {Greenfield}, {Droettboom}, {Bray}, {Aldcroft},
  {Davis}, {Ginsburg}, {Price-Whelan}, {Kerzendorf}, {Conley}, {Crighton},
  {Barbary}, {Muna}, {Ferguson}, {Grollier}, {Parikh}, {Nair}, {Unther},
  {Deil}, {Woillez}, {Conseil}, {Kramer}, {Turner}, {Singer}, {Fox}, {Weaver},
  {Zabalza}, {Edwards}, {Azalee Bostroem}, {Burke}, {Casey}, {Crawford},
  {Dencheva}, {Ely}, {Jenness}, {Labrie}, {Lim}, {Pierfederici}, {Pontzen},
  {Ptak}, {Refsdal}, {Servillat}, and {Streicher}]{Astropy2013}
{Astropy Collaboration}, T.~P. {Robitaille}, E.~J. {Tollerud}, P.~{Greenfield},
  M.~{Droettboom}, E.~{Bray}, T.~{Aldcroft}, M.~{Davis}, A.~{Ginsburg}, A.~M.
  {Price-Whelan}, W.~E. {Kerzendorf}, A.~{Conley}, N.~{Crighton}, K.~{Barbary},
  D.~{Muna}, H.~{Ferguson}, F.~{Grollier}, M.~M. {Parikh}, P.~H. {Nair}, H.~M.
  {Unther}, C.~{Deil}, J.~{Woillez}, S.~{Conseil}, R.~{Kramer}, J.~E.~H.
  {Turner}, L.~{Singer}, R.~{Fox}, B.~A. {Weaver}, V.~{Zabalza}, Z.~I.
  {Edwards}, K.~{Azalee Bostroem}, D.~J. {Burke}, A.~R. {Casey}, S.~M.
  {Crawford}, N.~{Dencheva}, J.~{Ely}, T.~{Jenness}, K.~{Labrie}, P.~L. {Lim},
  F.~{Pierfederici}, A.~{Pontzen}, A.~{Ptak}, B.~{Refsdal}, M.~{Servillat}, and
  O.~{Streicher}.
\newblock {Astropy: A community Python package for astronomy}.
\newblock \emph{\aap}, 558:\penalty0 A33, Oct. 2013.
\newblock \doi{10.1051/0004-6361/201322068}.

\bibitem[{Astropy Collaboration} et~al.(2018){Astropy Collaboration},
  {Price-Whelan}, {Sip{\H o}cz}, {G{\"u}nther}, {Lim}, {Crawford}, {Conseil},
  {Shupe}, {Craig}, {Dencheva}, {Ginsburg}, {VanderPlas}, {Bradley},
  {P{\'e}rez-Su{\'a}rez}, {de Val-Borro}, {Aldcroft}, {Cruz}, {Robitaille},
  {Tollerud}, {Ardelean}, {Babej}, {Bach}, {Bachetti}, {Bakanov}, {Bamford},
  {Barentsen}, {Barmby}, {Baumbach}, {Berry}, {Biscani}, {Boquien}, {Bostroem},
  {Bouma}, {Brammer}, {Bray}, {Breytenbach}, {Buddelmeijer}, {Burke},
  {Calderone}, {Cano Rodr{\'{\i}}guez}, {Cara}, {Cardoso}, {Cheedella},
  {Copin}, {Corrales}, {Crichton}, {D'Avella}, {Deil}, {Depagne}, {Dietrich},
  {Donath}, {Droettboom}, {Earl}, {Erben}, {Fabbro}, {Ferreira}, {Finethy},
  {Fox}, {Garrison}, {Gibbons}, {Goldstein}, {Gommers}, {Greco}, {Greenfield},
  {Groener}, {Grollier}, {Hagen}, {Hirst}, {Homeier}, {Horton}, {Hosseinzadeh},
  {Hu}, {Hunkeler}, {Ivezi{\'c}}, {Jain}, {Jenness}, {Kanarek}, {Kendrew},
  {Kern}, {Kerzendorf}, {Khvalko}, {King}, {Kirkby}, {Kulkarni}, {Kumar},
  {Lee}, {Lenz}, {Littlefair}, {Ma}, {Macleod}, {Mastropietro}, {McCully},
  {Montagnac}, {Morris}, {Mueller}, {Mumford}, {Muna}, {Murphy}, {Nelson},
  {Nguyen}, {Ninan}, {N{\"o}the}, {Ogaz}, {Oh}, {Parejko}, {Parley}, {Pascual},
  {Patil}, {Patil}, {Plunkett}, {Prochaska}, {Rastogi}, {Reddy Janga},
  {Sabater}, {Sakurikar}, {Seifert}, {Sherbert}, {Sherwood-Taylor}, {Shih},
  {Sick}, {Silbiger}, {Singanamalla}, {Singer}, {Sladen}, {Sooley},
  {Sornarajah}, {Streicher}, {Teuben}, {Thomas}, {Tremblay}, {Turner},
  {Terr{\'o}n}, {van Kerkwijk}, {de la Vega}, {Watkins}, {Weaver}, {Whitmore},
  {Woillez}, {Zabalza}, and {Astropy Contributors}]{Astropy2018}
{Astropy Collaboration}, A.~M. {Price-Whelan}, B.~M. {Sip{\H o}cz}, H.~M.
  {G{\"u}nther}, P.~L. {Lim}, S.~M. {Crawford}, S.~{Conseil}, D.~L. {Shupe},
  M.~W. {Craig}, N.~{Dencheva}, A.~{Ginsburg}, J.~T. {VanderPlas}, L.~D.
  {Bradley}, D.~{P{\'e}rez-Su{\'a}rez}, M.~{de Val-Borro}, T.~L. {Aldcroft},
  K.~L. {Cruz}, T.~P. {Robitaille}, E.~J. {Tollerud}, C.~{Ardelean},
  T.~{Babej}, Y.~P. {Bach}, M.~{Bachetti}, A.~V. {Bakanov}, S.~P. {Bamford},
  G.~{Barentsen}, P.~{Barmby}, A.~{Baumbach}, K.~L. {Berry}, F.~{Biscani},
  M.~{Boquien}, K.~A. {Bostroem}, L.~G. {Bouma}, G.~B. {Brammer}, E.~M. {Bray},
  H.~{Breytenbach}, H.~{Buddelmeijer}, D.~J. {Burke}, G.~{Calderone}, J.~L.
  {Cano Rodr{\'{\i}}guez}, M.~{Cara}, J.~V.~M. {Cardoso}, S.~{Cheedella},
  Y.~{Copin}, L.~{Corrales}, D.~{Crichton}, D.~{D'Avella}, C.~{Deil},
  {\'E}.~{Depagne}, J.~P. {Dietrich}, A.~{Donath}, M.~{Droettboom}, N.~{Earl},
  T.~{Erben}, S.~{Fabbro}, L.~A. {Ferreira}, T.~{Finethy}, R.~T. {Fox}, L.~H.
  {Garrison}, S.~L.~J. {Gibbons}, D.~A. {Goldstein}, R.~{Gommers}, J.~P.
  {Greco}, P.~{Greenfield}, A.~M. {Groener}, F.~{Grollier}, A.~{Hagen},
  P.~{Hirst}, D.~{Homeier}, A.~J. {Horton}, G.~{Hosseinzadeh}, L.~{Hu}, J.~S.
  {Hunkeler}, {\v Z}.~{Ivezi{\'c}}, A.~{Jain}, T.~{Jenness}, G.~{Kanarek},
  S.~{Kendrew}, N.~S. {Kern}, W.~E. {Kerzendorf}, A.~{Khvalko}, J.~{King},
  D.~{Kirkby}, A.~M. {Kulkarni}, A.~{Kumar}, A.~{Lee}, D.~{Lenz}, S.~P.
  {Littlefair}, Z.~{Ma}, D.~M. {Macleod}, M.~{Mastropietro}, C.~{McCully},
  S.~{Montagnac}, B.~M. {Morris}, M.~{Mueller}, S.~J. {Mumford}, D.~{Muna},
  N.~A. {Murphy}, S.~{Nelson}, G.~H. {Nguyen}, J.~P. {Ninan}, M.~{N{\"o}the},
  S.~{Ogaz}, S.~{Oh}, J.~K. {Parejko}, N.~{Parley}, S.~{Pascual}, R.~{Patil},
  A.~A. {Patil}, A.~L. {Plunkett}, J.~X. {Prochaska}, T.~{Rastogi}, V.~{Reddy
  Janga}, J.~{Sabater}, P.~{Sakurikar}, M.~{Seifert}, L.~E. {Sherbert},
  H.~{Sherwood-Taylor}, A.~Y. {Shih}, J.~{Sick}, M.~T. {Silbiger},
  S.~{Singanamalla}, L.~P. {Singer}, P.~H. {Sladen}, K.~A. {Sooley},
  S.~{Sornarajah}, O.~{Streicher}, P.~{Teuben}, S.~W. {Thomas}, G.~R.
  {Tremblay}, J.~E.~H. {Turner}, V.~{Terr{\'o}n}, M.~H. {van Kerkwijk}, A.~{de
  la Vega}, L.~L. {Watkins}, B.~A. {Weaver}, J.~B. {Whitmore}, J.~{Woillez},
  V.~{Zabalza}, and {Astropy Contributors}.
\newblock {The Astropy Project: Building an Open-science Project and Status of
  the v2.0 Core Package}.
\newblock \emph{\aj}, 156:\penalty0 123, Sept. 2018.
\newblock \doi{10.3847/1538-3881/aabc4f}.

\bibitem[{Babar} and {Weaver}(2015)]{Babar2015}
S.~{Babar} and J.~H. {Weaver}.
\newblock {Optical constants of Cu, Ag, and Au revisited}.
\newblock \emph{\ao}, 54\penalty0 (3):\penalty0 477, Jan 2015.
\newblock \doi{10.1364/AO.54.000477}.

\bibitem[{Batalha} et~al.(2011){Batalha}, {Borucki}, {Bryson}, {Buchhave},
  {Caldwell}, {Christensen-Dalsgaard}, {Ciardi}, {Dunham}, {Fressin}, and
  {Gautier}]{Batalha2011}
N.~M. {Batalha}, W.~J. {Borucki}, S.~T. {Bryson}, L.~A. {Buchhave}, D.~A.
  {Caldwell}, J.~{Christensen-Dalsgaard}, D.~{Ciardi}, E.~W. {Dunham},
  F.~{Fressin}, and I.~{Gautier}, Thomas~N.
\newblock {Kepler's First Rocky Planet: Kepler-10b}.
\newblock \emph{\apj}, 729\penalty0 (1):\penalty0 27, Mar 2011.
\newblock \doi{10.1088/0004-637X/729/1/27}.

\bibitem[Bennett and Porteus(1961)]{Bennett1961}
H.~E. Bennett and J.~O. Porteus.
\newblock Relation between surface roughness and specular reflectance at normal
  incidence.
\newblock \emph{J. Opt. Soc. Am.}, 51\penalty0 (2):\penalty0 123--129, Feb
  1961.
\newblock \doi{10.1364/JOSA.51.000123}.
\newblock URL
  \url{http://www.osapublishing.org/abstract.cfm?URI=josa-51-2-123}.

\bibitem[{Benz} et~al.(1988){Benz}, {Slattery}, and {Cameron}]{Benz1988}
W.~{Benz}, W.~L. {Slattery}, and A.~G.~W. {Cameron}.
\newblock {Collisional stripping of Mercury's mantle}.
\newblock \emph{\icarus}, 74\penalty0 (3):\penalty0 516--528, Jun 1988.
\newblock \doi{10.1016/0019-1035(88)90118-2}.

\bibitem[{Blander} et~al.(2009){Blander}, {Pelton}, and {Jung}]{Blander2009}
M.~{Blander}, A.~D. {Pelton}, and I.~H. {Jung}.
\newblock {A condensation model for the formation of chondrules in enstatite
  chondrites}.
\newblock \emph{Meteoritics and Planetary Science}, 44\penalty0 (4):\penalty0
  531--543, May 2009.
\newblock \doi{10.1111/j.1945-5100.2009.tb00749.x}.

\bibitem[{Bourrier} et~al.(2018){Bourrier}, {Dumusque}, {Dorn}, {Henry},
  {Astudillo-Defru}, {Rey}, {Benneke}, {H{\'e}brard}, {Lovis}, and
  {Demory}]{Bourrier2018(2)}
V.~{Bourrier}, X.~{Dumusque}, C.~{Dorn}, G.~W. {Henry}, N.~{Astudillo-Defru},
  J.~{Rey}, B.~{Benneke}, G.~{H{\'e}brard}, C.~{Lovis}, and B.~O. {Demory}.
\newblock {The 55 Cancri system reassessed}.
\newblock \emph{\aap}, 619:\penalty0 A1, Oct 2018.
\newblock \doi{10.1051/0004-6361/201833154}.

\bibitem[Bundy et~al.(1996)Bundy, Bassett, Weathers, Hemley, Mao, and
  Goncharov]{Bundy1996}
F.~Bundy, W.~Bassett, M.~Weathers, R.~Hemley, H.~Mao, and A.~Goncharov.
\newblock The pressure-temperature phase and transformation diagram for carbon;
  updated through 1994.
\newblock \emph{Carbon}, 34\penalty0 (2):\penalty0 141--153, 1996.
\newblock ISSN 0008-6223.

\bibitem[{Bundy}(1989)]{Bundy1989}
F.~P. {Bundy}.
\newblock {Pressure-temperature phase diagram of elemental carbon}.
\newblock \emph{Physica A Statistical Mechanics and its Applications},
  156\penalty0 (1):\penalty0 169--178, Mar 1989.
\newblock \doi{10.1016/0378-4371(89)90115-5}.

\bibitem[{Cameron}(1985)]{Cameron1985}
A.~G.~W. {Cameron}.
\newblock {The partial volatilization of Mercury}.
\newblock \emph{\icarus}, 64\penalty0 (2):\penalty0 285--294, Nov 1985.
\newblock \doi{10.1016/0019-1035(85)90091-0}.

\bibitem[{Chabot} et~al.(2014){Chabot}, {Wollack}, {Klima}, and
  {Minitti}]{Chabot2014}
N.~L. {Chabot}, E.~A. {Wollack}, R.~L. {Klima}, and M.~E. {Minitti}.
\newblock {Experimental constraints on Mercury's core composition}.
\newblock \emph{Earth and Planetary Science Letters}, 390:\penalty0 199--208,
  Mar 2014.
\newblock \doi{10.1016/j.epsl.2014.01.004}.

\bibitem[{Cierniewski} et~al.(2013){Cierniewski}, {Karnieli}, {Ku{\'s}nierek},
  {Goldberg}, and {Herrmann}]{Cierniewski2013}
J.~{Cierniewski}, A.~{Karnieli}, K.~{Ku{\'s}nierek}, A.~{Goldberg}, and
  I.~{Herrmann}.
\newblock {Approximating the average daily surface albedo with respect to soil
  roughness and latitude}.
\newblock \emph{International Journal of Remote Sensing}, 34\penalty0
  (9-10):\penalty0 3416--3424, May 2013.
\newblock \doi{10.1080/01431161.2012.716530}.

\bibitem[{Correa} et~al.(2006){Correa}, {Bonev}, and {Galli}]{Correa2006}
A.~A. {Correa}, S.~A. {Bonev}, and G.~{Galli}.
\newblock {Carbon under extreme conditions: Phase boundaries and electronic
  properties from first-principles theory}.
\newblock \emph{Proceedings of the National Academy of Science}, 103:\penalty0
  1204--1208, Jan. 2006.
\newblock \doi{10.1073/pnas.0510489103}.

\bibitem[{CRC Handbook}(2011)]{CRC92}
{CRC Handbook}.
\newblock \emph{CRC Handbook of Chemistry and Physics, 92nd Edition}.
\newblock 92 edition, Jun 2011.
\newblock ISBN 1439855110.

\bibitem[{Dai} et~al.(2019){Dai}, {Masuda}, {Winn}, and {Zeng}]{Dai2019}
F.~{Dai}, K.~{Masuda}, J.~N. {Winn}, and L.~{Zeng}.
\newblock {Homogeneous Analysis of Hot Earths: Masses, Sizes, and
  Compositions}.
\newblock \emph{\apj}, 883\penalty0 (1):\penalty0 79, Sep 2019.
\newblock \doi{10.3847/1538-4357/ab3a3b}.

\bibitem[Davies and Harte(1987)]{Davies1987}
R.~B. Davies and D.~S. Harte.
\newblock Tests for hurst effect.
\newblock \emph{Biometrika}, 74\penalty0 (1):\penalty0 95--101, 1987.
\newblock ISSN 00063444.

\bibitem[{de Wijs} et~al.(1998){de Wijs}, {Kresse}, {Vo{\v{c}}adlo}, {Dobson},
  {Alf{\`e}}, {Gillan}, and {Price}]{Alfe1998}
G.~A. {de Wijs}, G.~{Kresse}, L.~{Vo{\v{c}}adlo}, D.~{Dobson}, D.~{Alf{\`e}},
  M.~J. {Gillan}, and G.~D. {Price}.
\newblock {The viscosity of liquid iron at the physical conditions of the
  Earth's core}.
\newblock \emph{\nat}, 392\penalty0 (6678):\penalty0 805--807, Apr. 1998.
\newblock \doi{10.1038/33905}.

\bibitem[{Deli{\`e}ge} et~al.(2017){Deli{\`e}ge}, {Kleyntssens}, and
  {Nicolay}]{Deliege2017}
A.~{Deli{\`e}ge}, T.~{Kleyntssens}, and S.~{Nicolay}.
\newblock {Mars topography investigated through the wavelet leaders method: A
  multidimensional study of its fractal structure}.
\newblock \emph{\planss}, 136:\penalty0 46--58, Feb 2017.
\newblock \doi{10.1016/j.pss.2016.12.008}.

\bibitem[{Demin} et~al.(2017){Demin}, {Andreev}, {Demina}, and
  {Nefedyev}]{Demin2017}
S.~A. {Demin}, A.~O. {Andreev}, N.~Y. {Demina}, and Y.~A. {Nefedyev}.
\newblock {The fractal analysis of the gravitational field and topography of
  the Mars}.
\newblock In \emph{Journal of Physics Conference Series}, volume 929, page
  012002, Nov 2017.
\newblock \doi{10.1088/1742-6596/929/1/012002}.

\bibitem[{Demin} et~al.(2018){Demin}, {Andreev}, {Demina}, and
  {Nefedyev}]{Demin2018}
S.~A. {Demin}, A.~O. {Andreev}, N.~Y. {Demina}, and Y.~A. {Nefedyev}.
\newblock {The fractal analysis of the topography and gravitational field of
  Venus}.
\newblock In \emph{Journal of Physics Conference Series}, volume 1038, page
  012020, Jun 2018.
\newblock \doi{10.1088/1742-6596/1038/1/012020}.

\bibitem[{Demory}(2014)]{Demory2014}
B.-O. {Demory}.
\newblock {The Albedos of Kepler's Close-in Super-Earths}.
\newblock \emph{\apjl}, 789:\penalty0 L20, July 2014.
\newblock \doi{10.1088/2041-8205/789/1/L20}.

\bibitem[{Dingwell} et~al.(2004){Dingwell}, {Courtial}, {Giordano}, and
  {Nichols}]{Dingwell2004}
D.~B. {Dingwell}, P.~{Courtial}, D.~{Giordano}, and A.~R.~L. {Nichols}.
\newblock {Viscosity of peridotite liquid}.
\newblock \emph{Earth and Planetary Science Letters}, 226\penalty0
  (1-2):\penalty0 127--138, Sept. 2004.
\newblock \doi{10.1016/j.epsl.2004.07.017}.

\bibitem[{Djuri{\v{s}}i{\'c}} and {Li}(1999)]{Djurisic1999}
A.~B. {Djuri{\v{s}}i{\'c}} and E.~H. {Li}.
\newblock {Optical properties of graphite}.
\newblock \emph{Journal of Applied Physics}, 85\penalty0 (10):\penalty0
  7404--7410, May 1999.
\newblock \doi{10.1063/1.369370}.

\bibitem[{Dorn} et~al.(2017){Dorn}, {Venturini}, {Khan}, {Heng}, {Alibert},
  {Helled}, {Rivoldini}, and {Benz}]{Dorn2017}
C.~{Dorn}, J.~{Venturini}, A.~{Khan}, K.~{Heng}, Y.~{Alibert}, R.~{Helled},
  A.~{Rivoldini}, and W.~{Benz}.
\newblock {A generalized Bayesian inference method for constraining the
  interiors of super Earths and sub-Neptunes}.
\newblock \emph{\aap}, 597:\penalty0 A37, Jan. 2017.
\newblock \doi{10.1051/0004-6361/201628708}.

\bibitem[{Dostal} and {Mueller}(2004)]{Dostal2004}
J.~{Dostal} and W.~{Mueller}.
\newblock \emph{Komatiite Geochemistry}, volume~12, pages 290--298.
\newblock 01 2004.

\bibitem[{Dumusque} et~al.(2014){Dumusque}, {Bonomo}, {Haywood}, {Malavolta},
  {S{\'e}gransan}, {Buchhave}, {Collier Cameron}, {Latham}, {Molinari}, and
  {Pepe}]{Dumusque2014}
X.~{Dumusque}, A.~S. {Bonomo}, R.~D. {Haywood}, L.~{Malavolta},
  D.~{S{\'e}gransan}, L.~A. {Buchhave}, A.~{Collier Cameron}, D.~W. {Latham},
  E.~{Molinari}, and F.~{Pepe}.
\newblock {The Kepler-10 Planetary System Revisited by HARPS-N: A Hot Rocky
  World and a Solid Neptune-Mass Planet}.
\newblock \emph{\apj}, 789\penalty0 (2):\penalty0 154, Jul 2014.
\newblock \doi{10.1088/0004-637X/789/2/154}.

\bibitem[{Durst} et~al.(2011){Durst}, {Mason}, {McKinley}, and
  {Baylot}]{Durst2011}
P.~{Durst}, G.~{Mason}, B.~{McKinley}, and A.~{Baylot}.
\newblock {Predicting RMS surface roughness using fractal dimension and PSD
  parameters}.
\newblock \emph{Journal of Terramechanics}, 48\penalty0 (2):\penalty0 105 --
  111, Apr 2011.
\newblock \doi{https://doi.org/10.1016/j.jterra.2010.05.004}.

\bibitem[{Edgett} and {Rice}(1997)]{Edgett1997}
K.~S. {Edgett} and J.~{Rice}, J.~W.
\newblock {Geologic Signature of Life on Mars: Low-Albedo Lava Flows and the
  Search for ``Warm Havens''}.
\newblock In S.~M. {Clifford}, A.~H. {Treiman}, H.~E. {Newsom}, and J.~D.
  {Farmer}, editors, \emph{Early Mars: Geologic and Hydrologic Evolution,
  Physical and Chemical Environments, and the Implications for Life}, volume
  916, page~29, Jan. 1997.

\bibitem[{Egan} et~al.(1975){Egan}, {Hilgeman}, and {Pang}]{Egan1975}
W.~G. {Egan}, T.~{Hilgeman}, and K.~{Pang}.
\newblock {Ultraviolet Complex Refractive Index of Martian Dust: Laboratory
  Measurements of Terrestrial Analogs}.
\newblock \emph{\icarus}, 25\penalty0 (2):\penalty0 344--355, Jun 1975.
\newblock \doi{10.1016/0019-1035(75)90029-9}.

\bibitem[{Ehrenreich} and {D{\'e}sert}(2011)]{Ehrenreich2011}
D.~{Ehrenreich} and J.-M. {D{\'e}sert}.
\newblock {Mass-loss rates for transiting exoplanets}.
\newblock \emph{\aap}, 529:\penalty0 A136, May 2011.
\newblock \doi{10.1051/0004-6361/201016356}.

\bibitem[{Elkins-Tanton} and {Seager}(2008)]{Elkins2008}
L.~T. {Elkins-Tanton} and S.~{Seager}.
\newblock {Coreless Terrestrial Exoplanets}.
\newblock \emph{\apj}, 688\penalty0 (1):\penalty0 628--635, Nov 2008.
\newblock \doi{10.1086/592316}.

\bibitem[{Emanuel}(1994)]{Emanuel1994}
K.~A. {Emanuel}.
\newblock \emph{Atmospheric Convection}.
\newblock Oxford University Press, 1994.

\bibitem[{Essack} et~al.(2020){Essack}, {Seager}, and {Pajusalu}]{Essack2020}
Z.~{Essack}, S.~{Seager}, and M.~{Pajusalu}.
\newblock {Low-albedo Surfaces of Lava Worlds}.
\newblock \emph{\apj}, 898\penalty0 (2):\penalty0 160, Aug. 2020.
\newblock \doi{10.3847/1538-4357/ab9cba}.

\bibitem[{Esteves} et~al.(2015){Esteves}, {De Mooij}, and
  {Jayawardhana}]{Esteves2015}
L.~J. {Esteves}, E.~J.~W. {De Mooij}, and R.~{Jayawardhana}.
\newblock {Changing Phases of Alien Worlds: Probing Atmospheres of Kepler
  Planets with High-precision Photometry}.
\newblock \emph{\apj}, 804\penalty0 (2):\penalty0 150, May 2015.
\newblock \doi{10.1088/0004-637X/804/2/150}.

\bibitem[{Esteves} et~al.(2017){Esteves}, {de Mooij}, {Jayawardhana}, {Watson},
  and {de Kok}]{Esteves2017}
L.~J. {Esteves}, E.~J.~W. {de Mooij}, R.~{Jayawardhana}, C.~{Watson}, and
  R.~{de Kok}.
\newblock {A Search for Water in a Super-Earth Atmosphere: High-resolution
  Optical Spectroscopy of 55Cancri e}.
\newblock \emph{\aj}, 153:\penalty0 268, June 2017.
\newblock \doi{10.3847/1538-3881/aa7133}.

\bibitem[{Evans} et~al.(2012){Evans}, {Peplowski}, {Rhodes}, {Lawrence},
  {McCoy}, {Nittler}, {Solomon}, {Sprague}, {Stockstill-Cahill}, {Starr},
  {Weider}, {Boynton}, {Hamara}, and {Goldsten}]{Evans2012}
L.~G. {Evans}, P.~N. {Peplowski}, E.~A. {Rhodes}, D.~J. {Lawrence}, T.~J.
  {McCoy}, L.~R. {Nittler}, S.~C. {Solomon}, A.~L. {Sprague}, K.~R.
  {Stockstill-Cahill}, R.~D. {Starr}, S.~Z. {Weider}, W.~V. {Boynton}, D.~K.
  {Hamara}, and J.~O. {Goldsten}.
\newblock {Major-element abundances on the surface of Mercury: Results from the
  MESSENGER Gamma-Ray Spectrometer}.
\newblock \emph{Journal of Geophysical Research (Planets)}, 117:\penalty0
  E00L07, Nov 2012.
\newblock \doi{10.1029/2012JE004178}.

\bibitem[{Fabian} et~al.(2001){Fabian}, {Henning}, {J{\"a}ger}, {Mutschke},
  {Dorschner}, and {Wehrhan}]{Fabian2001}
D.~{Fabian}, T.~{Henning}, C.~{J{\"a}ger}, H.~{Mutschke}, J.~{Dorschner}, and
  O.~{Wehrhan}.
\newblock {Steps toward interstellar silicate mineralogy. VI. Dependence of
  crystalline olivine IR spectra on iron content and particle shape}.
\newblock \emph{\aap}, 378:\penalty0 228--238, Oct 2001.
\newblock \doi{10.1051/0004-6361:20011196}.

\bibitem[{Fairbairn}(2005)]{Fairbairn2005}
M.~B. {Fairbairn}.
\newblock {Planetary Photometry: The Lommel-Seeliger Law}.
\newblock \emph{\jrasc}, 99\penalty0 (3):\penalty0 92, June 2005.

\bibitem[{Fegley} and {Schaefer}(2009)]{Fegley2009}
B.~{Fegley} and L.~{Schaefer}.
\newblock {Silicate Atmosphere and Clouds of Hot Earth-like Exoplanets}.
\newblock \emph{Meteoritics and Planetary Science Supplement}, 72:\penalty0
  5032, Sep 2009.

\bibitem[{Fogtmann-Schulz} et~al.(2014){Fogtmann-Schulz}, {Hinrup}, {Van
  Eylen}, {Christensen-Dalsgaard}, {Kjeldsen}, {Silva Aguirre}, and
  {Tingley}]{Fogtmann2014}
A.~{Fogtmann-Schulz}, B.~{Hinrup}, V.~{Van Eylen}, J.~{Christensen-Dalsgaard},
  H.~{Kjeldsen}, V.~{Silva Aguirre}, and B.~o. {Tingley}.
\newblock {Accurate Parameters of the Oldest Known Rocky-exoplanet Hosting
  System: Kepler-10 Revisited}.
\newblock \emph{\apj}, 781\penalty0 (2):\penalty0 67, Feb 2014.
\newblock \doi{10.1088/0004-637X/781/2/67}.

\bibitem[{Fornasier} et~al.(2016){Fornasier}, {Lantz}, {Perna}, {Campins},
  {Barucci}, and {Nesvorny}]{Fornasier2016}
S.~{Fornasier}, C.~{Lantz}, D.~{Perna}, H.~{Campins}, M.~A. {Barucci}, and
  D.~{Nesvorny}.
\newblock {Spectral variability on primitive asteroids of the Themis and Beagle
  families: Space weathering effects or parent body heterogeneity?}
\newblock \emph{\icarus}, 269:\penalty0 1--14, May 2016.
\newblock \doi{10.1016/j.icarus.2016.01.002}.

\bibitem[Fulcher(1925{\natexlab{a}})]{Fulcher1925}
G.~S. Fulcher.
\newblock Analysis of recent measurements of the viscosity of glasses: I.
\newblock \emph{Journal of the American Ceramic Society}, 8\penalty0
  (6):\penalty0 339--355, 1925{\natexlab{a}}.
\newblock \doi{10.1111/j.1151-2916.1925.tb16731.x}.
\newblock URL
  \url{https://ceramics.onlinelibrary.wiley.com/doi/abs/10.1111/j.1151-2916.1925.tb16731.x}.

\bibitem[Fulcher(1925{\natexlab{b}})]{Fulcher1925b}
G.~S. Fulcher.
\newblock Analysis of recent measurements of the viscosity of glasses: Ii.
\newblock \emph{Journal of the American Ceramic Society}, 8\penalty0
  (12):\penalty0 789--794, 1925{\natexlab{b}}.
\newblock \doi{10.1111/j.1151-2916.1925.tb18582.x}.
\newblock URL
  \url{https://ceramics.onlinelibrary.wiley.com/doi/abs/10.1111/j.1151-2916.1925.tb18582.x}.

\bibitem[{Fulton} et~al.(2017){Fulton}, {Petigura}, {Howard}, {Isaacson},
  {Marcy}, {Cargile}, {Hebb}, {Weiss}, {Johnson}, {Morton}, {Sinukoff},
  {Crossfield}, and {Hirsch}]{Fulton2017}
B.~J. {Fulton}, E.~A. {Petigura}, A.~W. {Howard}, H.~{Isaacson}, G.~W. {Marcy},
  P.~A. {Cargile}, L.~{Hebb}, L.~M. {Weiss}, J.~A. {Johnson}, T.~D. {Morton},
  E.~{Sinukoff}, I.~J.~M. {Crossfield}, and L.~A. {Hirsch}.
\newblock {The California-Kepler Survey. III. A Gap in the Radius Distribution
  of Small Planets}.
\newblock \emph{\aj}, 154:\penalty0 109, Sept. 2017.
\newblock \doi{10.3847/1538-3881/aa80eb}.

\bibitem[{Ghosh}(1999)]{Ghosh1999}
G.~{Ghosh}.
\newblock {Dispersion-equation coefficients for the refractive index and
  birefringence of calcite and quartz crystals}.
\newblock \emph{Optics Communications}, 163\penalty0 (1-3):\penalty0 95--102,
  May 1999.
\newblock \doi{10.1016/S0030-4018(99)00091-7}.

\bibitem[{Giordano} et~al.(2008){Giordano}, {Russell}, and
  {Dingwell}]{Giordano2008}
D.~{Giordano}, J.~K. {Russell}, and D.~B. {Dingwell}.
\newblock {Viscosity of magmatic liquids: A model}.
\newblock \emph{Earth and Planetary Science Letters}, 271\penalty0
  (1-4):\penalty0 123--134, July 2008.
\newblock \doi{10.1016/j.epsl.2008.03.038}.

\bibitem[{Grainger} et~al.(2014){Grainger}, {Peters}, {Clarisse}, and
  {Herbin}]{Grainger2008}
D.~{Grainger}, D.~{Peters}, L.~{Clarisse}, and H.~{Herbin}.
\newblock Aerosol refractive index archive, 2014.
\newblock URL \url{http://eodg.atm.ox.ac.uk/ARIA/}.

\bibitem[Hagemann et~al.(1975)Hagemann, Gudat, and Kunz]{Hagemann1975}
H.-J. Hagemann, W.~Gudat, and C.~Kunz.
\newblock Optical constants from the far infrared to the x-ray region: Mg, al,
  cu, ag, au, bi, c, and al2o3.
\newblock \emph{J. Opt. Soc. Am.}, 65\penalty0 (6):\penalty0 742--744, Jun
  1975.
\newblock \doi{10.1364/JOSA.65.000742}.
\newblock URL
  \url{http://www.osapublishing.org/abstract.cfm?URI=josa-65-6-742}.

\bibitem[{Hakim} et~al.(2018){Hakim}, {van Westrenen}, and
  {Dominik}]{Hakim2018}
K.~{Hakim}, W.~{van Westrenen}, and C.~{Dominik}.
\newblock {Capturing the oxidation of silicon carbide in rocky exoplanetary
  interiors}.
\newblock \emph{\aap}, 618:\penalty0 L6, Oct 2018.
\newblock \doi{10.1051/0004-6361/201833942}.

\bibitem[{Hakim} et~al.(2019){Hakim}, {Spaargaren}, {Grewal}, {Rohrbach},
  {Berndt}, {Dominik}, and {van Westrenen}]{Hakim2019}
K.~{Hakim}, R.~{Spaargaren}, D.~S. {Grewal}, A.~{Rohrbach}, J.~{Berndt},
  C.~{Dominik}, and W.~{van Westrenen}.
\newblock {Mineralogy, Structure, and Habitability of Carbon-Enriched Rocky
  Exoplanets: A Laboratory Approach}.
\newblock \emph{Astrobiology}, 19\penalty0 (7):\penalty0 867--884, Jul 2019.
\newblock \doi{10.1089/ast.2018.1930}.

\bibitem[Haltrin et~al.(2001)Haltrin, McBride~III, and Arnone]{Haltrin2001}
V.~I. Haltrin, W.~E. McBride~III, and R.~A. Arnone.
\newblock Spectral approach to calculate specular reflection of light from wavy
  water surface.
\newblock pages 133--138, 2001.

\bibitem[{Hamano} et~al.(2015){Hamano}, {Kawahara}, {Abe}, {Onishi}, and
  {Hashimoto}]{Hamano2015}
K.~{Hamano}, H.~{Kawahara}, Y.~{Abe}, M.~{Onishi}, and G.~L. {Hashimoto}.
\newblock {Lifetime and Spectral Evolution of a Magma Ocean with a Steam
  Atmosphere: Its Detectability by Future Direct Imaging}.
\newblock \emph{\apj}, 806\penalty0 (2):\penalty0 216, Jun 2015.
\newblock \doi{10.1088/0004-637X/806/2/216}.

\bibitem[Hapke(2012)]{Hapke2012}
B.~Hapke.
\newblock \emph{Theory of Reflectance and Emittance Spectroscopy}.
\newblock Cambridge University Press, 2 edition, 2012.
\newblock \doi{10.1017/CBO9781139025683}.

\bibitem[{Hauck} et~al.(2013){Hauck}, {Margot}, {Solomon}, {Phillips},
  {Johnson}, {Lemoine}, {Mazarico}, {McCoy}, {Padovan}, {Peale}, {Perry},
  {Smith}, and {Zuber}]{Hauck2013}
S.~A. {Hauck}, J.-L. {Margot}, S.~C. {Solomon}, R.~J. {Phillips}, C.~L.
  {Johnson}, F.~G. {Lemoine}, E.~{Mazarico}, T.~J. {McCoy}, S.~{Padovan}, S.~J.
  {Peale}, M.~E. {Perry}, D.~E. {Smith}, and M.~T. {Zuber}.
\newblock {The curious case of Mercury's internal structure}.
\newblock \emph{Journal of Geophysical Research (Planets)}, 118\penalty0
  (6):\penalty0 1204--1220, Jun 2013.
\newblock \doi{10.1002/jgre.20091}.

\bibitem[{Haughton} et~al.(1974){Haughton}, {Roeder}, and
  {Skinner}]{Haughton1974}
D.~R. {Haughton}, P.~L. {Roeder}, and B.~J. {Skinner}.
\newblock {Solubility of Sulfur in Mafic Magmas}.
\newblock \emph{Economic Geology}, 69\penalty0 (4):\penalty0 451--467, 07 1974.
\newblock ISSN 0361-0128.
\newblock \doi{10.2113/gsecongeo.69.4.451}.
\newblock URL \url{https://doi.org/10.2113/gsecongeo.69.4.451}.

\bibitem[{Henning} et~al.(1995){Henning}, {Begemann}, {Mutschke}, and
  {Dorschner}]{Henning1995}
T.~{Henning}, B.~{Begemann}, H.~{Mutschke}, and J.~{Dorschner}.
\newblock {Optical properties of oxide dust grains.}
\newblock \emph{\aaps}, 112:\penalty0 143, Jul 1995.

\bibitem[{Hoppe} et~al.(2010){Hoppe}, {Leitner}, {Gr{\"o}ner}, {Marhas},
  {Meyer}, and {Amari}]{Hoppe2010}
P.~{Hoppe}, J.~{Leitner}, E.~{Gr{\"o}ner}, K.~K. {Marhas}, B.~S. {Meyer}, and
  S.~{Amari}.
\newblock {NanoSIMS Studies of Small Presolar SiC Grains: New Insights into
  Supernova Nucleosynthesis, Chemistry, and Dust Formation}.
\newblock \emph{\apj}, 719\penalty0 (2):\penalty0 1370--1384, Aug 2010.
\newblock \doi{10.1088/0004-637X/719/2/1370}.

\bibitem[{Hu} et~al.(2012){Hu}, {Ehlmann}, and {Seager}]{Hu2012}
R.~{Hu}, B.~L. {Ehlmann}, and S.~{Seager}.
\newblock {Theoretical Spectra of Terrestrial Exoplanet Surfaces}.
\newblock \emph{\apj}, 752\penalty0 (1):\penalty0 7, Jun 2012.
\newblock \doi{10.1088/0004-637X/752/1/7}.

\bibitem[{Hunten} et~al.(1987){Hunten}, {Pepin}, and {Walker}]{Hunten1987}
D.~M. {Hunten}, R.~O. {Pepin}, and J.~C.~G. {Walker}.
\newblock {Mass fractionation in hydrodynamic escape}.
\newblock \emph{\icarus}, 69:\penalty0 532--549, Mar. 1987.
\newblock \doi{10.1016/0019-1035(87)90022-4}.

\bibitem[{Ito} et~al.(2015){Ito}, {Ikoma}, {Kawahara}, {Nagahara}, {Kawashima},
  and {Nakamoto}]{Ito2015}
Y.~{Ito}, M.~{Ikoma}, H.~{Kawahara}, H.~{Nagahara}, Y.~{Kawashima}, and
  T.~{Nakamoto}.
\newblock {Theoretical Emission Spectra of Atmospheres of Hot Rocky
  Super-Earths}.
\newblock \emph{\apj}, 801\penalty0 (2):\penalty0 144, Mar 2015.
\newblock \doi{10.1088/0004-637X/801/2/144}.

\bibitem[{Jaeger} et~al.(1998){Jaeger}, {Molster}, {Dorschner}, {Henning},
  {Mutschke}, and {Waters}]{Jaeger1998}
C.~{Jaeger}, F.~J. {Molster}, J.~{Dorschner}, T.~{Henning}, H.~{Mutschke}, and
  L.~B.~F.~M. {Waters}.
\newblock {Steps toward interstellar silicate mineralogy. IV. The crystalline
  revolution}.
\newblock \emph{\aap}, 339:\penalty0 904--916, Nov 1998.

\bibitem[{Javoy} et~al.(2010){Javoy}, {Kaminski}, {Guyot}, {Andrault},
  {Sanloup}, {Moreira}, {Labrosse}, {Jambon}, {Agrinier}, {Davaille}, and
  {Jaupart}]{Javoy2010}
M.~{Javoy}, E.~{Kaminski}, F.~{Guyot}, D.~{Andrault}, C.~{Sanloup},
  M.~{Moreira}, S.~{Labrosse}, A.~{Jambon}, P.~{Agrinier}, A.~{Davaille}, and
  C.~{Jaupart}.
\newblock {The chemical composition of the Earth: Enstatite chondrite models}.
\newblock \emph{Earth and Planetary Science Letters}, 293\penalty0
  (3-4):\penalty0 259--268, May 2010.
\newblock \doi{10.1016/j.epsl.2010.02.033}.

\bibitem[{Jin} and {Mordasini}(2018)]{Jin2018}
S.~{Jin} and C.~{Mordasini}.
\newblock {Compositional Imprints in Density-Distance-Time: A Rocky Composition
  for Close-in Low-mass Exoplanets from the Location of the Valley of
  Evaporation}.
\newblock \emph{\apj}, 853\penalty0 (2):\penalty0 163, Feb 2018.
\newblock \doi{10.3847/1538-4357/aa9f1e}.

\bibitem[{Jin} et~al.(2014){Jin}, {Mordasini}, {Parmentier}, {van Boekel},
  {Henning}, and {Ji}]{Jin2014}
S.~{Jin}, C.~{Mordasini}, V.~{Parmentier}, R.~{van Boekel}, T.~{Henning}, and
  J.~{Ji}.
\newblock {Planetary Population Synthesis Coupled with Atmospheric Escape: A
  Statistical View of Evaporation}.
\newblock \emph{\apj}, 795:\penalty0 65, Nov. 2014.
\newblock \doi{10.1088/0004-637X/795/1/65}.

\bibitem[Kischkat et~al.(2012)Kischkat, Peters, Gruska, Semtsiv, Chashnikova,
  Klinkm\"{u}ller, Fedosenko, Machulik, Aleksandrova, Monastyrskyi, Flores, and
  Masselink]{Kischkat2012}
J.~Kischkat, S.~Peters, B.~Gruska, M.~Semtsiv, M.~Chashnikova,
  M.~Klinkm\"{u}ller, O.~Fedosenko, S.~Machulik, A.~Aleksandrova,
  G.~Monastyrskyi, Y.~Flores, and W.~T. Masselink.
\newblock Mid-infrared optical properties of thin films of aluminum oxide,
  titanium dioxide, silicon dioxide, aluminum nitride, and silicon nitride.
\newblock \emph{Appl. Opt.}, 51\penalty0 (28):\penalty0 6789--6798, Oct 2012.
\newblock \doi{10.1364/AO.51.006789}.
\newblock URL \url{http://ao.osa.org/abstract.cfm?URI=ao-51-28-6789}.

\bibitem[{Kite} et~al.(2016){Kite}, {Fegley}, {Schaefer}, and
  {Gaidos}]{Kite2016}
E.~S. {Kite}, B.~{Fegley}, Jr., L.~{Schaefer}, and E.~{Gaidos}.
\newblock {Atmosphere-interior Exchange on Hot, Rocky Exoplanets}.
\newblock \emph{\apj}, 828:\penalty0 80, Sept. 2016.
\newblock \doi{10.3847/0004-637X/828/2/80}.

\bibitem[{Kubyshkina} et~al.(2018){Kubyshkina}, {Fossati}, {Erkaev},
  {Johnstone}, {Cubillos}, {Kislyakova}, {Lammer}, {Lendl}, and
  {Odert}]{Kubyshkina2018(2)}
D.~{Kubyshkina}, L.~{Fossati}, N.~V. {Erkaev}, C.~P. {Johnstone}, P.~E.
  {Cubillos}, K.~G. {Kislyakova}, H.~{Lammer}, M.~{Lendl}, and P.~{Odert}.
\newblock {Grid of upper atmosphere models for 1-40 M$_{\oplus}$ planets:
  application to CoRoT-7 b and HD 219134 b,c}.
\newblock \emph{\aap}, 619:\penalty0 A151, Nov. 2018.
\newblock \doi{10.1051/0004-6361/201833737}.

\bibitem[{Kuzmenko} et~al.(2008){Kuzmenko}, {van Heumen}, {Carbone}, and {van
  der Marel}]{Kuzmenko2008}
A.~B. {Kuzmenko}, E.~{van Heumen}, F.~{Carbone}, and D.~{van der Marel}.
\newblock {Universal Optical Conductance of Graphite}.
\newblock \emph{\prl}, 100\penalty0 (11):\penalty0 117401, Mar 2008.
\newblock \doi{10.1103/PhysRevLett.100.117401}.

\bibitem[{Lammer} et~al.(2013){Lammer}, {Erkaev}, {Odert}, {Kislyakova},
  {Leitzinger}, and {Khodachenko}]{Lammer2013}
H.~{Lammer}, N.~V. {Erkaev}, P.~{Odert}, K.~G. {Kislyakova}, M.~{Leitzinger},
  and M.~L. {Khodachenko}.
\newblock {Probing the blow-off criteria of hydrogen-rich `super-Earths'}.
\newblock \emph{\mnras}, 430:\penalty0 1247--1256, Apr. 2013.
\newblock \doi{10.1093/mnras/sts705}.

\bibitem[Lazányi and Szirmay-Kalos(2005)]{Lazanyi2005}
I.~Lazányi and L.~Szirmay-Kalos.
\newblock Fresnel term approximations for metals.
\newblock In \emph{WSCG’2005: Short Papers Proceedings}, pages 77--80,
  University of West Bohemia, Plzen, Czech Republic, 01 2005. University of
  West Bohemia.

\bibitem[{Le Maitre} et~al.(2002){Le Maitre}, {Streckeisen}, {Zanettin}, {Le
  Bas}, {Bonin}, and {Bateman}]{Lemaitre2002}
R.~{Le Maitre}, A.~{Streckeisen}, B.~{Zanettin}, M.~{Le Bas}, B.~{Bonin}, and
  P.~{Bateman}.
\newblock \emph{Igneous Rocks: A Classification and Glossary of Terms:
  Recommendations of the International Union of Geological Sciences
  Subcommission on the Systematics of Igneous Rocks}.
\newblock Cambridge University Press, 2 edition, 2002.
\newblock \doi{10.1017/CBO9780511535581}.

\bibitem[{Lecavelier Des Etangs}(2007)]{Lecavelier2007}
A.~{Lecavelier Des Etangs}.
\newblock {A diagram to determine the evaporation status of extrasolar
  planets}.
\newblock \emph{\aap}, 461:\penalty0 1185--1193, Jan. 2007.
\newblock \doi{10.1051/0004-6361:20065014}.

\bibitem[{Lewis}(1972)]{Lewis1972}
J.~S. {Lewis}.
\newblock {Metal/silicate fractionation in the Solar System}.
\newblock \emph{Earth and Planetary Science Letters}, 15\penalty0 (3):\penalty0
  286--290, Jul 1972.
\newblock \doi{10.1016/0012-821X(72)90174-4}.

\bibitem[{Lewis}(1974)]{Lewis1974}
J.~S. {Lewis}.
\newblock {The Temperature Gradient in the Solar Nebula}.
\newblock \emph{Science}, 186\penalty0 (4162):\penalty0 440--443, Nov 1974.
\newblock \doi{10.1126/science.186.4162.440}.

\bibitem[{Lhermitte} et~al.(2014){Lhermitte}, {Abermann}, and
  {Kinnard}]{Lhermitte2014}
S.~{Lhermitte}, J.~{Abermann}, and C.~{Kinnard}.
\newblock {Albedo over rough snow and ice surfaces}.
\newblock \emph{The Cryosphere}, 8\penalty0 (3):\penalty0 1069--1086, Jun 2014.
\newblock \doi{10.5194/tc-8-1069-2014}.

\bibitem[{Li}(1980)]{Li1980}
H.~H. {Li}.
\newblock {Refractive index of silicon and germanium and its wavelength and
  temperature derivatives}.
\newblock \emph{Journal of Physical and Chemical Reference Data}, 9\penalty0
  (3):\penalty0 561--658, Jul 1980.
\newblock \doi{10.1063/1.555624}.

\bibitem[{Liu} and {Sieckmann}(1966)]{Liu1966}
C.~J. {Liu} and E.~F. {Sieckmann}.
\newblock {Refractive Index of Calcium Oxide}.
\newblock \emph{Journal of Applied Physics}, 37\penalty0 (6):\penalty0
  2450--2452, May 1966.
\newblock \doi{10.1063/1.1708835}.

\bibitem[{Locci} et~al.(2019){Locci}, {Cecchi-Pestellini}, and
  {Micela}]{Locci2019}
D.~{Locci}, C.~{Cecchi-Pestellini}, and G.~{Micela}.
\newblock {Photo-evaporation of close-in gas giants orbiting around G and M
  stars}.
\newblock \emph{\aap}, 624:\penalty0 A101, Apr 2019.
\newblock \doi{10.1051/0004-6361/201834491}.

\bibitem[{Lodders} and {Fegley}(1995)]{Lodders1995}
K.~{Lodders} and J.~{Fegley}, B.
\newblock {The origin of circumstellar silicon carbide grains found in
  meteorites}.
\newblock \emph{Meteoritics}, 30\penalty0 (6):\penalty0 661, Nov 1995.
\newblock \doi{10.1111/j.1945-5100.1995.tb01164.x}.

\bibitem[{Lopez}(2017)]{Lopez2017}
E.~D. {Lopez}.
\newblock {Born dry in the photoevaporation desert: Kepler's ultra-short-period
  planets formed water-poor}.
\newblock \emph{\mnras}, 472\penalty0 (1):\penalty0 245--253, Nov. 2017.
\newblock \doi{10.1093/mnras/stx1558}.

\bibitem[{Lorenz} and {Hayes}(2012)]{Lorenz2012}
R.~D. {Lorenz} and A.~G. {Hayes}.
\newblock {The growth of wind-waves in Titan's hydrocarbon seas}.
\newblock \emph{\icarus}, 219\penalty0 (1):\penalty0 468--475, May 2012.
\newblock \doi{10.1016/j.icarus.2012.03.002}.

\bibitem[{Luger} and {Barnes}(2015)]{Luger2015}
R.~{Luger} and R.~{Barnes}.
\newblock {Extreme Water Loss and Abiotic O$_{2}$ Buildup On Planets Throughout
  the Habitable Zones of M Dwarfs}.
\newblock In \emph{American Astronomical Society Meeting Abstracts \#225},
  volume 225 of \emph{American Astronomical Society Meeting Abstracts}, page
  407.04, Jan 2015.

\bibitem[{Madhusudhan} et~al.(2012){Madhusudhan}, {Lee}, and
  {Mousis}]{Madhusudhan2012}
N.~{Madhusudhan}, K.~K.~M. {Lee}, and O.~{Mousis}.
\newblock {A Possible Carbon-rich Interior in Super-Earth 55 Cancri e}.
\newblock \emph{\apj}, 759\penalty0 (2):\penalty0 L40, Nov 2012.
\newblock \doi{10.1088/2041-8205/759/2/L40}.

\bibitem[{Mahapatra} et~al.(2017){Mahapatra}, {Helling}, and
  {Miguel}]{Mahapatra2017}
G.~{Mahapatra}, C.~{Helling}, and Y.~{Miguel}.
\newblock {Cloud formation in metal-rich atmospheres of hot super-Earths like
  55 Cnc e and CoRoT7b}.
\newblock \emph{\mnras}, 472\penalty0 (1):\penalty0 447--464, Nov 2017.
\newblock \doi{10.1093/mnras/stx1666}.

\bibitem[{Malavergne} et~al.(2010){Malavergne}, {Toplis}, {Berthet}, and
  {Jones}]{Malavergne2010}
V.~{Malavergne}, M.~J. {Toplis}, S.~{Berthet}, and J.~{Jones}.
\newblock {Highly reducing conditions during core formation on Mercury:
  Implications for internal structure and the origin of a magnetic field}.
\newblock \emph{\icarus}, 206\penalty0 (1):\penalty0 199--209, Mar 2010.
\newblock \doi{10.1016/j.icarus.2009.09.001}.

\bibitem[{Malavergne} et~al.(2014){Malavergne}, {Cordier}, {Righter}, {Brunet},
  {Zanda}, {Addad}, {Smith}, {Bureau}, {Surbl{\'e}}, {Raepsaet}, {Charon}, and
  {Hewins}]{Malavergne2014}
V.~{Malavergne}, P.~{Cordier}, K.~{Righter}, F.~{Brunet}, B.~{Zanda},
  A.~{Addad}, T.~{Smith}, H.~{Bureau}, S.~{Surbl{\'e}}, C.~{Raepsaet},
  E.~{Charon}, and R.~H. {Hewins}.
\newblock {How Mercury can be the most reduced terrestrial planet and still
  store iron in its mantle}.
\newblock \emph{Earth and Planetary Science Letters}, 394:\penalty0 186--197,
  May 2014.
\newblock \doi{10.1016/j.epsl.2014.03.028}.

\bibitem[{Malavolta}(2018)]{Malavolta2018}
L.~{Malavolta}.
\newblock {Ultra-short Period Rocky Super-Earths}.
\newblock \emph{European Planetary Science Congress}, 12:\penalty0
  EPSC2018-1115, Sept. 2018.

\bibitem[{Malyuk}(1986)]{Malyuk1986}
B.~{Malyuk}.
\newblock Major-element series and types of komatiites.
\newblock \emph{All-Union Institute of Scientific and Technical Information,
  Moscow}, 37:\penalty0 16--12, 12 1986.

\bibitem[{Mandelbrot}(1975)]{Mandelbrot1975}
B.~B. {Mandelbrot}.
\newblock {Stochastic Models for the Earth's Relief, the Shape and the Fractal
  Dimension of the Coastlines, and the Number-Area Rule for Islands}.
\newblock \emph{Proceedings of the National Academy of Science}, 72\penalty0
  (10):\penalty0 3825--3828, Oct 1975.
\newblock \doi{10.1073/pnas.72.10.3825}.

\bibitem[{Marcus} et~al.(2010){Marcus}, {Sasselov}, {Hernquist}, and
  {Stewart}]{Marcus2010}
R.~A. {Marcus}, D.~{Sasselov}, L.~{Hernquist}, and S.~T. {Stewart}.
\newblock {Minimum Radii of Super-Earths: Constraints from Giant Impacts}.
\newblock \emph{\apjl}, 712\penalty0 (1):\penalty0 L73--L76, Mar 2010.
\newblock \doi{10.1088/2041-8205/712/1/L73}.

\bibitem[{Marley} et~al.(1999){Marley}, {Gelino}, {Stephens}, {Lunine}, and
  {Freedman}]{Marley1999}
M.~S. {Marley}, C.~{Gelino}, D.~{Stephens}, J.~I. {Lunine}, and R.~{Freedman}.
\newblock {Reflected Spectra and Albedos of Extrasolar Giant Planets. I. Clear
  and Cloudy Atmospheres}.
\newblock \emph{\apj}, 513\penalty0 (2):\penalty0 879--893, Mar. 1999.
\newblock \doi{10.1086/306881}.

\bibitem[{Matthias} et~al.(2000){Matthias}, {Fimbres}, {Sano}, {Post},
  {Accioly}, {Batchily}, and {Ferreira}]{Matthias2000}
A.~D. {Matthias}, A.~{Fimbres}, E.~E. {Sano}, D.~F. {Post}, L.~{Accioly}, A.~K.
  {Batchily}, and L.~G. {Ferreira}.
\newblock {Surface Roughness Effects on Soil Albedo}.
\newblock \emph{Soil Science Society of America Journal}, 64\penalty0
  (3):\penalty0 1035, Jan 2000.
\newblock \doi{10.2136/sssaj2000.6431035x}.

\bibitem[{McCoy} et~al.(1999){McCoy}, {Dickinson}, and {Lofgren}]{Mccoy1999}
T.~J. {McCoy}, T.~L. {Dickinson}, and G.~E. {Lofgren}.
\newblock {Partial melting of the Indarch (EH4) Meteorite: A textural, chemical
  and phase relations view of melting and melt migration}.
\newblock \emph{Meteoritics and Planetary Science}, 34\penalty0 (5):\penalty0
  735--746, Sep 1999.
\newblock \doi{10.1111/j.1945-5100.1999.tb01386.x}.

\bibitem[{Met Office}(2010)]{MetOffice2010}
{Met Office}.
\newblock {National Meteorological Library and Archive Fact Sheet 6 - The
  Beaufort Scale}, Mar 2010.
\newblock URL
  \url{https://www.metoffice.gov.uk/binaries/content/assets/metofficegovuk/pdf/research/library-and-archive/library/publications/factsheets/factsheet_6-the-beaufort-scale.pdf}.

\bibitem[{Miozzi} et~al.(2018){Miozzi}, {Morard}, {Antonangeli}, {Clark},
  {Mezouar}, {Dorn}, {Rozel}, and {Fiquet}]{Miozzi2018}
F.~{Miozzi}, G.~{Morard}, D.~{Antonangeli}, A.~N. {Clark}, M.~{Mezouar},
  C.~{Dorn}, A.~{Rozel}, and G.~{Fiquet}.
\newblock {Equation of State of SiC at Extreme Conditions: New Insight Into the
  Interior of Carbon-Rich Exoplanets}.
\newblock \emph{Journal of Geophysical Research (Planets)}, 123\penalty0
  (9):\penalty0 2295--2309, Sep 2018.
\newblock \doi{10.1029/2018JE005582}.

\bibitem[Modirrousta-Galian et~al.(2020{\natexlab{a}})Modirrousta-Galian,
  Locci, and Micela]{Modirrousta2020b}
D.~Modirrousta-Galian, D.~Locci, and G.~Micela.
\newblock The bimodal distribution in exoplanet radii: Considering varying core
  compositions and h2 envelope's sizes.
\newblock \emph{The Astrophysical Journal}, 891\penalty0 (2):\penalty0 158, mar
  2020{\natexlab{a}}.
\newblock \doi{10.3847/1538-4357/ab7379}.
\newblock URL \url{https://doi.org/10.3847%2F1538-4357%2Fab7379}.

\bibitem[Modirrousta-Galian et~al.(2020{\natexlab{b}})Modirrousta-Galian,
  Locci, Tinetti, and Micela]{Modirrousta2020}
D.~Modirrousta-Galian, D.~Locci, G.~Tinetti, and G.~Micela.
\newblock Hot super-earths with hydrogen atmospheres: A model explaining their
  paradoxical existence.
\newblock \emph{The Astrophysical Journal}, 888\penalty0 (2):\penalty0 87, jan
  2020{\natexlab{b}}.
\newblock \doi{10.3847/1538-4357/ab616b}.
\newblock URL \url{https://doi.org/10.3847%2F1538-4357%2Fab616b}.

\bibitem[{Moriarty} et~al.(2014){Moriarty}, {Madhusudhan}, and
  {Fischer}]{Moriarty2014}
J.~{Moriarty}, N.~{Madhusudhan}, and D.~{Fischer}.
\newblock {Chemistry in an Evolving Protoplanetary Disk: Effects on Terrestrial
  Planet Composition}.
\newblock \emph{\apj}, 787\penalty0 (1):\penalty0 81, May 2014.
\newblock \doi{10.1088/0004-637X/787/1/81}.

\bibitem[{Namur} et~al.(2016){Namur}, {Charlier}, {Holtz}, {Cartier}, and
  {McCammon}]{Namur2016}
O.~{Namur}, B.~{Charlier}, F.~{Holtz}, C.~{Cartier}, and C.~{McCammon}.
\newblock {Sulfur solubility in reduced mafic silicate melts: Implications for
  the speciation and distribution of sulfur on Mercury}.
\newblock \emph{Earth and Planetary Science Letters}, 448:\penalty0 102--114,
  Aug 2016.
\newblock \doi{10.1016/j.epsl.2016.05.024}.

\bibitem[{Nittler} and {Weider}(2019)]{Nittler2019}
L.~R. {Nittler} and S.~Z. {Weider}.
\newblock {The Surface Composition of Mercury}.
\newblock \emph{Elements}, 15\penalty0 (1):\penalty0 33--38, 02 2019.
\newblock ISSN 1811-5209.
\newblock \doi{10.2138/gselements.15.1.33}.
\newblock URL \url{https://doi.org/10.2138/gselements.15.1.33}.

\bibitem[{Nittler} et~al.(2011){Nittler}, {Starr}, {Weider}, {McCoy},
  {Boynton}, {Ebel}, {Ernst}, {Evans}, {Goldsten}, {Hamara}, {Lawrence},
  {McNutt}, {Schlemm}, {Solomon}, and {Sprague}]{Nittler2011}
L.~R. {Nittler}, R.~D. {Starr}, S.~Z. {Weider}, T.~J. {McCoy}, W.~V. {Boynton},
  D.~S. {Ebel}, C.~M. {Ernst}, L.~G. {Evans}, J.~O. {Goldsten}, D.~K. {Hamara},
  D.~J. {Lawrence}, R.~L. {McNutt}, C.~E. {Schlemm}, S.~C. {Solomon}, and A.~L.
  {Sprague}.
\newblock {The Major-Element Composition of Mercury{\textquoteright}s Surface
  from MESSENGER X-ray Spectrometry}.
\newblock \emph{Science}, 333\penalty0 (6051):\penalty0 1847, Sep 2011.
\newblock \doi{10.1126/science.1211567}.

\bibitem[{Oguntunde} et~al.(2006){Oguntunde}, {Ajayi}, and {van de
  Giesen}]{Oguntunde2006}
P.~{Oguntunde}, A.~{Ajayi}, and N.~{van de Giesen}.
\newblock {Tillage and surface moisture effects on bare-soil albedo of a
  tropical loamy sand}.
\newblock \emph{Soil and Tillage Research}, 85\penalty0 (1):\penalty0 107 --
  114, Jan 2006.
\newblock \doi{https://doi.org/10.1016/j.still.2004.12.009}.
\newblock URL
  \url{http://www.sciencedirect.com/science/article/pii/S0167198705000462}.

\bibitem[{O'Neill} and {Palme}(1998)]{Oneill1998}
H.~S.~C. {O'Neill} and H.~{Palme}.
\newblock {Composition of the silicate earth: implications for accretion of and
  core formation}.
\newblock In I.~{Jackson}, editor, \emph{The Earth's mantle; composition,
  structure, and evolution}, volume 264, pages 3--126, Cambridge, 1998.
  Cambridge University Press.

\bibitem[{Ordal} et~al.(1988){Ordal}, {Bell}, {Alexander}, {Newquist}, and
  {Querry}]{Ordal1988}
M.~A. {Ordal}, R.~J. {Bell}, J.~{Alexander}, Ralph~W., L.~A. {Newquist}, and
  M.~R. {Querry}.
\newblock {Optical properties of Al, Fe, Ti, Ta, W, and Mo at submillimeter
  wavelengths}.
\newblock \emph{\ao}, 27\penalty0 (6):\penalty0 1203--1209, Mar 1988.
\newblock \doi{10.1364/AO.27.001203}.

\bibitem[{Owen} and {Wu}(2013)]{Owen2013}
J.~E. {Owen} and Y.~{Wu}.
\newblock {Kepler Planets: A Tale of Evaporation}.
\newblock \emph{\apj}, 775:\penalty0 105, Oct. 2013.
\newblock \doi{10.1088/0004-637X/775/2/105}.

\bibitem[{Owen} and {Wu}(2017)]{Owen2017}
J.~E. {Owen} and Y.~{Wu}.
\newblock {The Evaporation Valley in the Kepler Planets}.
\newblock \emph{\apj}, 847:\penalty0 29, Sept. 2017.
\newblock \doi{10.3847/1538-4357/aa890a}.

\bibitem[{Papoular} and {Papoular}(2014)]{Papoular2014}
R.~J. {Papoular} and R.~{Papoular}.
\newblock {Some optical properties of graphite from IR to millimetric
  wavelengths}.
\newblock \emph{\mnras}, 443\penalty0 (4):\penalty0 2974--2982, Oct 2014.
\newblock \doi{10.1093/mnras/stu1348}.

\bibitem[{Philipp}(1977)]{Philipp1977}
H.~R. {Philipp}.
\newblock {Infrared optical properties of graphite}.
\newblock \emph{\prb}, 16\penalty0 (6):\penalty0 2896--2900, Sep 1977.
\newblock \doi{10.1103/PhysRevB.16.2896}.

\bibitem[{Pluriel} et~al.(2019){Pluriel}, {Marcq}, and {Turbet}]{Pluriel2019}
W.~{Pluriel}, E.~{Marcq}, and M.~{Turbet}.
\newblock {Modeling the albedo of Earth-like magma ocean planets with
  H$_{2}$O-CO$_{2}$ atmospheres}.
\newblock \emph{\icarus}, 317:\penalty0 583--590, Jan 2019.
\newblock \doi{10.1016/j.icarus.2018.08.023}.

\bibitem[{Pollack} et~al.(1973){Pollack}, {Toon}, and {Khare}]{Pollack1973}
J.~B. {Pollack}, O.~B. {Toon}, and B.~N. {Khare}.
\newblock {Optical properties of some terrestrial rocks and glasses}.
\newblock \emph{\icarus}, 19\penalty0 (3):\penalty0 372--389, Jul 1973.
\newblock \doi{10.1016/0019-1035(73)90115-2}.

\bibitem[Polyanskiy(2010)]{Polyanskiy2008}
M.~N. Polyanskiy.
\newblock Refractive index database, 2010.
\newblock URL \url{https://refractiveindex.info}.

\bibitem[{Posch} et~al.(2007){Posch}, {Baier}, {Mutschke}, and
  {Henning}]{Posch2007}
T.~{Posch}, A.~{Baier}, H.~{Mutschke}, and T.~{Henning}.
\newblock {Carbonates in Space: The Challenge of Low-Temperature Data}.
\newblock \emph{\apj}, 668\penalty0 (2):\penalty0 993--1000, Oct 2007.
\newblock \doi{10.1086/521390}.

\bibitem[{Querry}(1985)]{Querry1985}
M.~R. {Querry}.
\newblock {Optical constants}.
\newblock Technical report, Jun 1985.

\bibitem[{Querry}(1987)]{Querry1987}
M.~R. {Querry}.
\newblock \emph{{Optical Constants of Minerals and Other Materials from the
  Millimeter to the Ultraviolet}}.
\newblock PhD thesis, University of Missouri-Kansas City, 1987.

\bibitem[{Raki{\'c}} and {Majewski}(1996)]{Rakic1996}
A.~D. {Raki{\'c}} and M.~L. {Majewski}.
\newblock {Modeling the optical dielectric function of GaAs and AlAs: Extension
  of Adachi's model}.
\newblock \emph{Journal of Applied Physics}, 80\penalty0 (10):\penalty0
  5909--5914, Nov 1996.
\newblock \doi{10.1063/1.363586}.

\bibitem[Raki\'{c} et~al.(1998)Raki\'{c}, Djuri\v{s}i\'{c}, Elazar, and
  Majewski]{Rakic1998}
A.~D. Raki\'{c}, A.~B. Djuri\v{s}i\'{c}, J.~M. Elazar, and M.~L. Majewski.
\newblock Optical properties of metallic films for vertical-cavity
  optoelectronic devices.
\newblock \emph{Appl. Opt.}, 37\penalty0 (22):\penalty0 5271--5283, Aug 1998.
\newblock \doi{10.1364/AO.37.005271}.
\newblock URL \url{http://ao.osa.org/abstract.cfm?URI=ao-37-22-5271}.

\bibitem[{Ridden-Harper} et~al.(2016){Ridden-Harper}, {Snellen}, {Keller}, {de
  Kok}, {Di Gloria}, {Hoeijmakers}, {Brogi}, {Fridlund}, {Vermeersen}, and {van
  Westrenen}]{Ridden2016}
A.~R. {Ridden-Harper}, I.~A.~G. {Snellen}, C.~U. {Keller}, R.~J. {de Kok},
  E.~{Di Gloria}, H.~J. {Hoeijmakers}, M.~{Brogi}, M.~{Fridlund}, B.~L.~A.
  {Vermeersen}, and W.~{van Westrenen}.
\newblock {Search for an exosphere in sodium and calcium in the transmission
  spectrum of exoplanet 55 Cancri e}.
\newblock \emph{\aap}, 593:\penalty0 A129, Oct. 2016.
\newblock \doi{10.1051/0004-6361/201628448}.

\bibitem[{Riner} et~al.(2008){Riner}, {Bina}, {Robinson}, and
  {Desch}]{Riner2008}
M.~A. {Riner}, C.~R. {Bina}, M.~S. {Robinson}, and S.~J. {Desch}.
\newblock {Internal structure of Mercury: Implications of a molten core}.
\newblock \emph{Journal of Geophysical Research (Planets)}, 113\penalty0
  (E8):\penalty0 E08013, Aug 2008.
\newblock \doi{10.1029/2007JE002993}.

\bibitem[{Rouan} et~al.(2011){Rouan}, {Deeg}, {Demangeon}, {Samuel},
  {Cavarroc}, {Fegley}, and {L{\'e}ger}]{Rouan2011}
D.~{Rouan}, H.~J. {Deeg}, O.~{Demangeon}, B.~{Samuel}, C.~{Cavarroc},
  B.~{Fegley}, and A.~{L{\'e}ger}.
\newblock {The Orbital Phases and Secondary Transits of Kepler-10b. A Physical
  Interpretation Based on the Lava-ocean Planet Model}.
\newblock \emph{\apjl}, 741\penalty0 (2):\penalty0 L30, Nov 2011.
\newblock \doi{10.1088/2041-8205/741/2/L30}.

\bibitem[{Russell} et~al.(2004){Russell}, {Giordano}, and
  {Dingwell}]{Russell2004}
J.~K. {Russell}, D.~{Giordano}, and D.~B. {Dingwell}.
\newblock {High-temperature limits on viscosity of non-Arrhenian silicate
  melts}.
\newblock \emph{American Mineralogist}, 88\penalty0 (8-9):\penalty0 1390--1394,
  Aug. 2004.
\newblock \doi{10.2138/am-2003-8-924}.

\bibitem[{Schaefer} and {Fegley}(2009)]{Schaefer2009}
L.~{Schaefer} and B.~{Fegley}.
\newblock {Chemistry of Silicate Atmospheres of Evaporating Super-Earths}.
\newblock \emph{\apjl}, 703:\penalty0 L113--L117, Oct. 2009.
\newblock \doi{10.1088/0004-637X/703/2/L113}.

\bibitem[{Schaefer} et~al.(2012){Schaefer}, {Lodders}, and
  {Fegley}]{Schaefer2012}
L.~{Schaefer}, K.~{Lodders}, and B.~{Fegley}.
\newblock {Vaporization of the Earth: Application to Exoplanet Atmospheres}.
\newblock \emph{\apj}, 755:\penalty0 41, Aug. 2012.
\newblock \doi{10.1088/0004-637X/755/1/41}.

\bibitem[{Schlick}(1994)]{Schlick1994}
C.~{Schlick}.
\newblock An inexpensive brdf model for physically-based rendering.
\newblock \emph{Computer Graphics Forum}, 13\penalty0 (3):\penalty0 233--246,
  1994.
\newblock \doi{10.1111/1467-8659.1330233}.

\bibitem[Seeliger(1884)]{Seeliger1884}
H.~Seeliger.
\newblock Zur photometrie des saturnringes.
\newblock \emph{Astronomische Nachrichten}, 109\penalty0 (20):\penalty0
  305--314, 1884.
\newblock \doi{10.1002/asna.18841092002}.
\newblock URL
  \url{https://onlinelibrary.wiley.com/doi/abs/10.1002/asna.18841092002}.

\bibitem[{Shelberg} et~al.(1983){Shelberg}, {Lam}, and
  {Moellering}]{Shelberg1983}
M.~{Shelberg}, N.~{Lam}, and H.~{Moellering}.
\newblock Measuring the fractal dimensions of surfaces.
\newblock page~11, 10 1983.

\bibitem[{Shestopalov} et~al.(2013){Shestopalov}, {Golubeva}, and
  {Cloutis}]{Shestopalov2013}
D.~I. {Shestopalov}, L.~F. {Golubeva}, and E.~A. {Cloutis}.
\newblock {Optical maturation of asteroid surfaces}.
\newblock \emph{\icarus}, 225\penalty0 (1):\penalty0 781--793, Jul 2013.
\newblock \doi{10.1016/j.icarus.2013.05.002}.

\bibitem[Siefke et~al.(2016)Siefke, Kroker, Pfeiffer, Puffky, Dietrich, Franta,
  Ohlídal, Szeghalmi, Kley, and Tünnermann]{Siefke2016}
T.~Siefke, S.~Kroker, K.~Pfeiffer, O.~Puffky, K.~Dietrich, D.~Franta,
  I.~Ohlídal, A.~Szeghalmi, E.-B. Kley, and A.~Tünnermann.
\newblock Materials pushing the application limits of wire grid polarizers
  further into the deep ultraviolet spectral range.
\newblock \emph{Advanced Optical Materials}, 4\penalty0 (11):\penalty0
  1780--1786, 2016.
\newblock \doi{10.1002/adom.201600250}.
\newblock URL
  \url{https://onlinelibrary.wiley.com/doi/abs/10.1002/adom.201600250}.

\bibitem[{Stephens} and {Malitson}(1952)]{Stephens1952}
R.~E. {Stephens} and I.~H. {Malitson}.
\newblock {Index of Refraction of Magnesium Oxide}.
\newblock \emph{Journal of Research of the National Bureau of Standards},
  49\penalty0 (4):\penalty0 249--252, Oct 1952.

\bibitem[{Stiassnie}(1991)]{Stiassnie1991}
M.~{Stiassnie}.
\newblock The fractal dimension of the ocean surface.
\newblock pages 633--647, 01 1991.

\bibitem[{Stiassnie} et~al.(1991){Stiassnie}, {Agnon}, and
  {Shemer}]{Stiassnie1991b}
M.~{Stiassnie}, Y.~{Agnon}, and L.~{Shemer}.
\newblock {Fractal dimensions of random water surfaces}.
\newblock \emph{Physica D Nonlinear Phenomena}, 47\penalty0 (3):\penalty0
  341--352, Jan 1991.
\newblock \doi{10.1016/0167-2789(91)90034-7}.

\bibitem[{Swain} et~al.(2019){Swain}, {Estrela}, {Sotin}, {Roudier}, and
  {Zellem}]{Swain2019}
M.~R. {Swain}, R.~{Estrela}, C.~{Sotin}, G.~M. {Roudier}, and R.~T. {Zellem}.
\newblock {Two Terrestrial Planet Families with Different Origins}.
\newblock \emph{\apj}, 881\penalty0 (2):\penalty0 117, Aug 2019.
\newblock \doi{10.3847/1538-4357/ab2714}.

\bibitem[{Taft} and {Philipp}(1965)]{Taft1965}
E.~A. {Taft} and H.~R. {Philipp}.
\newblock {Optical Properties of Graphite}.
\newblock \emph{Physical Review}, 138\penalty0 (1A):\penalty0 197--202, Apr
  1965.
\newblock \doi{10.1103/PhysRev.138.A197}.

\bibitem[Tammann and Hesse(1926)]{Tammann1926}
G.~Tammann and W.~Hesse.
\newblock Die abhängigkeit der viscosität von der temperatur bie
  unterkühlten flüssigkeiten.
\newblock \emph{Zeitschrift für anorganische und allgemeine Chemie},
  156\penalty0 (1):\penalty0 245--257, 1926.
\newblock \doi{10.1002/zaac.19261560121}.
\newblock URL
  \url{https://onlinelibrary.wiley.com/doi/abs/10.1002/zaac.19261560121}.

\bibitem[{Tsiaras} et~al.(2016){Tsiaras}, {Rocchetto}, {Waldmann}, {Venot},
  {Varley}, {Morello}, {Damiano}, {Tinetti}, {Barton}, {Yurchenko}, and
  {Tennyson}]{Tsiaras2016}
A.~{Tsiaras}, M.~{Rocchetto}, I.~P. {Waldmann}, O.~{Venot}, R.~{Varley},
  G.~{Morello}, M.~{Damiano}, G.~{Tinetti}, E.~J. {Barton}, S.~N. {Yurchenko},
  and J.~{Tennyson}.
\newblock {Detection of an Atmosphere Around the Super-Earth 55 Cancri e}.
\newblock \emph{\apj}, 820:\penalty0 99, Apr. 2016.
\newblock \doi{10.3847/0004-637X/820/2/99}.

\bibitem[{Valencia} et~al.(2013){Valencia}, {Guillot}, {Parmentier}, and
  {Freedman}]{Valencia2013}
D.~{Valencia}, T.~{Guillot}, V.~{Parmentier}, and R.~S. {Freedman}.
\newblock {Bulk Composition of GJ 1214b and Other Sub-Neptune Exoplanets}.
\newblock \emph{\apj}, 775:\penalty0 10, Sept. 2013.
\newblock \doi{10.1088/0004-637X/775/1/10}.

\bibitem[Viswanath and Natarajan(1989)]{Viswanath1989}
D.~Viswanath and G.~Natarajan.
\newblock \emph{Data Book on the Viscosity of Liquids}.
\newblock Hemisphere Publishing Corporation, 1989.
\newblock ISBN 0-89116-778-1.

\bibitem[{Vogel}(1921)]{Vogel1921}
H.~{Vogel}.
\newblock {Das Temperaturabhängigkeitsgesetz der Viskosität von
  Flüssigkeiten}.
\newblock \emph{Physikalische Zeitschrift}, 22\penalty0 (28):\penalty0
  645--646, 1921.

\bibitem[{Wang} et~al.(2013){Wang}, {Zhan}, {Wang}, {Xuan}, {Zhang}, {Liu},
  {Xu}, {Liu}, {Wei}, and {Chen}]{Wang2013}
S.~{Wang}, M.~{Zhan}, G.~{Wang}, H.~{Xuan}, W.~{Zhang}, C.~{Liu}, C.~{Xu},
  Y.~{Liu}, Z.~{Wei}, and X.~{Chen}.
\newblock {4H-SiC: a new nonlinear material for midinfrared lasers}.
\newblock \emph{Laser \& Photonics Review}, 7\penalty0 (5):\penalty0 831--838,
  Sep 2013.
\newblock \doi{10.1002/lpor.201300068}.

\bibitem[{Wasson} and {Kallemeyn}(1988)]{Wasson1988}
J.~T. {Wasson} and G.~W. {Kallemeyn}.
\newblock {Compositions of Chondrites}.
\newblock \emph{Philosophical Transactions of the Royal Society of London
  Series A}, 325\penalty0 (1587):\penalty0 535--544, Jul 1988.
\newblock \doi{10.1098/rsta.1988.0066}.

\bibitem[{Weider} et~al.(2014){Weider}, {Nittler}, {Starr}, {McCoy}, and
  {Solomon}]{Weider2014}
S.~Z. {Weider}, L.~R. {Nittler}, R.~D. {Starr}, T.~J. {McCoy}, and S.~C.
  {Solomon}.
\newblock {Variations in the abundance of iron on Mercury's surface from
  MESSENGER X-Ray Spectrometer observations}.
\newblock \emph{\icarus}, 235:\penalty0 170--186, Jun 2014.
\newblock \doi{10.1016/j.icarus.2014.03.002}.

\bibitem[{Weiss} et~al.(2016){Weiss}, {Rogers}, {Isaacson}, {Agol}, {Marcy},
  {Rowe}, {Kipping}, {Fulton}, {Lissauer}, and {Howard}]{Weiss2016}
L.~M. {Weiss}, L.~A. {Rogers}, H.~T. {Isaacson}, E.~{Agol}, G.~W. {Marcy},
  J.~F. {Rowe}, D.~{Kipping}, B.~J. {Fulton}, J.~J. {Lissauer}, and A.~W.
  {Howard}.
\newblock {Revised Masses and Densities of the Planets around Kepler-10}.
\newblock \emph{\apj}, 819\penalty0 (1):\penalty0 83, Mar 2016.
\newblock \doi{10.3847/0004-637X/819/1/83}.

\bibitem[{Woitke} et~al.(2018){Woitke}, {Helling}, {Hunter}, {Millard},
  {Turner}, {Worters}, {Blecic}, and {Stock}]{Woitke2018}
P.~{Woitke}, C.~{Helling}, G.~H. {Hunter}, J.~D. {Millard}, G.~E. {Turner},
  M.~{Worters}, J.~{Blecic}, and J.~W. {Stock}.
\newblock {Equilibrium chemistry down to 100 K. Impact of silicates and
  phyllosilicates on the carbon to oxygen ratio}.
\newblock \emph{\aap}, 614:\penalty0 A1, Jun 2018.
\newblock \doi{10.1051/0004-6361/201732193}.

\bibitem[Wood and Chan(1994)]{Wood1994}
A.~T.~A. Wood and G.~Chan.
\newblock Simulation of stationary gaussian processes in [0,1]d.
\newblock \emph{Journal of Computational and Graphical Statistics}, 3\penalty0
  (4):\penalty0 409--432, 1994.

\bibitem[{Zahnle} and {Kasting}(1986)]{Zahnle1986}
K.~J. {Zahnle} and J.~F. {Kasting}.
\newblock {Mass fractionation during transonic escape and implications for loss
  of water from Mars and Venus}.
\newblock \emph{\icarus}, 68\penalty0 (3):\penalty0 462--480, Dec 1986.
\newblock \doi{10.1016/0019-1035(86)90051-5}.

\bibitem[{Zeidler} et~al.(2011){Zeidler}, {Posch}, {Mutschke}, {Richter}, and
  {Wehrhan}]{Zeidler2011}
S.~{Zeidler}, T.~{Posch}, H.~{Mutschke}, H.~{Richter}, and O.~{Wehrhan}.
\newblock {Near-infrared absorption properties of oxygen-rich stardust analogs.
  The influence of coloring metal ions}.
\newblock \emph{\aap}, 526:\penalty0 A68, Feb 2011.
\newblock \doi{10.1051/0004-6361/201015219}.

\bibitem[{Zeng} and {Sasselov}(2013)]{Zeng2013}
L.~{Zeng} and D.~{Sasselov}.
\newblock {A Detailed Model Grid for Solid Planets from 0.1 through 100 Earth
  Masses}.
\newblock \emph{\pasp}, 125\penalty0 (925):\penalty0 227, Mar 2013.
\newblock \doi{10.1086/669163}.

\bibitem[{Zeng} et~al.(2016){Zeng}, {Sasselov}, and {Jacobsen}]{Zeng2016}
L.~{Zeng}, D.~D. {Sasselov}, and S.~B. {Jacobsen}.
\newblock {Mass-Radius Relation for Rocky Planets Based on PREM}.
\newblock \emph{\apj}, 819:\penalty0 127, Mar. 2016.
\newblock \doi{10.3847/0004-637X/819/2/127}.

\bibitem[{Zeng} et~al.(2018){Zeng}, {Jacobsen}, {Sasselov}, {Vand erburg},
  {Lopez-Morales}, {Perez-Mercader}, {Petaev}, {Mattsson}, {Li}, and
  {Matthew}]{Zeng2018}
L.~{Zeng}, S.~B. {Jacobsen}, D.~D. {Sasselov}, A.~M. {Vand erburg},
  M.~{Lopez-Morales}, J.~{Perez-Mercader}, M.~I. {Petaev}, T.~R. {Mattsson},
  G.~{Li}, and H.~{Matthew}.
\newblock {Growth Model Interpretation of Planet Size Distribution}.
\newblock In \emph{AGU Fall Meeting Abstracts}, volume 2018, pages P53C--2985,
  Dec 2018.

\end{thebibliography}
\bibliographystyle{abbrvnat}

\appendix

\section{A Comparison of Schlick's Approximation With Other Models}
\label{sec:comparison_of_models}

\begin{figure}[ht]
  		\centering
   		\includegraphics[scale=0.55]{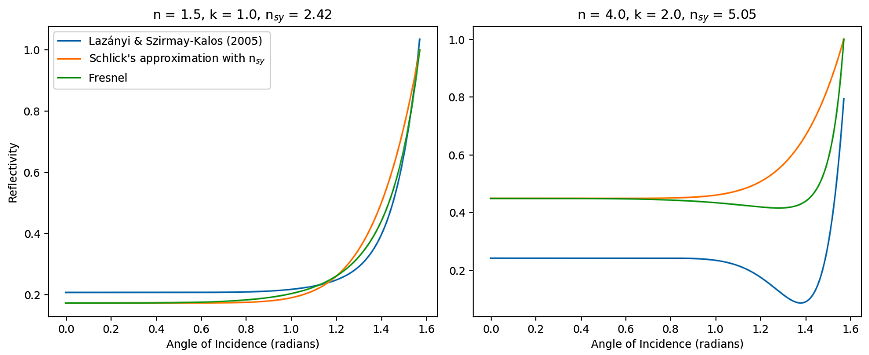}
   		\caption{A comparison of the model by \citet{Lazanyi2005}, Schlick's approximation with $n_{sy}$, and the Fresnel equations.}
   		\centering
\label{fig:Correction}
\end{figure}

Schlick's approximation with a correction for the complex index of refraction matches the Fresnel equations well. This approach has been adopted by other models such as \citet{Lazanyi2005}, who use an \textit{ad hoc} correction to account for the behaviour of the complex refractive index. In spite of this, their model is only applicable for low $n$ and $k$ values. In our model we created $n_{sy}$ in order to account for the effects of the imaginary refractive index (i.e. $k$) whilst in their paper they added a term to Schlick's approximation which includes $k$. Having said this, we decided not to use the model by \citet{Lazanyi2005} as it overestimates the reflectance at small angles of incidence and it becomes inaccurate at high indices of refraction (see Fig.~\ref{fig:Correction}). In addition, their model has \textit{ad-hoc} corrections whilst ours is Schlick's approximation with an implementation of the imaginary index of refraction. Our method is slightly more computationally intensive than the one proposed by \citet{Lazanyi2005} but we believe that it is the appropriate balance of accuracy and efficiency for our purposes.

Another commonly used model for planetary photometry is the Lommel-Seeliger Law \citep{Seeliger1884}. This is a simple analytical approach that well approximates diffuse reflection. Unfortunately, this model fails for spherical albedo values larger than $\simeq 0.20$ \citep{Fairbairn2005}, which falls short of the remit of our study. Another alternative model is the Hapke model \citep{Hapke2012}, which is a semi-empirical analytical approach for estimating the spherical albedo of airless bodies. We decided not to use this method as it is adapted for regolith surfaces such as asteroids, but not fluids like magmas.

In any case, even for unusually high indices of refraction our approach closely matches the results predicted from the Fresnel equations (see Fig.~\ref{fig:Correction}) for most angles of incidence. Only at very high angles of incidence do the deviations become noticeable. To further emphasise the suitability of our approach, we ran a ray tracing simulation with $n_{sy} = 6$ and $H = 0.01$ (very rough) with Schlick's approximation and the Fresnel equations to give a spherical albedo of 0.495 and 0.480 respectively. As one can see, the two models provide almost identical results. We are aware that the Fresnel equations are more accurate but we had to make this compromise in order to keep the simulations inexpensive.

\section{Derivation of the Synthetic Refractive Index}
\label{sec:derivation_n_sy}

\begin{subequations}
From classical electromagnetic theory \citep{Abraham1950}, the reflectance of a system in contact with a vacuum (or an optically thin medium) is:
\begin{equation}
\label{Eq:nkR}
    R = \left(\frac{ \left|n - 1 \right|^{2} + \left|k \right|^{2} }{ \left|n + 1 \right|^{2} + \left|k \right|^{2} } \right)
\end{equation}
Let us now define the synthetic refractive index, $n_{sy}$, as an index which is real and able to reproduce an identical reflectance:
\begin{equation}
\label{Eq:nsyR}
    R = \left(\frac{ n_{sy} - 1 }{ n_{sy} + 1 } \right)^{2}
\end{equation}
Since we have defined $n_{sy}$ as a real index that is able to produce the same reflectance, we can equate Eq.~\ref{Eq:nkR} with Eq.~\ref{Eq:nsyR}:
\begin{equation}
    \left(\frac{ n_{sy} - 1 }{ n_{sy} + 1 } \right)^{2} = \left(\frac{ \left|n - 1 \right|^{2} + \left|k \right|^{2} }{ \left|n + 1 \right|^{2} + \left|k \right|^{2} } \right)
\end{equation}
Solving for $n_{sy}$ gives:
\begin{equation}
    n_{sy} = \frac{ \left(\frac{ \left|n - 1 \right|^{2} + \left|k \right|^{2} }{ \left|n + 1 \right|^{2} + \left|k \right|^{2} } \right)^{1/2} + 1 }{ 1 - \left(\frac{ \left|n - 1 \right|^{2} + \left|k \right|^{2} }{ \left|n + 1 \right|^{2} + \left|k \right|^{2} } \right)^{1/2} }
\end{equation}
\end{subequations}

\section{Estimating the Hurst Exponent of Kepler-10b}
\label{sec:hurst}

\begin{table}
	\centering
	\caption{Assumptions used to calculate the Hurst exponent of Kepler-10b}
	\label{tab:Hurstcalculation}
	\begin{tabular}{P{6cm}P{3cm}P{7cm}} 
		\hline
		\hline                     
		Variable & Approximation & Explanation \\
		\hline
		Day-side temperature ($T_{day}$) &$\sim\rm 2750 \: K$& Sub-stellar point temperature for a blackbody.\\
		Mineral atmosphere wind speeds ($v_{atm}$) & $\rm \sim \rm 100 \: ms^{-1} $ & This is hard to model and it is not within the objectives of our study so for simplicity we will adopt the values used by \citet{Kite2016}. \\
		Mineral atmosphere pressure ($P_{atm}$) & $\rm \sim \rm 7000 \: Pa $ & We used the data from \citet{Ito2015} with a best-fit equation of the form given by \citet{Fegley2009}. \\
		Density of magma ($\rho_{l}$) & $\rm \sim 2000 \: kg  \:  m^{-3}$ & We adopt the density of an Earth-like rhyolitic magma as at the high temperatures found on Kepler-10b's day-side, relatively low magma densities would be expected. \\
		\hline  
	\end{tabular}
\end{table}
Using the ideal gas equation in conjunction with the data from Table~\ref{tab:Hurstcalculation}, the density of the atmosphere can be estimated at $\rm \sim 6 \times 10^{-3} \: kg \: m^{-3}$. Furthermore, using Eq.~1 from \citet{Kite2016}, the diameter of the magma ocean is calculated to be approximately $\rm L\approx 3 \times 10^{7} \: m$. With this information we can now estimate the atmospheric-ocean energy transfer \citep{Lorenz2012,Kite2016}:
\begin{equation}
E \sim \frac{1}{2} C_{d} \rho_{atm} \varepsilon v^{2}_{atm} L
\label{eq:atmosphereocean}
\end{equation}
Where $C_{d} \sim 2\times10^{-3}$ is the equivalent of a drag coefficient \citep{Emanuel1994} and $\varepsilon\sim0.01$ is an efficiency factor for winds near the ocean surface \citep{Kite2016}. Inputting the above mentioned data into Eq.~\ref{eq:atmosphereocean} gives $\rm E\sim2 \times 10^{4} \: J \: m^{-2}$. Finally, we can use Eq.~2 from \citet{Lorenz2012} to estimate the wave height:
\begin{equation}
E \sim \frac{1}{8} \rho_{l} g_{P} h^{2}
\label{eq:waveheight}
\end{equation}
Where $g_{P}$ is the gravitational acceleration and $h$ is the wave height. We know that the wave height (peak to trough) is twice the RMS height, and we also know how the RMS is related to the Hurst exponent which is shown in Eq.~\ref{eq:DurstH2}. Therefore, with a few trivial calculations the Hurst exponent can be estimated to be approximately $H\sim0.8$.

\section{spherical albedo Values of Different Materials}
\label{sec:spectral_albedo_materials}

Here we show the spherical albedo values for different materials and a constant Hurst exponent of 0.8:
\begin{figure}[!htb]
	\centering
	\includegraphics[scale=0.5]{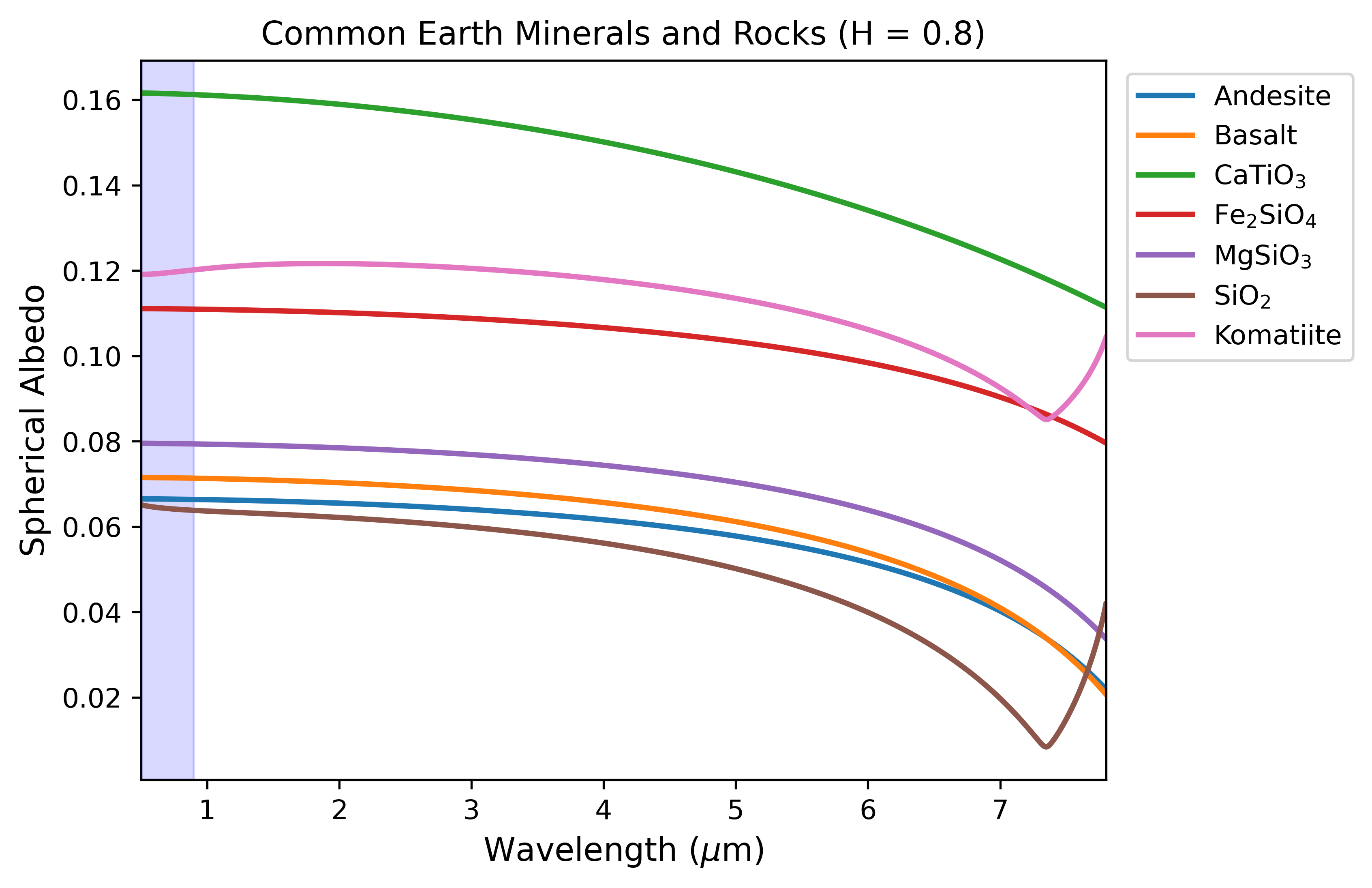}
	\includegraphics[scale=0.5]{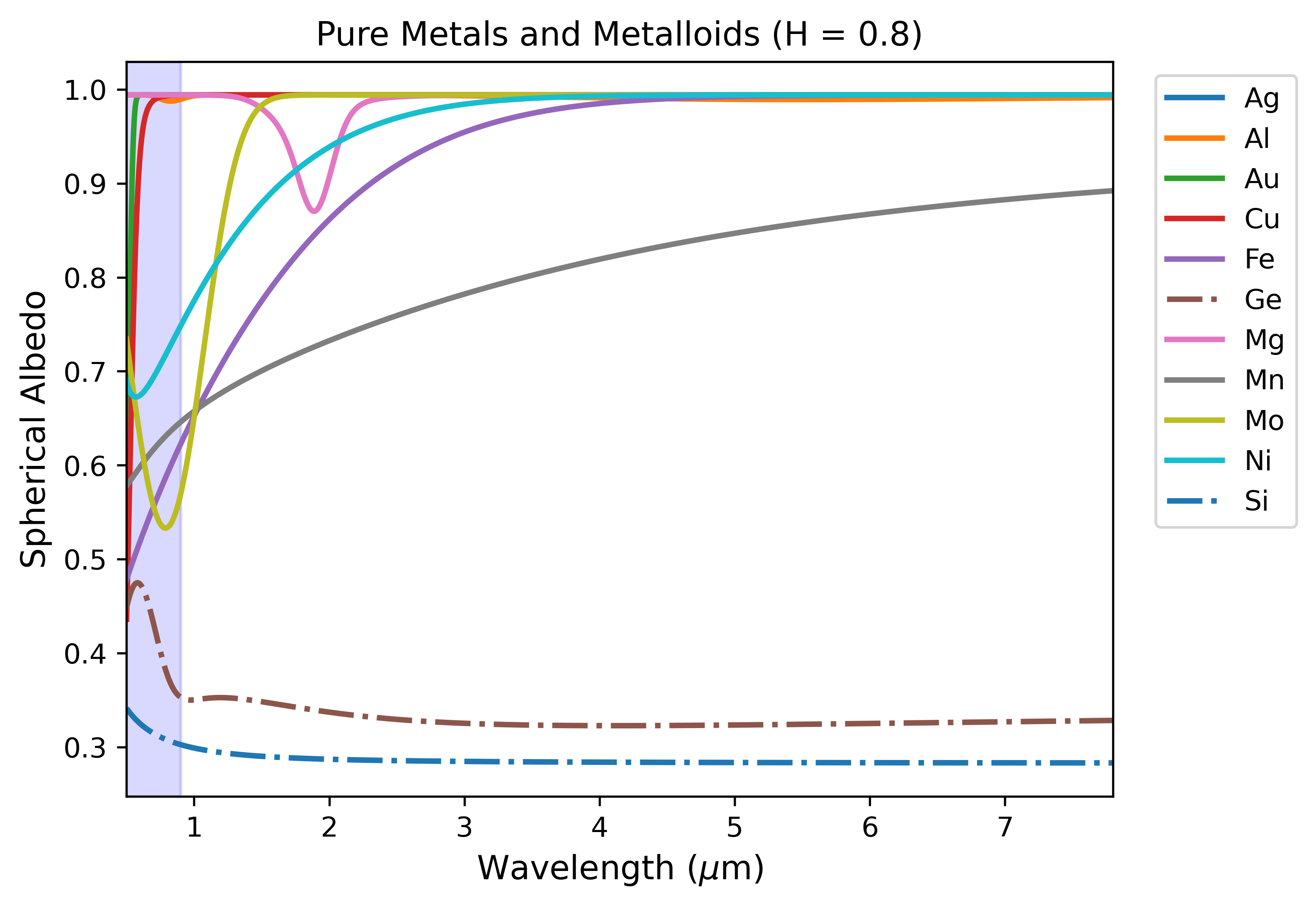}
	\includegraphics[scale=0.5]{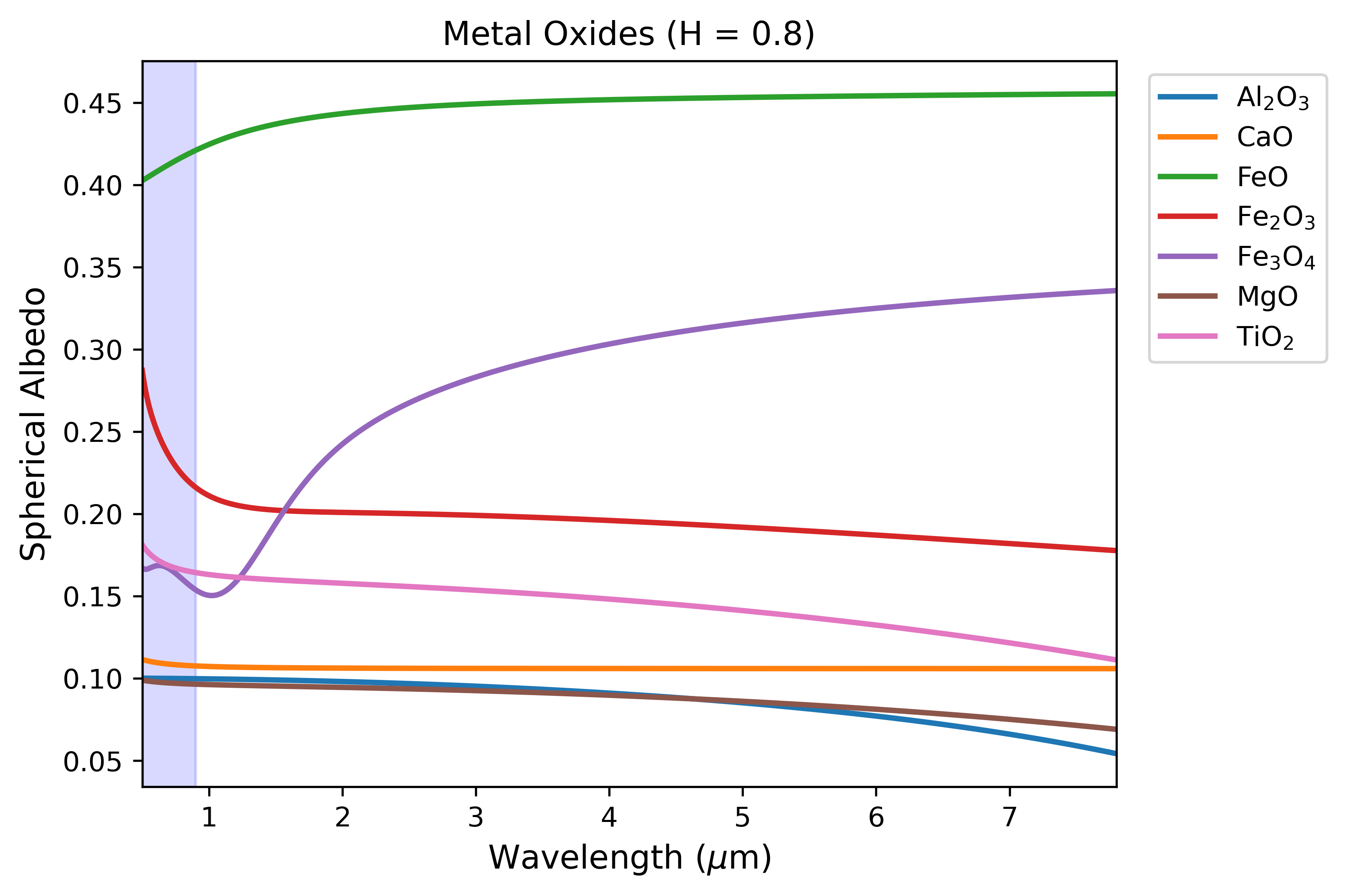}
	\includegraphics[scale=0.5]{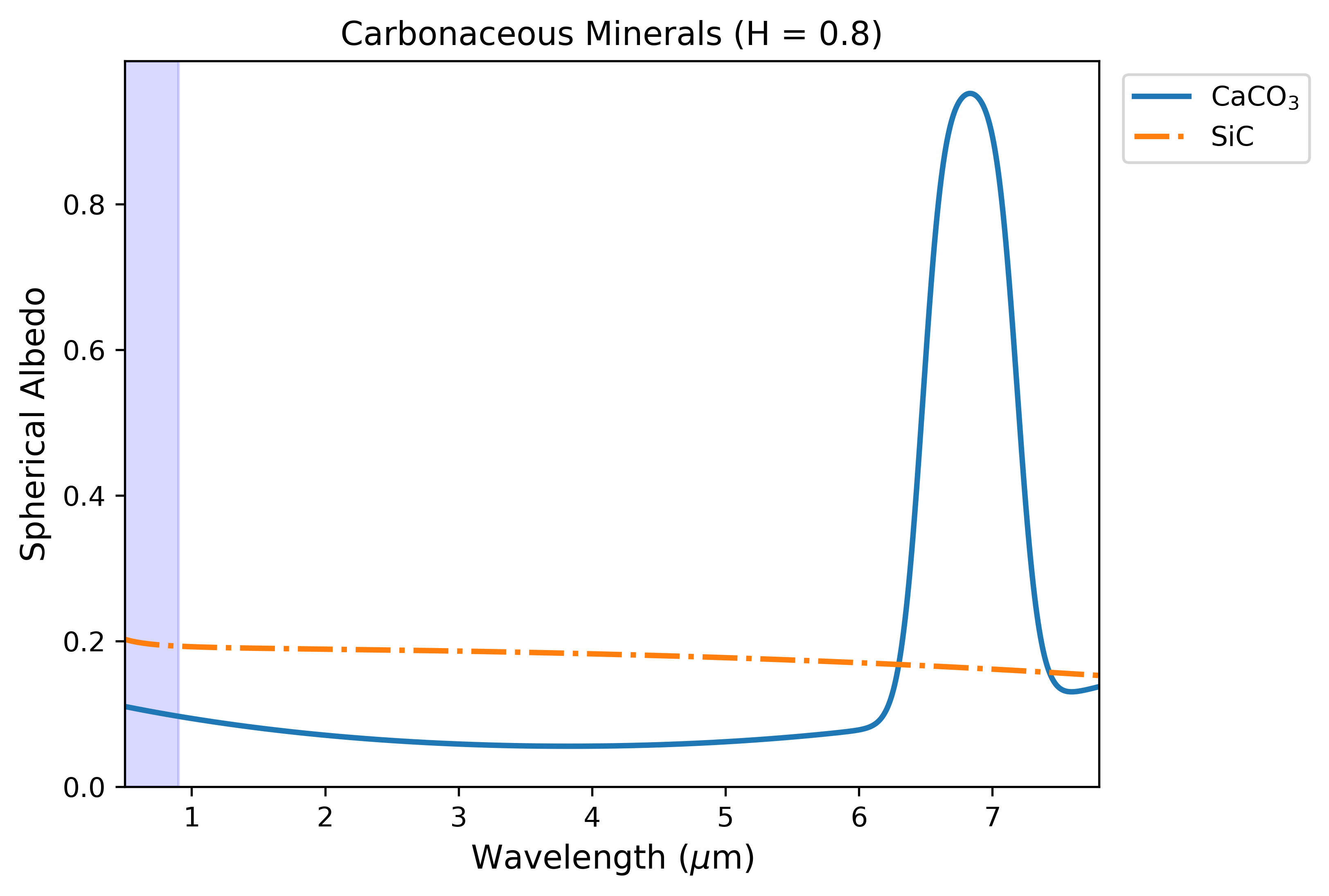}
	\caption{The spherical albedo values of common Earth minerals and rocks, metals and metalloids, metal oxides, and two carbonaceous minerals, for a Hurst exponent of 0.8. The light-blue section is \textit{Kepler's} band-pass. This figure was generated with Eq.~(\ref{eq:A_S}), (\ref{eq:Albedo}), (\ref{eq:psiconstant}), (\ref{eq:phase_function}), and (\ref{eq:phase_function_constants}).}
	\centering
	\label{fig:materials}
\end{figure}

\section{List of Equations for the Different Materials}
\label{sec:listofequations}

\begin{longtable}{P{2.5cm}P{10cm}P{5cm}}
	\caption{Best-fit equations for the relationship between the refractive index and the wavelength of light. For komatiite (shown in Fig.~\ref{fig:materials}) we used 9:4:2 of $\rm SiO_{2}$, $\rm MgO$ and $\rm FeO$ \citep{Malyuk1986,Dostal2004}.}\label{tab:bestfitequations}\\
	\hline
	\hline                     
	Material & Best-fit Equation for $\rm n, k$ or $\rm n_{sy}$ & Reference \\
	\hline
	\endhead
	\hline
	\endfoot
	Ag (n) & $ 0.0387594 - 0.0191537\lambda + 0.0560141\lambda^{2} - 0.000858264\lambda^{3} $ & \citet{Babar2015} \\
	Ag (k) & $-0.564253 + 7.70979\lambda -0.050914\lambda^{2} $ & \citet{Babar2015} \\
	Al (n) & $ 0.312451 + \frac{33.7164\lambda^{2}}{66.9468 + \lambda^{2}} + \lambda^{6.49463 - 12.476\lambda^{1.51686}}$ & \citet{Ordal1988} \\
	Al (k) & $ 5.13215 + \frac{9.6749\lambda^{2}}{1.7134 + \lambda^{2}} + \lambda^{1.4507 - 2.77918\lambda^{-0.854905}} $ & \citet{Ordal1988} \\
	$\rm Al_{2}O_{3}$ (n) & $\left(3.086+\frac{4.90423\lambda^{2}}{\lambda^{2}-307.228}\right)^{1/2}$ & \citet{Querry1987} \\
	$\rm Al_{2}O_{3}$ (k) & $\left(0.000319568-\frac{0.00116775\lambda^{2}}{\lambda^{2}-90.2538}\right)^{1/2}$ & \citet{Querry1987} \\
	Andesite ($\rm n_{sy}$) & $\left( 2.1743 + \frac{0.648257\lambda^{2}}{\lambda^{2} - 103.233}\right)^{1/2} $ & \citet{Pollack1973} \\
	Au (n) & $(0.040181 + 0.010122\lambda + 0.0525666\lambda^{2} + 0.000539972\lambda^{3} + \frac{0.000157248}{\lambda^{12.1332}}$ & \citet{Babar2015} \\
	Au (k) & $\left(-10.9924 + 5.09542\lambda + 46.1863\lambda^{2}\right)^{1/2} $ & \citet{Babar2015} \\
	Basalt ($\rm n_{sy}$) & $\left( 2.298649 - \frac{0.00148953\lambda^{2}}{\lambda^{2} + 4.37075} + \frac{0.841871\lambda^{2}}{\lambda^{2} - 108.512}\right)^{1/2} $ & \citet{Pollack1973} \\
	$\rm CaCO_{3}$ ($\rm n_{sy}$) & $1.38352e^{-\frac{\left(\lambda-3.82577\right)^{2}}{-38.7232}}+23.3249e^{-\frac{\left(\lambda-6.83304\right)^{2}}{0.0826662}}$ & \citet{Ghosh1999,Posch2007} \\
	CaO ($\rm n_{sy}$) & $ \left( 1 + \frac{2.260277\lambda^{2}}{\lambda^{2} - 0.01804729} \right)^{1/2} $ & \citet{Liu1966} \\
	$\rm CaTiO_{3}$ ($\rm n_{sy}$) & $ \left( 5.29944 + \frac{51.8911\lambda^{2}}{\lambda^{2} - 1752.85} \right)^{1/2} $ & \citet{Zeidler2011} \\
	Cu (n) & $ \left(0.0933843 + \frac{1098.64 \lambda^{2}}{1727.94 + \lambda^{2}} + \frac{-20.7799 \lambda^{2}}{37.8 + \lambda^{2}} + \frac{0.000030363}{\lambda^{15.601}}\right)^{1/2}  $ & \citet{Babar2015} \\
	Cu (k) & $ \left( -5.80213 - \frac{13691.8\lambda^{2}}{-560.774 + \lambda^{2}} + \frac{1097.43 \lambda^{2}}{66.0935 + \lambda^{2}} \right)^{1/2} $ & \citet{Babar2015} \\
	Fe (n) & $\left( 5.8345 + \frac{10.4512\lambda^{2}}{\lambda^{2} + 1.62692} \frac{221.262\lambda^{2}}{\lambda^{2} + 580.508} \right)^{1/2}$ & \citet{Querry1987} \\
	Fe (k) & $\left(1.09676 + 11.6663\lambda + 5.19826\lambda^{2} + 0.256431\lambda^{3} \right)^{1/2}$ & \citet{Querry1987} \\
	FeO (n) & $\left( 5.4211 + \frac{177291\lambda^{2}}{\lambda^{2} + 9.98168} - \frac{177290\lambda^{2}}{\lambda^{2} + 9.9821}\right)^{1/2}$ & \citet{Henning1995} \\
	FeO (k) & $\left(5.02746 + \frac{2.38467\lambda^{2}}{0.988753 \lambda^{2}} \right)^{1/2} $ & \citet{Henning1995} \\
	$\rm Fe_{2}O_{3}$ (n) & $\left( 24.0927 - \frac{42.0941\lambda^{2}}{0.23055+\lambda^{2}} - \frac{4.66637\lambda^{2}}{142.27+\lambda^{2}} + \frac{25.4114\lambda^{2}}{0.426954+\lambda^{2}} \right)^{1/2}$ & \citet{Querry1987} \\
	$\rm Fe_{2}O_{3}$ (k) & $ 1.81355 - \frac{1.7828}{\lambda^{0.00612776}} + \frac{0.00612776}{\lambda^{6.74851}} $ & \citet{Querry1987} \\
	$\rm Fe_{3}O_{4}$ (n) & $\left(191.094-\frac{564.131}{\lambda}+\frac{646.954}{\lambda^{2}}-\frac{308.721}{\lambda^{3}}+\frac{52.8087}{\lambda^{4}}\right)^{0.263078}$ & \citet{Querry1987} \\
	$\rm Fe_{3}O_{4}$ (k) & $\left(0.56873 + \frac{1.141}{\lambda} - \frac{0.933493}{\lambda^{2}} - \frac{4.37513}{\lambda^{3}} + \frac{8.59397}{\lambda^{4}} \right)^{-0.713551}$ & \citet{Querry1987} \\
	$\rm Fe_{2}SiO_{4}$ ($\rm n_{sy}$) & $\left(3.42567+\frac{0.981637\lambda^{2}}{\lambda^{2}-125.692}\right)^{1/2}$ & \citet{Fabian2001} \\
	GaAs (n) & $ \left(14.2857 + \frac{1.31107\lambda^{2}}{\lambda^{2} - 0.19523} + \frac{4266.07\lambda^{2}}{\lambda^{2} + 1.03145} + \frac{-4270.94\lambda^{2}}{\lambda^{2} + 1.03024} \right)^{1/2} $ & \citet{Rakic1996} \\
	GaAs (k) & $10^{\left(-4.13226 + \frac{2.44252}{\lambda^{0.678051}}\right) }$ & \citet{Rakic1996} \\
	Ge (n) & $ 9.27088 - \frac{1}{0.342728\lambda} - \frac{1}{24.1123\lambda^{2}} - \frac{5.00253\lambda^{2}}{\lambda^{2} + 1.41971} $ & \citet{Amotchkina2020} \\
	Ge (k) & $ \frac{1.12327}{(1.01186 - 0.0522531\lambda + 0.0452528\lambda^{2})^{290.1425}}  $ & \citet{Amotchkina2020} \\
	Mg (n) & $ \left|-1.48883 + 2.57035\lambda + 0.213623\lambda^{-9.93907\lambda^{2} + 27.2749\lambda - 10.8071} \right| $ & \citet{Hagemann1975} \\
	Mg (k) & $ \left|-3.08158 + 1.74441\lambda + 11.9179\lambda^{-0.000808172\lambda^{13.8785} - 542.183\lambda^{0.000433637} + 542.183} \right.$ $\left. + \frac{745.947\lambda^{2}}{1.994 + \lambda^{2}} - \frac{717.65\lambda^{2}}{1.87182+\lambda^{2}}\right| $ & \citet{Hagemann1975} \\
	MgO ($\rm n_{sy}$) & $\left(2.956362 + \frac{0.02195770}{\lambda^{2} - 0.01428322} - 0.01062387\lambda^{2} - 0.0000204968\lambda^{4} \right)^{1/2}$ & \citet{Stephens1952} \\
	$\rm MgSiO_{3}$ ($\rm n_{sy}$) & $ 1.58257 + \frac{0.219036\lambda^{2}}{\lambda^{2} - 96.5198} $ & \citet{Jaeger1998} \\
	Mn (n) & $\left(-1.41011 + \frac{41.9909\lambda^{2}}{\lambda^{2} + 2.52168} + \frac{281.743\lambda^{2}}{\lambda^{2} + 134.924}\right)^{1/2} $ & \citet{Querry1987} \\
	Mn (k) & $\left(-0.499692 + \frac{95.9185\lambda^{2}}{47.4233 + \lambda^{2}} + \frac{26.3224\lambda^{2}}{0.460497 + \lambda^{2}} \right)^{1/2}$ & \citet{Querry1987} \\
	Mo (n) & $\left(-74264.2 + \frac{74262\lambda^{2}}{-0.000113513 + \lambda^{2}} + \frac{-40.6174\lambda^{2}}{-102.076 + \lambda^{2}}\right)^{1/2} $ & \citet{Ordal1988} \\
	Mo (k) & $\left(-32.634 + \frac{38213\lambda^{2}}{1092.74 + \lambda^{2}} + \frac{13.6911}{\lambda^{2}}\right)^{1/2}$ & \citet{Ordal1988} \\
	Ni (n) & $\left(-36.2204 + \frac{23.8522\lambda^{2}}{7.4471 + \lambda^{2}} - \frac{9.67652\lambda^{2}}{\lambda^{2} - 90.5159} + \frac{42.2902\lambda^{2}}{0.0310738 + \lambda^{2}}\right)^{1/2}$ & \citet{Rakic1998} \\
	Ni (k) & $ 1.74046 + 3.38213\lambda $ & \citet{Rakic1998} \\
	Si ($\rm n_{sy}$) & $ \left( 1381.62 - \frac{1369.35\lambda^{2}}{\lambda^{2} + 0.000919293}\right)^{1/2} $ & \citet{Li1980} \\
	SiC ($\rm n_{sy}$) & $\left(1 + \frac{0.20075\lambda^{2}}{\lambda^{2} + 12.07224} + \frac{5.54861\lambda^{2}}{\lambda^{2} - 0.02641} + \frac{35.65066\lambda^{2}}{\lambda^{2} - 1268.24708} \right)^{1/2}$ & \citet{Wang2013} \\
	$\rm SiO_{2}$ (n) & $\left( 1 + \frac{0.6961663\lambda^{2}}{\lambda^{2}-0.0684043^2}+\frac{0.4079426\lambda^{2}}{\lambda^{2}-0.1162414^2}+\frac{0.8974794\lambda^{2}}{\lambda^{2}-9.896161^2} \right)^{1/2} $ & \citet{Kischkat2012} \\
	$\rm SiO_{2}$ (k) & $ 0.00813575 - 0.00164212\lambda - \frac{0.00216415\lambda^{2}}{\lambda^{2} - 62.6425} $ & \citet{Kischkat2012} \\
	$\rm TiO_{2}$ ($\rm n_{sy}$) & $\left(3.36314 + \frac{1.85104\lambda^{2}}{\lambda^{2}-0.0852835}-\frac{78.7336\lambda^{2}}{2617.1 + \lambda^{2}}\right)^{1/2}$ & \citet{Siefke2016} \\
\end{longtable}

\section{Atmospheric Effects}
\label{sec:Atmosphere}

\begin{figure}[h]
	\centering
	\includegraphics[scale=0.9]{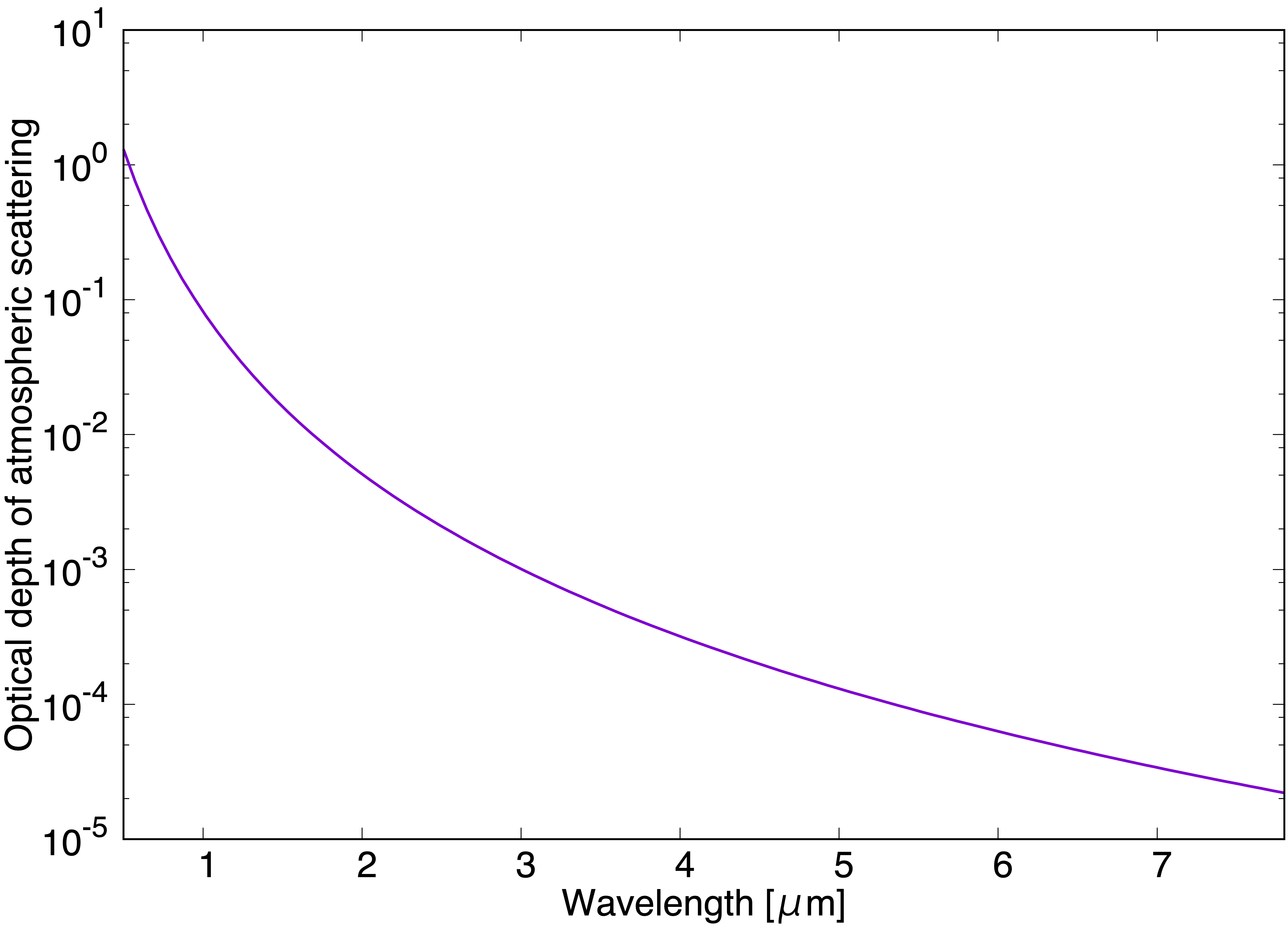}
	\caption{The vertical optical depth of Rayleigh scattering as a function of wavelength for a pure Na atmosphere on Kepler-10b. The optical depth is derived assuming a pressure of 7$\times$10$^4$~dyne/cm$^2$, 1700~cm/s$^2$ as the surface gravity, and Na as the sole atmospheric constituent.}
	\centering
	\label{fig:AS}
\end{figure}

\end{document}